\documentclass[12pt]{article}
\usepackage{amsmath}
\usepackage{amsthm}
\usepackage{mathtools}
\usepackage{graphicx}
\usepackage{gensymb}
\usepackage{url} % not crucial - just used below for the URL 
\usepackage{mwe}
\usepackage{soul}
\usepackage{comment}
\usepackage{algorithm}
\usepackage{enumerate}
\usepackage{algcompatible}
\usepackage{tikz}
\usepackage[font=small]{caption}
\usepackage{caption}
\usepackage{amssymb}
\usepackage{capt-of}
 \PassOptionsToPackage{hyphens}{url}\usepackage{hyperref}
\usepackage[noabbrev]{cleveref}
\usepackage{caption}
\usepackage{silence}
\usepackage{natbib}
\usepackage{bm}
\usepackage{comment}
\usepackage[normalem]{ulem}
 \usepackage{relsize}
 \usepackage{amsmath}
\newcommand\norm[1]{\left\lVert#1\right\rVert}
\tikzset{cross/.style={cross out, draw=black, minimum size=2*(#1-\pgflinewidth), inner sep=0pt, outer sep=0pt},
cross/.default={1pt}}
\DeclareMathOperator*{\argmax}{arg\,max}
\DeclareMathOperator*{\argmin}{arg\,min}
\providecommand{\abs}[1]{\lvert#1\rvert}
\renewcommand{\d}[1]{\ensuremath{\operatorname{d}\!{#1}}}
\newcommand{\defeq}{\mathrel{\mathop:}=}
\newcommand{\ra}{\right \rangle}
\newcommand{\la}{\left \langle}
\newcommand\smallO{
  \mathchoice
    {{\scriptstyle\mathcal{O}}}% \displaystyle
    {{\scriptstyle\mathcal{O}}}% \textstyle
    {{\scriptscriptstyle\mathcal{O}}}% \scriptstyle
    {\scalebox{.7}{$\scriptscriptstyle\mathcal{O}$}}%\scriptscriptstyle
  }

\newcommand{\beginsupplement}{%
        \setcounter{table}{0}
        \renewcommand{\thetable}{S\arabic{table}}%
        \setcounter{section}{0}%
        \renewcommand{\thesection}{S\arabic{section}}
        \setcounter{subsection}{0}%
        \renewcommand{\thesubsection}{S\arabic{subsection}}
        \setcounter{figure}{0}
        \renewcommand{\thefigure}{S\arabic{figure}}%
        \setcounter{equation}{0}%
        \renewcommand{\theequation}{S\arabic{equation}}
        \setcounter{algorithm}{0}%
        \renewcommand{\thealgorithm}{S\arabic{algorithm}}
     }

\newcommand\remove{\bgroup\markoverwith{\textcolor{orange}{\rule[.5ex]{2pt}{1pt}}}\ULon}
\newcommand{\slfrac}[2]{\left.#1\middle/#2\right.}

\newcommand{\blind}{0}

\begin{document}

\def\spacingset#1{\renewcommand{\baselinestretch}%
{#1}\small\normalsize} \spacingset{1}

%%%%%%%%%%%%%%%%%%%%%%%%%%%%%%%%%%%%%%%%%%%%%%%%%%%%%%%%%%%%%%%%%%%%%%%%%%%%%%

\if0\blind
{
  \title{\vspace{-2cm}\bf Neural Conditional Simulation for Complex Spatial Processes}
  \author{Julia Walchessen
    %\hspace{.2cm}\\
    \\Department of Statistics, North Carolina State University\\
    \\
    Andrew Zammit-Mangion\\
    School of Mathematics and Statistics, University of New South Wales Sydney\\
    \\
    Rapha\"{e}l Huser\\
    Statistics Program,  King Abdullah University of Science and Technology\\
    \\
    Mikael Kuusela\\
    Department of Statistics and Data Science, Carnegie Mellon University
    }
    \date{}
  \maketitle
} \fi

\if1\blind
{
  \begin{center}
    {\bf Neural Conditional Simulation for Complex Spatial Processes}
\end{center}
} \fi

\begin{abstract}
A key objective in spatial statistics is to simulate from the distribution of a spatial process at a selection of unobserved locations conditional on observations (i.e., a predictive distribution) to enable spatial prediction and uncertainty quantification. However, exact conditional simulation from this predictive distribution is intractable or inefficient for many spatial process models. In this paper, we propose neural conditional simulation (NCS), a general method for spatial conditional simulation that is based on neural diffusion models. Specifically, using spatial masks, we implement a conditional score-based diffusion model that evolves Gaussian noise into samples from a predictive distribution when given a partially observed spatial field and spatial process parameters as inputs. The diffusion model relies on a neural network that only requires unconditional samples from the spatial process for training. Once trained, the diffusion model is amortized with respect to the observations in the partially observed field, the number and locations of those observations, and the spatial process parameters, and can therefore be used to conditionally simulate from a broad class of predictive distributions without retraining the neural network. We assess the NCS-generated simulations against simulations from the true conditional distribution of a Gaussian process model, and against Markov chain Monte Carlo (MCMC) simulations from a Brown--Resnick process model for spatial extremes. In the latter case, we show that it is more efficient and accurate to conditionally simulate using NCS than classical MCMC techniques implemented in standard software. We also demonstrate the use of NCS to predict spatial extremes in the Red Sea. We conclude that NCS enables efficient and accurate conditional simulation from spatial predictive distributions that are challenging to sample from using traditional methods.
\end{abstract}

\noindent%
\vfill

\newpage
\spacingset{2}
\allowdisplaybreaks
\section{Introduction}
\label{sec:intro}

Spatial process models are widely used across various disciplines, including environmental science, epidemiology, and social science. However, their practical implementation comes with several challenges, particularly in computation: fitting a spatial process model to data and using it to generate predictions and prediction intervals can be computationally demanding. While numerous approximations exist for Gaussian spatial process models when exact methods are unavailable, options remain more limited for complex non-Gaussian spatial models.

In recent years, neural networks have been increasingly used to tackle computational challenges in the first step of data analysis: model fitting through parameter inference \citep{Louppe2016, Cramner, Dalmasso2020, Dalmasso2024}. These techniques have been applied in various ways to the spatial setting. For instance, \citet{Walchessen2024} leverage neural networks to efficiently evaluate the likelihood function of a spatial process model, while \citet{Lenzi2021} and \citet{SD2024,SD2025} employ neural networks as Bayes estimators for parameter estimation. Both approaches rely on simulation-based inference (SBI) paired with machine learning---the practice of utilizing simulations from a statistical model to train a machine learning model for statistical inference. These methods preclude the need for an explicitly known or computationally tractable likelihood function \citep{Cramner, Dalmasso2020}. In contrast, less work has focused on the subsequent---yet equally important---spatial inference task: making predictions and quantifying uncertainty at unobserved locations when the predictive distributions are not available in closed form. Existing methods, which typically do not use SBI and machine learning, include Markov chain Monte Carlo (MCMC) \citep{Gelfand} and variational inference \citep{Jordan1999, Banerjee}. However, these approaches often involve a trade-off between computational efficiency and statistical accuracy and usually become impractical when dealing with a large number of observations or prediction locations.

Here, we propose a new method using SBI to train a neural generative model that can simulate from predictive distributions in a spatial setting. Specifically, we employ state-of-the-art diffusion models that are used for image generation \citep{Yang2024}, because regularly gridded spatial data can effectively be viewed as image data. These models leverage both forward diffusion and backward evolution to generate samples from a target distribution in a computationally efficient manner. In the forward diffusion process, samples from a complex target distribution are progressively transformed into samples from a simpler reference distribution. This forward process, described by a stochastic differential equation (SDE), is designed to ensure the existence of a corresponding reverse SDE that evolves samples from the reference distribution back into samples from the target distribution \citep{Song2021}. This reverse SDE is characterized by a combination of user-defined terms from the forward SDE and an unknown score function, which is approximated using a neural network trained via a specifically designed loss function. Once the neural network is trained, a sample from the complex target distribution can be drawn efficiently by simulating from the reference distribution and applying the learned reverse SDE.

There are several approaches to incorporating conditional information into a diffusion process, many of which originate from the inpainting problem, a topic extensively studied in computer vision \citep[e.g.,][]{Lugmayr2022, Quan2024, Simkus2025}. Generally, in image inpainting, the objective is to produce realistic, yet diverse images from a partial image. These approaches usually make no assumptions about an underlying statistical model and, consequently, are not designed to approximate any true predictive distributions. In contrast, \citet{cardoso2025} adopt an approach that first trains a diffusion model to unconditionally sample from a spatial process and then performs conditional simulation from the predictive distribution via training-free guidance. However, this approach has not been applied to non-Gaussian processes and relies on several approximations during conditional simulation, which can lead to slower and potentially less accurate generation. A few inpainting methods, such as twisted diffusion \citep{Wu2023} and controllable generation \citep{Song2021}, have asymptotic guarantees for exact conditional simulation. Yet, these methods have not thus far been validated in a statistically rigorous manner and may not scale easily to high-dimensional data. 

A recent contribution by \citet{Gloeckler2024} provides a masking approach to incorporate conditional information into the score-based diffusion process that is designed to accurately simulate from conditional distributions of parameters and data including posterior distributions. However, their approach was not designed for high-dimensional conditioning, including spatial prediction and uncertainty quantification. A more explicit masking approach for handling missing or censored data is also well-established for estimating model parameters in SBI. \citet{Wang2024}, \citet{Richards2024}, and \citet{SD2025} include a mask labeling missing or censored data as input to the neural network for parameter inference tasks.

Here, we combine the masking approach of \citet{Gloeckler2024} with the more explicit masking approach of \citet{Richards2024} for the spatial process model setting. The result is a conditional score-based diffusion model within the SDE framework which can quickly and accurately conditionally simulate from high-dimensional predictive distributions. Specifically, we train a neural network that approximates the time-evolving conditional score function, taking as inputs the observed locations, the observed values, the spatial process model parameters, and the diffusion timestep. Importantly, training the neural network only requires \emph{unconditional} simulation from the spatial process model, which is feasible for a broad class of models. Once trained, the neural network can be used to evaluate the conditional score function in the conditional reverse process, enabling conditional sampling for any observations---regardless of value, number, and location of these observations---and any model parameters without retraining the neural network. Our approach to conditional simulation, which we refer to as \emph{neural conditional simulation} (NCS), is said to be \emph{amortized} with respect to these quantities.

We carry out experiments comparing NCS to classical methods for conditional simulation with two spatial processes: (i) a Gaussian process and (ii) a Brown--Resnick process \citep{BrownResnick1977, Kabluchko2009}, which is a non-Gaussian max-stable process commonly used to model spatial extremes \citep{Huser2014, Engelke2015, Davison2015}. In the former case, exact conditional simulation is both tractable and computationally efficient; this case study serves to validate NCS and verify that it correctly samples from the true predictive distribution. For the latter case, exact conditional simulation is computationally and analytically intractable in moderate to high dimensions \citep{Dombry2012}; the standard approximation involving Markov Chain Monte Carlo (MCMC) has been implemented in the \texttt{SpatialExtremes} package \citep{SpatialExtremes}. We show that NCS is both more accurate and computationally efficient than this standard approximation across a range of observed patterns and spatial process model parameters. From these case studies, we conclude that NCS is a viable methodology for sampling from intractable predictive distributions of spatial processes with both high accuracy and computational efficiency.

The paper is organized as follows. Section~\ref{sec:method} provides an overview of score-based diffusion models for generative modeling, and describes how they can be adapted for conditional simulation of spatial processes. In Section~\ref{sec:casestudies}, we evaluate NCS on simulations from a Brown--Resnick spatial process and show that the NCS-generated simulations are representative of the true predictive distributions via a variety of validation metrics. In Section~\ref{sec:da}, we apply our method to predict spatial extremes in the Red Sea. Finally, in Section~\ref{sec:discuss}, we discuss the limitations of NCS as well as possible extensions.

\section{Methodology}
\label{sec:method}
\subsection{General Setting}
Consider a finite-variance spatial process $h(\cdot)$ dependent on parameter $\bm{\theta}$ and defined on a spatial domain $\mathcal{D}\subset \mathbb{R}^{d}$ (typically $d=2,3)$. Denote by $\mathcal{S}\defeq \{\bm{s}_i\in \mathcal{D} : i \in [n]\}$ a collection of $n$ locations, where $[n]$ is shorthand notation for the set $\{1,\dots,n\}$. Further, let $\mathring{\mathcal{S}}\subset \mathcal{S}$  be the set of spatial locations at which we observe $h(\cdot)$, and let $\mathcal{\tilde{S}} \defeq\mathcal{S}\setminus \mathring{\mathcal{S}}$ be the set of spatial locations at which we wish to predict $h(\cdot)$. The task of conditional simulation can be summarized as follows:
\begin{subequations}
\begin{align}
\text{Process:} &~h(\cdot) \sim \textrm{SpatialProcess}(\bm{\theta}), \label{eqn:simprob1}\\
\text{Observe:} &~\bm{\mathring{x}}_{0} := h(\mathring{\mathcal{S}}), \label{eqn:simprob2}\\
\text{Conditionally simulate:} &~\tilde{\bm{x}}_{0} \sim p(\tilde{\bm{x}}_{0} \mid \bm{\mathring{x}}_{0}; \bm{\theta}), \label{eqn:simprob3}
\end{align}
\end{subequations}
where the target conditional distribution $p(\tilde{\bm{x}}_{0} \mid \bm{\mathring{x}}_{0};\bm{\theta})$ in \eqref{eqn:simprob3}, which we also refer to as the predictive distribution, is induced by the stochastic process $h(\cdot)$. Note that there are multiple predictive distributions because $\bm{\theta}$, $\mathring{\mathcal{S}}$, and $\bm{\mathring{x}}_{0}$ can vary.
 We assume $\bm{\theta}$ is known when simulating from a predictive distribution; in practice, one often replaces $\bm{\theta}$ with an estimate $\bm{\hat{\theta}}$. We assume that, for any $\bm{\theta}$ and $\mathring{\mathcal{S}}$, it is straightforward to unconditionally simulate observations from the spatial process according to \eqref{eqn:simprob2} in which $\mathring{\mathcal{S}}=\mathcal{S}$. Note that although we focus on spatial processes, our methodology can be used for any continuous stochastic process defined on a fixed, finite, Euclidean domain.

In Section~\ref{sec:unconditionalsde}, we describe score-based diffusion models for simulating unconditionally from $h(\mathcal{S})$; in Section~\ref{sec:conditionalsde}, we modify the score-based diffusion model to conditionally simulate from \eqref{eqn:simprob3} for any $\mathring{\mathcal {S}}\subset {\mathcal S}$. In Section~\ref{sec:discrete}, we describe the discretizations of the forward and reverse SDEs characterizing the conditional score-based diffusion. In Section~\ref{sec:nnarchitecture}, we describe the neural network architecture we use for approximating the conditional score function of a spatial process. Finally, in Section~\ref{sec:validation}, we discuss validation methods. For ease of exposition, we omit the dependence of the spatial process and the induced quantities on the parameter~$\bm {\theta}$ in the remainder of Section~\ref{sec:method}.

\subsection{Unconditional Score-Based Diffusion}
\label{sec:unconditionalsde}
Here, we first consider the case in which we want to sample $\bm{x}_{0} := h(\mathcal{S}) \sim p(\bm{x}_{0})$, ${\bm{x}_0 \in \mathbb{R}^{\abs{\mathcal{S}}}}$, where $\abs{\mathcal{S}}=n$ is the number of locations, via an unconditional score-based diffusion model. In score-based diffusion, one first defines an SDE that diffuses $\bm{x}_0$ into a sample $\bm{x}_{T}\sim p(\bm{x}_{T})$, where $p(\bm{x}_{T})$ closely approximates a fixed, predetermined reference distribution $p_{\textrm{ref}}(\bm{x})$ selected for ease of sampling \citep{Pedrotti2024}. Provided the reverse process exists, one can then approximately sample from $p(\bm{x}_{0})$ by first sampling $\bm{x}_{T}\sim p_{\textrm{ref}}(\bm{x})$ and then evolving the sample through the corresponding reverse SDE \citep{Song2021}. 

In our approach, we specify the forward process as a sequence of Gaussian perturbations, such that the reference distribution $p_{\textrm{ref}}(\bm{x})$ is Gaussian. One can verify that, as $T\to \infty$, the distribution $p(\bm{x}_{T})$ converges to the reference distribution. The forward SDE we use has the following form, with $t\in [0,T]$ denoting time and $T$ the time horizon: 
\begin{subequations}
\begin{align}
\text{Initial condition:} &~\bm{x}_{0} \sim p(\bm{x}_{0}),  \\
\text{Forward evolution:} &~\mathrm{d}\bm{x}_{t} = \bm{f}(\bm{x}_{t},t)\mathrm{d}t + g(t)\mathrm{d}\bm{B}_{t},\quad t\in [0,T],
\end{align}
\label{eq:forwardsde}
\end{subequations} where $\bm{B}_{t}$ is an $n$-dimensional Brownian motion, and where the drift and diffusion coefficients $\bm{f}(\cdot,\cdot):\mathbb{R}^{\abs{\mathcal{S}}}\times [0,T]\to \mathbb{R}^{\abs{\mathcal{S}}}$ and $g(\cdot):[0,T]\to \mathbb{R}^{+}$ are user-specified and fixed. Note that time here is artificial and distinct from any temporal dimension in the variables being diffused. The reverse SDE is also a diffusion process given by 
\begin{subequations}
\begin{align}
\text{Initial condition:} &~\bm{x}_{T} \sim p_{\textrm{ref}}(\bm{x}),\\
\text{Reverse evolution:} &~\mathrm{d}\bm{x}_{t} = \big(\bm{f}(\bm{x}_{t},t)-g(t)^{2} \nabla_{\bm{x}_{t}} \log(p(\bm{x}_{t}))\big) \mathrm{d} t + g(t)\mathrm{d} \bm{R}_{t}, \quad t\in [0,T], \label{eqn:reverseevolution}
\end{align}
\label{eq:reversesde}
\end{subequations} where $\bm{R}_{t}$ is an $n$-dimensional reverse Brownian motion (i.e., a Brownian motion that evolves backwards in time) and $p(\bm{x}_{t})$ is the distribution of the samples $\bm{x}_{t}$ from the perturbed target distribution at time $t$ \citep{Anderson1982}. Note that the drift coefficient in the reverse SDE differs from that in the forward SDE as it includes the \emph{score function} $\nabla_{\bm{x}_{t}} \log(p(\bm{x}_{t}))$ evaluated at the sample $\bm{x}_{t}$, for time $t\in [0,T]$. If the target distribution $p(\bm{x}_{0})$ is non-Gaussian, $p(\bm{x}_{t})$ cannot be adequately approximated by a Gaussian density for $t$ sufficiently smaller than $T$, and hence the score function is generally  analytically or computationally intractable.

Indirect simulation from $p(\bm{x}_{0})$ via the reverse process is possible by replacing the score function in \eqref{eqn:reverseevolution} with an accurate surrogate $\bm{s}_{\bm{\phi}}(\bm{x}_{t},t)$---a neural network with trainable parameters $\bm{\phi}$ in this case. One way to ensure the surrogate sufficiently approximates the score function for all $\bm{x}_{t}\in \mathbb{R}^{\abs{\mathcal{S}}}$ and $t\in [0,T]$ is by score matching, which involves solving the following minimization problem \citep{Song2021}:
\begin{equation}
\begin{aligned}
\bm{\phi}^{*}& = \argmin_{\bm{\phi}} \mathbb{E}_{t\sim U(0,T)}\Big(\lambda(t) \mathbb{E}_{\bm{x}_{t}} \big[ \norm{\bm{s}_{\bm{\phi}}(\bm{x}_{t},t)-\nabla_{\bm{x}_{t}} \log(p(\bm{x}_{t}))}_{2}^{2}\big]\Big) \\ 
& = \argmin_{\bm{\phi}} \mathbb{E}_{t\sim U(0,T)}\Big(\lambda(t) \mathbb{E}_{\bm{x}_{0}} \mathbb{E}_{\bm{x}_{t}\mid \bm{x}_{0}} \big[ \norm{\bm{s}_{\bm{\phi}}(\bm{x}_{t},t)-\nabla_{\bm{x}_{t}} \log(p(\bm{x}_{t}\mid \bm{x}_{0}))}_{2}^{2}\big]\Big),
\end{aligned}
\label{eqn:unconditionalsdelossfunction}
\end{equation} where  $\lambda:[0,T]\to \mathbb{R}^{+}$ is a time-varying weighting function, and the loss function includes an expectation over time to ensure the surrogate $\bm{s}_{\bm{\phi}}(\bm{x}_{t},\cdot)$ learns the score function for any $t \in (0, T)$ in the reverse SDE. For completeness, we provide in the Supplementary Material, Section~\ref{supp:supplossfunction}, a proof of the second equality in \eqref{eqn:unconditionalsdelossfunction} derived from that in \citet{Vincent2011}.

Since the forward SDE \eqref{eq:forwardsde} only involves perturbations of $\bm{x}_{0}$ with Gaussian noise, the transition kernel $p(\bm{x}_{t} \mid \bm{x}_{0})$ is Gaussian provided that the forward drift coefficient $\bm{f}(\bm{x}_{t},t)$ is an affine function of $\bm{x}_{t}$ for all $t\in [0,T]$. Consequently, the score function of the transition kernel $\nabla_{\bm{x}_{t}} \log(p(\bm{x}_{t} \mid \bm{x}_{0}))$ can be easily computed, allowing us to learn the unconditional score $\nabla_{\bm{x}_{t}} \log p(\bm{x}_{t})$ in \eqref{eqn:unconditionalsdelossfunction} \citep{Song2021}. Further, for affine $f(\cdot,t)$, as $T\to \infty$, $p(\bm{x}_{T})$ approaches a multivariate Gaussian distribution, which we can use as a reference distribution. In this paper, we specifically use a variance-preserving SDE (VPSDE) in which the forward process employs an affine drift coefficient to diffuse $\bm{x}_{0}$ into $\bm{x}_{T}~\dot{\sim}~\mathcal{N}(\bm{0}, \bm{I})$ (where $\dot{\sim}$ denotes ``approximately distributed as''), and we therefore set $p_{\textrm{ref}}(\bm{x}):=\mathcal{N}(\bm{0},\bm{I})$ such that $\bm{I}$ is an $n\times n$ identity matrix. See Section~\ref{sec:discrete} for more details.

\subsection{Conditional Score-Based Diffusion}
\label{sec:conditionalsde}
In this section, we modify the unconditional score-based diffusion approach of Section~\ref{sec:unconditionalsde} to generate samples from the predictive distribution at $\tilde{\mathcal{S}}$ using observations at $\mathring{\mathcal{S}}$, such that $\mathcal{S} = {\tilde{\mathcal{S}}} \cup {\mathring{\mathcal{S}}}$ and $\tilde{\mathcal{S}}\cap \mathring{\mathcal{S}}=\emptyset$. We adapt the approach of \citet{Gloeckler2024} to allow conditional simulation from predictive distributions involving arbitrary partitionings of $\mathcal{S}$ in a spatial setting. We modify the forward and reverse SDEs in \eqref{eq:forwardsde} and \eqref{eq:reversesde}, respectively, so that only $\tilde{\bm{x}}_{t}$, the unobserved component of $\bm{x}_{t}=(\tilde{\bm{x}}'_{t},\bm{\mathring{x}}'_{t})'$, is diffused. Specifically, we use the following conditional forward SDE: \begin{subequations}
\begin{align}
\text{Initial condition:}&~\tilde{\bm{x}}_{0} \sim p(\tilde{\bm{x}}_{0} \mid \bm{\mathring{x}}_{0}), \quad \bm{\mathring{x}}_{0} \defeq h(\mathring{\mathcal{S}}), \label{eq:conditionalforwardfirst}\\
\text{Forward evolution:}&~\mathrm{d} \tilde{\bm{x}}_{t}=
\tilde{\bm{f}}(\tilde{\bm{x}}_{t},t) \mathrm{d} t + g(t)\mathrm{d}\tilde{\bm{B}}_{t}, \quad t\in [0,T], \label{eq:conditionalforwardsecond}
\end{align}
\label{eq:conditionalforwardsde}
\end{subequations} and the following conditional reverse-time SDE: \begin{subequations}
\begin{align}
\text{Initial condition:} &~\tilde{\bm{x}}_{T}\sim p_{\textrm{ref}}(\tilde{\bm{x}}), \quad \bm{\mathring{x}}_{0} := h(\mathring{\mathcal{S}}), \label{eq:condreverseinitial}\\
\text{Reverse evolution:} &~\mathrm{d} \tilde{\bm{x}}_{t} =
\big(\tilde{\bm{f}}(\tilde{\bm{x}}_{t},t)-g(t)^{2} \nabla_{\tilde{\bm{x}}_{t}} \log(p(\tilde{\bm{x}}_{t} \mid \bm{\mathring{x}}_{0}))\big)\mathrm{d} t +
g(t)\mathrm{d}\tilde{\bm{R}}_{t}, \quad t \in [0,T], \label{eq:conditionalreversesde}
\end{align}
\end{subequations} where $\tilde{\bm{f}}(\cdot,\cdot)$ is the component of $\bm{f}(\cdot,\cdot)$ corresponding to $\tilde{\mathcal{S}}$, and where $\tilde{\bm{B}}_{t}$ and $\tilde{\bm{R}}_{t}$ are forward- and reverse-time Brownian motions defined on $\mathbb{R}^{\abs{\tilde{\mathcal{S}}}}$, respectively. Note that there are generally different conditional forward and reverse SDEs for different initial conditions, $\mathring{\mathcal{S}}$ and $\bm{\mathring{x}}_{0}$, in \eqref{eq:conditionalforwardfirst} and \eqref{eq:condreverseinitial}.

Consider a distribution over the observed locations, $p(\mathring{\mathcal{S}})$, that has positive mass over the set of all possible partitions of $\mathcal{S}$. By the average optimality property for non-negative loss functions \citep{Zammit2024}, the modified score matching algorithm is given by:
\small
\begin{subequations}
\begin{align}
\bm{\phi}^{*}& = \argmin_{\bm{\phi}} \mathlarger{\mathbb{E}_{t\sim U(0,T)}} \Big( \lambda(t) \mathlarger{\mathbb{E}_{\mathring{\mathcal{S}}}} \mathlarger{\mathbb{E}_{\bm{\mathring{x}}_{0}}} \mathlarger{\mathbb{E}_{\tilde{\bm{x}}_{t}\mid \bm{\mathring{x}}_{0}}} \Big[ \norm{\tilde{\bm{s}}_{\bm{\phi}}\big({\bm{x}}_{t},\mathring{\mathcal{S}},t\big)-\nabla_{\tilde{\bm{x}}_{t}} \log(p(\tilde{\bm{x}}_{t} \mid \bm{\mathring{x}}_{0}))}_{2}^{2}\Big]\Big) \notag \\ 
& = \argmin_{\bm{\phi}} \mathlarger{\mathbb{E}_{t\sim U(0,T)}}\Big( \lambda(t) \mathlarger{\mathbb{E}_{\mathring{\mathcal{S}}}} \mathlarger{\mathbb{E}_{\bm{\mathring{x}}_{0}}} \mathlarger{\mathbb{E}_{\tilde{\bm{x}}_{0} \mid \bm{\mathring{x}}_{0}}} \mathlarger{\mathbb{E}_{\tilde{\bm{x}}_{t} \mid \bm{\mathring{x}}_{0},\tilde{\bm{x}}_{0}}} \Big[ \norm{\tilde{\bm{s}}_{\bm{\phi}}\big({\bm{x}}_{t},\mathring{\mathcal{S}},t\big) -\nabla_{\tilde{\bm{x}}_{t}} \log(p(\tilde{\bm{x}}_{t} \mid \bm{\mathring{x}}_{0}, \tilde{\bm{x}}_{0}))}_{2}^{2} \Big] \Big) \notag \\
& = \argmin_{\bm{\phi}} \mathlarger{\mathbb{E}_{t\sim U(0,T)}}\Big( \lambda(t) \mathlarger{\mathbb{E}_{\mathring{\mathcal{S}}}} \mathlarger{\mathbb{E}_{\bm{x}_{0}}} \mathlarger{\mathbb{E}_{\tilde{\bm{x}}_{t} \mid \bm{x}_{0}}} \Big[ \norm{\tilde{\bm{s}}_{\bm{\phi}}\big({\bm{x}}_{t},\mathring{\mathcal{S}},t\big) -\nabla_{\tilde{\bm{x}}_{t}} \log(p(\tilde{\bm{x}}_{t} \mid \bm{x}_{0} ))}_{2}^{2}\Big]\Big), \label{eq:conditionallossfunction}
\end{align}
\end{subequations} where $\tilde{\bm{s}}_{\bm{\phi}}(\cdot,\cdot,\cdot)$ is the output of the surrogate conditional score function corresponding to $\tilde{\mathcal{S}}$, $\bm{x}_{t}=(\bm{\mathring{x}}_{0}',\tilde{\bm{x}}_{t}')'$, and the second line follows from the same argument as that of \eqref{eqn:unconditionalsdelossfunction}.

Crucially, \eqref{eq:conditionallossfunction} reveals that one does not need to sample from or have knowledge of the predictive distributions $p(\tilde{\bm{x}}_{0} \mid \bm{\mathring{x}}_{0})$ to train the conditional score-matching surrogate---one only needs to simulate from the user-specified $p(\mathring{\mathcal{S}})$ and the \emph{unconditional} distribution of the spatial process model before running the forward evolution \eqref{eq:conditionalforwardsde}. Thus, our diffusion model can be used with a wide class of process models that are straightforward to sample from unconditionally but difficult to sample from conditionally. Note that the output dimension of the surrogate $\tilde{\bm{s}}_{\bm{\phi}}(\cdot,\cdot,\cdot)$ changes with $\abs{\tilde{\mathcal{S}}}$; Section~\ref{sec:nnarchitecture} describes how to accommodate this in practice. As in Section~\ref{sec:unconditionalsde}, we use a VPSDE, as detailed in the next section.

\subsection{Discretization}
\label{sec:discrete}
A popular SDE used for score-based diffusion models is the VPSDE, which, in the conditional setting, sets $\tilde{\bm{f}}(\tilde{\bm{x}}_{t},t)=\frac{-\beta(t)}{2}\tilde{\bm{x}}_{t}$ and $g(t)=\sqrt{\beta(t)}$ in \eqref{eq:conditionalforwardsecond}, where $\beta(t):[0,T]\to \mathbb{R}^{+}$ is selected in such a manner that the process has bounded variance as $T\to \infty$ \citep{Song2021}. To evolve $\tilde{\bm{x}}_{t}$ with the conditional forward and reverse SDEs, we discretize time. The discretized versions of \eqref{eq:conditionalforwardsecond} and \eqref{eq:conditionalreversesde} are given by
\begin{equation}
\tilde{\bm{x}}_{t+1} = \sqrt{1-\beta_{t}}\tilde{\bm{x}}_{t} + \sqrt{\beta_{t}} \tilde{\bm{\epsilon}}_{t}, \quad \tilde{\bm{\epsilon}}_{t}\sim \mathcal{N}(\tilde{\bm{0}}, \tilde{\bm{I}}), \quad t = 0,\dots,T-1,
\label{eq:forwardconditionalsdediscretized}
\end{equation}
and
\begin{equation}
\tilde{\bm{x}}_{t-1} = (1-\beta_{t})^{-\frac{1}{2}}\big(\tilde{\bm{x}}_{t} + \beta_{t}\nabla_{\tilde{\bm{x}}_{t}} \log(p(\tilde{\bm{x}}_{t} \mid \bm{\mathring{x}}_{0}))\big)+\sqrt{\beta_{t}} \tilde{\bm{\epsilon}}_{t}, \quad \tilde{\bm{\epsilon}}_{t}\sim \mathcal{N}(\tilde{\bm{0}}, \tilde{\bm{I}}), \quad t = T,\dots,1,
\label{eq:reverseconditionalsdediscretized}
\end{equation}
where $\beta_{t}=\beta(t)$ and $\tilde{\bm{\epsilon}}_{t}$ is defined on $\tilde{\mathcal{S}}$. See the Supplementary Material, Section~\ref{supp:suppdiscretization}, and Appendix B in \citet{Song2021} for derivation details. The corresponding transition kernels in the discretized conditional forward SDE have the form
\begin{equation}
p(\tilde{\bm{x}}_{t} \mid \bm{x}_{0}):=\mathcal{N}(\sqrt{\underline{\alpha}_t}\tilde{\bm{x}}_{0}, {\underline{\sigma}}_{t}^{2}\tilde{\bm{I}}),
\label{eq:transitionkernel}
\end{equation}
where $\underline{\sigma}_{t}=1-\underline{\alpha}_{t}$, $\underline{\alpha}_{t}=\prod_{s=1}^{t-1} \alpha_{s}$ and $\alpha_{s}=1-\beta_{s}$. To sample from the target distribution $p(\tilde{\bm{x}}_{0} \mid \bm{\mathring{x}}_{0})$ using \eqref{eq:reverseconditionalsdediscretized}, we replace the true score function $\nabla_{\tilde{\bm{x}}_{t}} \log(p(\tilde{\bm{x}}_{t} \mid \bm{\mathring{x}}_{0}))$ with the approximation $\tilde{\bm{s}}_{\bm{\phi}}(\bm{x}_{t}, \mathring{\mathcal{S}}, t)$. Note that in \eqref{eq:conditionallossfunction}, $t$ now needs to be equipped with a probability mass function on $\{1,\dots,T\}$ instead of a continuous one. The VPSDE framework for score-based diffusion in \eqref{eq:forwardconditionalsdediscretized} and \eqref{eq:reverseconditionalsdediscretized} is sometimes also referred to as a \emph{denoising diffusion probabilistic model} (DDPM); Chapter 4 in \citet{Chan2025} clarifies the connection between these two models.

\subsection{Architecture for Conditional Score Neural Network}
\label{sec:nnarchitecture}
A key consideration in our methodology is the architecture of the neural network surrogate for the conditional score function $\tilde{\bm{s}}_{\bm{\phi}}(\cdot,\cdot,\cdot)$ which must be flexible enough to accurately approximate highly non-linear score functions. Here, we restrict $\mathcal{S}$ to points on a regular grid, which enables the use of convolutional neural network (CNN) architectures \citep{Lecun1998}. In what follows, we regard each realization of the spatial process on $\mathcal{S}$ as an image, and $\tilde{\mathcal{S}}$ corresponds to pixels within the image that are treated as unobserved. To cater to score functions of variable dimension, we train a neural network whose input and output at time $t\in [0,T]$ has dimension $\mathbb{R}^{\abs{\mathcal{S}}}$ and which outputs $0$ for all observed locations. That is, our neural network,
\begin{equation}
\bm{s}_{\bm{\phi}}(\cdot,\mathring{\mathcal{S}},t)=(\tilde{\bm{s}}_{\bm{\phi}}(\cdot,\mathring{\mathcal{S}},t)', \bm{\mathring{0}}_{\mathring{\mathcal{S}}}')'\in \mathbb{R}^{\abs{\mathcal{S}}},
\label{eqn:nns}
\end{equation}
yields the appropriate subvector $\tilde{\bm{s}}_{\bm{\phi}}(\cdot,\mathring{\mathcal{S}},t)$ for use in \eqref{eq:conditionallossfunction} and \eqref{eq:reverseconditionalsdediscretized} as part of its output. For brevity, we assume \eqref{eqn:nns} places the values at the correct locations to avoid cumbersome notation.

For $\bm{s}_{\bm{\phi}}(\cdot,\cdot, \cdot)$, we use a specific type of CNN, referred to as U-Net, which consists of two main parts: a contracting path (or encoder) which reduces the image dimension and yields a large number of feature maps through multiple layers, and an expanding path (or decoder), that uses several layers of transposed convolutions, skip connections, and upsampling stages to restore the  original spatial resolution; see \citet{Ronneberger2015} for more details. There are many variants of U-Nets. In our implementation, we employ the noise conditional score network++ model (NCSN++) of \citet{Song2021} with minor modifications to reduce the U-Net size and thus, decrease training and evaluation time. Specifically, we reduced the number of blocks in both the contracting and expanding paths of the U-Net from $4$ to $2$ and the number of Gaussian Fourier projections determining the sinusoidal time embeddings from $128$ to $64$.

Recall from \eqref{eq:conditionallossfunction} that the conditional score depends on the observed locations, and therefore $\mathring{\mathcal{S}}$ needs to be an input to the score matching neural network. Since CNNs can only handle inputs of fixed dimension, we instead input a mask $\bm{M}(\mathring{\mathcal{S}})\in \{0,1\}^{\abs{S}}$ where 

\spacingset{1}
\begin{equation}
M_{i}(\mathring{\mathcal{S}})=\begin{cases}
    1, & {\bm{s}}_i\in \mathcal{\mathring{\mathcal{S}}},\\
    0, & \bm{s}_{i} \in \mathcal{\tilde{S}}.
    \end{cases}
\label{eqn:mask}
\end{equation}\spacingset{2}\noindent Specifically, we input $\bm{x}_{t}\in \mathbb{R}^{\abs{\mathcal{S}}}$ and ${\bm M}(\mathring{\mathcal{S}})\in \{0,1\}^{\abs{\mathcal{S}}}$ as two input channels, enabling the U-Net to extract separate spatial information from both inputs.

\subsection{Validation}
\label{sec:validation}
Validating whether the neural conditional simulations come from the true predictive distributions is difficult due to the high-dimensional setting, the assumed intractability of the predictive distributions, as well as the variable conditioning set. To address these issues, we propose a sample-based validation method derived from \citet{Linhart2023}, which compares the NCS-approximated unconditional distribution (described below) to the true unconditional distribution. This approach has multiple benefits: First, this method is tractable because simulation via NCS and from the true unconditional distribution is computationally efficient. Second, the NCS-approximated unconditional distribution does not depend on a single conditioning set, and thus, we can sidestep the issue of verifying the predictive distribution for each and every possible conditioning set. Finally, this validation technique enables the comparison of multivariate metrics, such as the spatial minimum and maximum of each unconditional simulation, which will only follow the correct distribution if the corresponding high-dimensional characteristics of the true unconditional distributions are accurately captured. However, as with any validation technique, passing these checks is a necessary but not sufficient condition for correct conditional simulation.

In addition to these multivariate metrics, we also use low-dimensional metrics. For the low-dimensional metrics, we created a validation data set involving replicates generated via NCS and either the true predictive distribution or an MCMC approximation of the true predictive distribution. Since these low-dimensional metrics are straightforward summaries, here we only discuss how to generate the NCS-approximated unconditional distribution in depth, and defer the technical details for the low-dimensional validation metrics to the Supplementary Material, Section~\ref{supp:validation}.

\paragraph*{NCS-Approximated Unconditional Distribution}

Consider the setting in which parameters are fixed and the conditioning sets are drawn from a fixed distribution $p(\mathring{\mathcal{S}})$. We simulate $m$ replicates $\{\bm{x}_{0,i}\}_{i\in [m]}$ from the true unconditional distribution and $m$ corresponding conditioning sets $\{\mathring{\mathcal{S}}_{i}\}_{i\in [m]}$ and masks $\{\bm{M}(\mathring{\mathcal{S}}_{i})\}_{i\in [m]}$ according to $p(\mathring{\mathcal{S}})$. For each of the $m$ replicates $\bm{x}_{0,i}$ and masks $\bm{M}(\mathring{\mathcal{S}}_{0,i})$, we obtain the corresponding partially observed spatial fields $\bm{\mathring{x}}_{0,i}$, generate a single NCS simulation $\tilde{\bm{x}}_{0,i}$ conditional on $\bm{\mathring{x}}_{0,i}$, and concatenate the NCS simulation with the partially observed field to produce an ``unconditional" simulation $(\bm{\mathring{x}}_{0,i}',\tilde{\bm{x}}_{0,i}')'$. If NCS accurately samples from $p(\tilde{\bm{x}}_{0,i} \mid \bm{\mathring{x}}_{0,i})$, then $({\bm{\mathring{x}}}_{0,i}',\tilde{\bm{x}}_{0,i}')'$ is a sample from $p(\bm{x}_{0})$, and can therefore be validated against other samples drawn from $p(\bm{x}_{0})$. See Algorithms~\ref{algo:linhartmethodsmall} and~\ref{algo:linhartmethod} in the Supplementary Material for pseudocode of this procedure.

\section{Simulation Experiments}
\label{sec:casestudies}
To demonstrate the accuracy of NCS, we use it to conditionally simulate from two spatial processes: a Gaussian process and a Brown--Resnick process. The first Gaussian case study verifies that our method is able to generate samples from true (known) predictive distributions; the second Brown--Resnick case study serves to demonstrate the utility of our method in settings where fast and efficient unconditional simulation is possible, but from which the predictive distributions are difficult or intractable to sample. For brevity, here we only present the results for the Brown--Resnick case study; see the Supplementary Material, Section~\ref{supp:GPcasestudy}, for the Gaussian case study.

\subsection{Experimental Setup}
\label{sec:casestudymethodology}
We let the spatial locations in $\mathcal{S}$ be a 32 $\times$ 32 gridding on $\mathcal{D}=[-10,10]\times [-10,10]$. The conditioning sets $\mathring{\mathcal{S}}$ and corresponding masks $\bm{M}(\mathring{\mathcal{S}})$ were constructed either by sampling at random a fixed number of locations from the $32^{2}$ grid points (Section~\ref{sec:results_small}), or through $32^{2}$ independent Bernoulli draws with probability of success (corresponding to whether a pixel is observed or not) equal to $\rho$ (Section~\ref{sec:brvalidation}). For the diffusion processes in \eqref{eq:forwardconditionalsdediscretized} and \eqref{eq:reverseconditionalsdediscretized},  we set $T=1000$, $\beta_{0}=0.0001$, $\beta_{T}=0.02$ and we let $ \beta_{t}=\beta(t)$ increase linearly with $t$ (specifically, we let $\beta(t)=0.0001+0.0199 t/T$). We used a weighting function $\lambda(t)=1-\underline{\alpha}_{t}$ for $\underline{\alpha}_{t} = \prod_{s=1}^{t} (1-\beta_{s})$ in the conditional loss function \eqref{eq:conditionallossfunction}; see \citet{Song2021} and the Supplementary Material, Sections~\ref{supp:train} and~\ref{supp:brtrain}, for details. Training and validation data were generated ``on the fly" \citep{Chan2018}.  

The Brown--Resnick process that we use in this paper is a two-parameter max-stable process with $\bm{\theta} \equiv (\lambda, \nu)'$, with range parameter $\lambda \in \mathbb{R}^{+}$ and smoothness parameter $\nu \in (0,2]$. It admits a likelihood function that is extremely costly to evaluate, and predictive distributions that are computationally challenging to sample from \citep{Padoan2010, Davison, Dombry2012, Huser2019}. See the Supplementary Material, Section~\ref{supp:spatialprocessdescription}, for greater detail on the Brown--Resnick process.

For the purpose of assessing NCS, we consider the Gibbs sampler \citep{Dombry2012} implemented in the \texttt{SpatialExtremes} R package \citep{SpatialExtremes} as an alternative conditional simulation method. We found that full conditional simulation (FCS) via the \texttt{SpatialExtremes} package is only computationally feasible when the conditioning set contains seven or fewer observations. For conditioning sets with more than seven observed locations, we resort to local conditional simulation (LCS) by only using the the seven observed locations closest to a prediction location for conditional simulation. In contrast to FCS and NCS, LCS targets a low-dimensional approximation of a predictive distribution and will only be used here for univariate and bivariate approximations of the predictive distributions. The Gibbs sampler is computationally intensive, and requires a computing wall time that depends on several factors, such as the number of observed locations and model parameters. See the Supplementary Material, Section~\ref{supp:brapprox}, for more details.

In contrast, as we show in Section~\ref{sec:results_small}, NCS allows for consistent computational efficiency, regardless of the number of observed locations or process-model parameters. We implemented NCS and LCS/FCS on an Nvidia Tesla T4 GPU and an AMD EPYC 7V12 processor (with 64 cores, no multithreading) respectively. For the Brown--Resnick process, we log-transformed the simulated spatial fields from the unit Fr{\'e}chet scale to the standard Gumbel scale before training the U-Nets for estimating the score function.

\subsection{Results: Small Number of Observed Locations}
\label{sec:results_small}
In this section, we compare FCS and NCS for the case when the number of observed locations $|\mathring{\mathcal{S}}| \in \{1,\dots,7\}$ (where FCS is computationally feasible) and where the range parameter  $\lambda \in \{1,\dots,5\}$ (weak to strong spatial dependence) varies. For each range value, we trained and implemented separate U-Nets that are each amortized for conditioning sets of $1$ to $7$ locations. See the Supplementary Material, Section~\ref{supp:train}, for details. To assess NCS, we examine the extremal correlation function $\chi(h)=2-\zeta(h)$, where the spatial extremal coefficient function $\zeta(h):\mathbb{R}^{+}\to [1,2]$ quantifies the extremal dependence between pairs of locations a fixed distance $h$ apart \citep{deHaan1985, Smith1990}. We compute the empirical extremal correlation functions for the true unconditional distributions, the FCS-approximated unconditional distributions, and the NCS-approximated unconditional distributions using the approach described in Section~\ref{sec:validation} \citep{Linhart2023}. In addition to the extremal correlation functions, we empirically estimate the unconditional densities of the minimum, maximum, and the absolute sum of the process over $\mathcal{S}$. See the Supplementary Material, Section~\ref{supp:br}, for training and validation details as well as additional results.

\begin{figure}[t!]
 \centering
    \includegraphics[scale = .066]{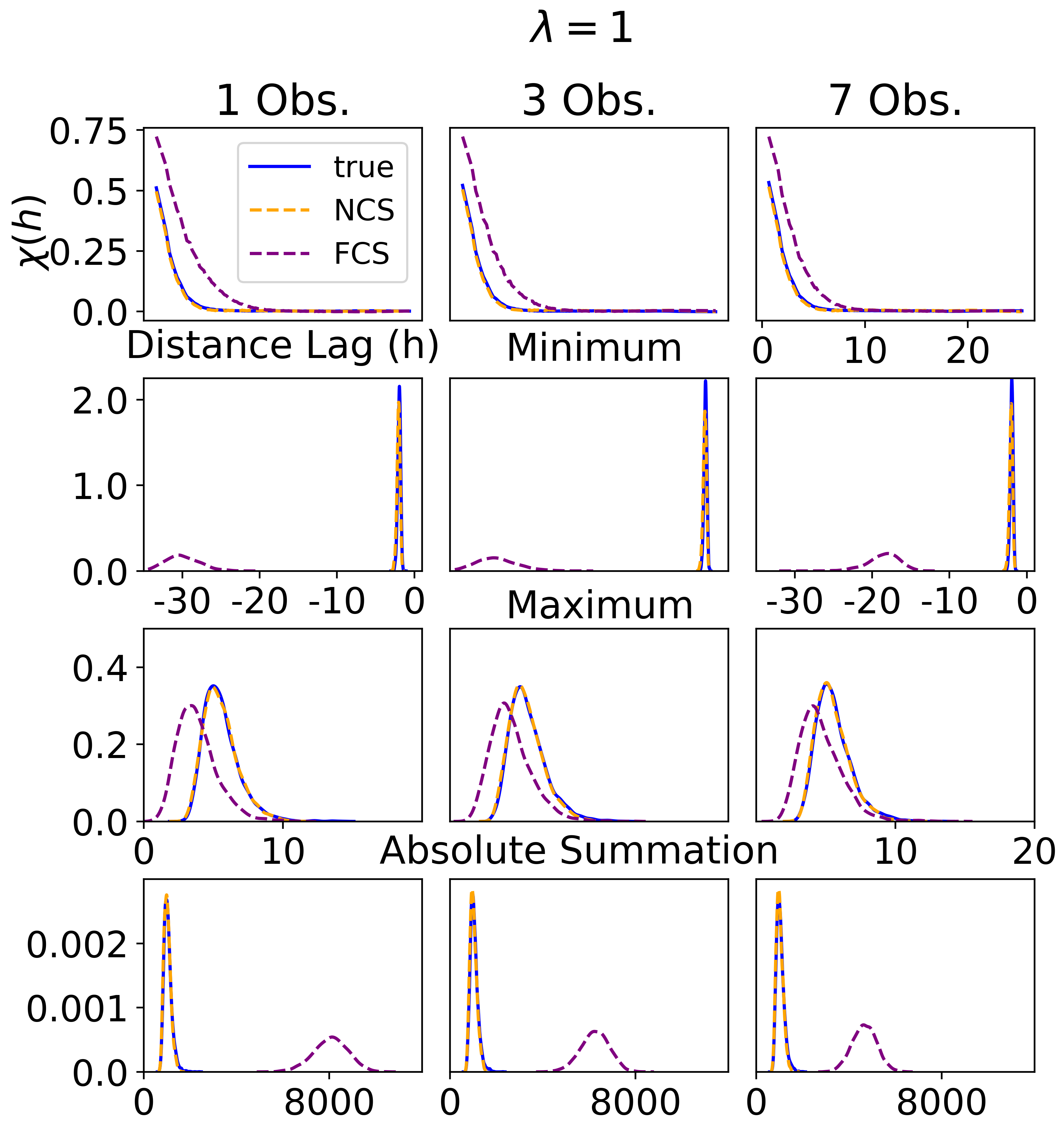}
    \includegraphics[scale = .066]{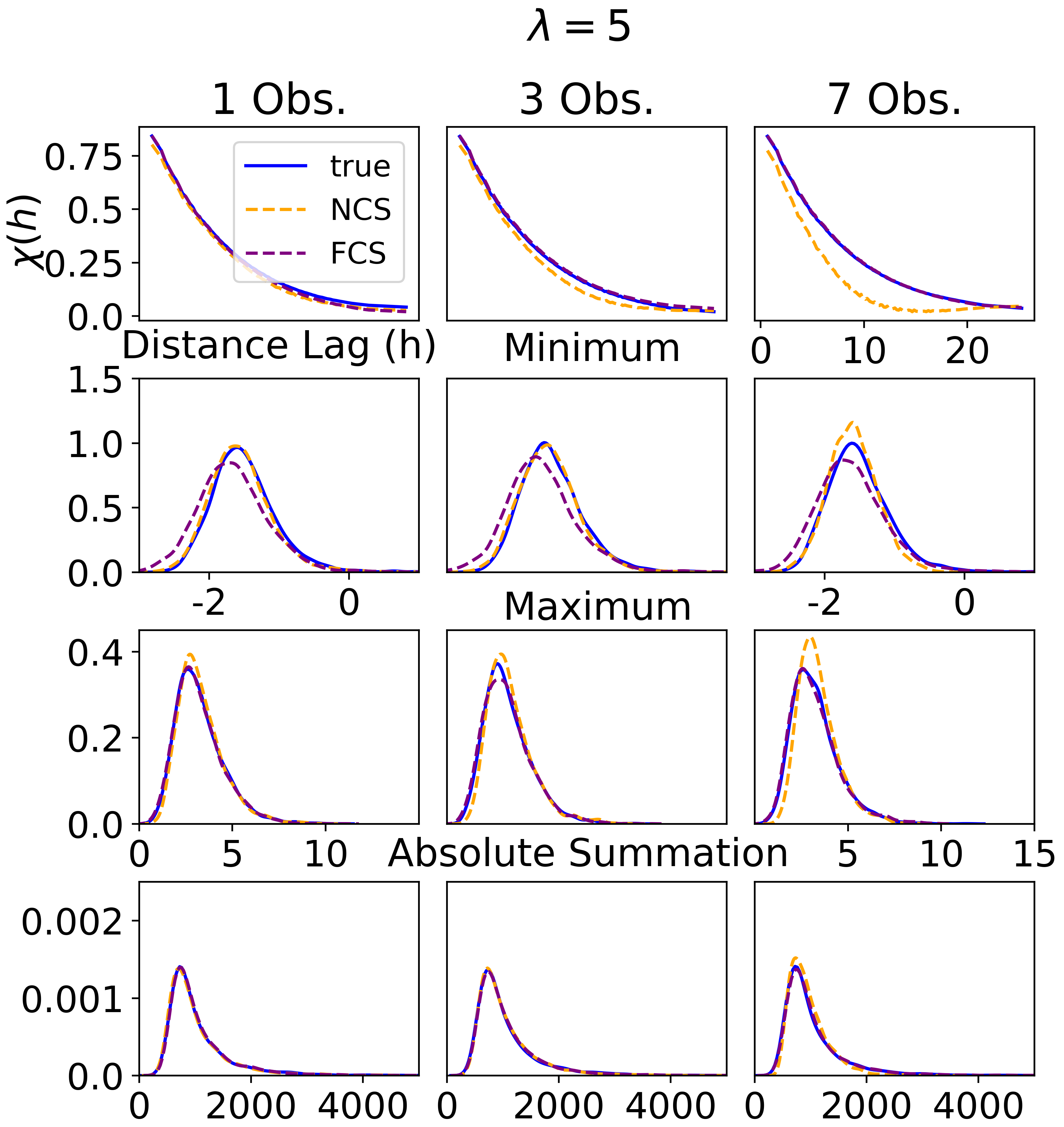}
    \caption{Extremal correlation $\chi(h)$ as a function of distance (top row), and the empirical distributions of the spatial minimum (second row), maximum (third row), and absolute sum (fourth row) for the true (blue), NCS-approximated (orange), and FCS-approximated (purple) unconditional distributions of a Brown--Resnick process with smoothness $\nu = 1.5$ and range $\lambda = 1$ (left panel) and $\lambda = 5$ (right panel) for one, three, or seven observed locations on the Gumbel scale.}
\label{fig:brfcsextremalcoefficient}
\end{figure}

In Figure~\ref{fig:brfcsextremalcoefficient}, we show the true, NCS, and FCS extremal correlation functions for the case in which only one to seven locations are observed, when the range parameter is $\lambda = 1$ (left panel) and $\lambda = 5$ (right panel). When $\lambda = 1$, the true and NCS extremal correlation functions match exactly, but the FCS extremal correlation function is reflective of a spatial dependence that is too long-range. As shown in Figure~\ref{fig:brfcsvisualization}, the FCS simulations are negatively biased at unobserved locations far from the conditioning sites when the range parameter is small. In contrast, when $\lambda = 5$, the true and FCS extremal correlation functions match exactly, whereas the NCS extremal correlation function is slightly inaccurate in the case of seven observed locations. In this case, the NCS simulations appear more diffuse than the FCS simulations. However, this discrepancy is not as stark as the negative bias in the FCS simulations when the range parameter is small.

The empirical distributions of the spatial minimum and maximum computed from the true, FCS-approximated, and NCS-approximated unconditional simulations are shown in the second and third rows of Figure~\ref{fig:brfcsextremalcoefficient}. We see that the distributions of the minima and maxima from NCS match the true ones (except for some slight inaccuracy in the $\lambda =5$ case with seven observations), while those from FCS are clearly biased for the $\lambda=1$ case. In the fourth row of Figure \ref{fig:brfcsextremalcoefficient}, we show the distribution of the absolute sum of the process over $\mathcal{S}$. Again, we see that the true and NCS distributions of this quantity match almost exactly. As with the distributions of the spatial minimum and maximum, FCS does not capture the absolute summation distributions well for $\lambda=1$.

\begin{figure}[t!]
    \centering
    \includegraphics[scale = .5]{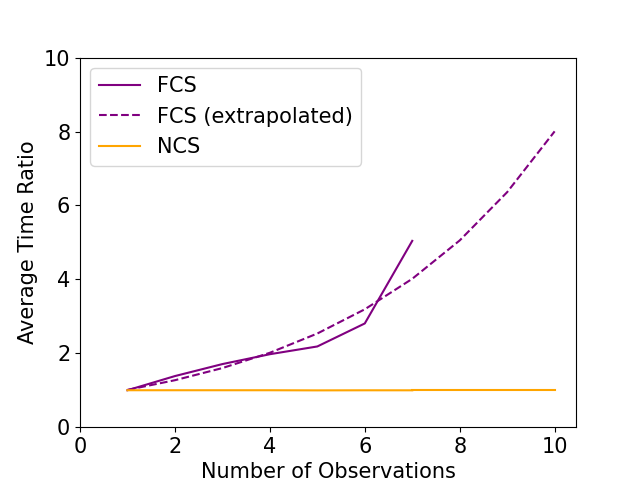}
    \caption{The ratio of the time needed for conditional simulation with a varying number of observations to the time needed for conditional simulation with one observation,  using FCS (purple) and NCS (orange).}
\label{fig:brtime}
\end{figure}

To highlight the computational inefficiency and ultimately infeasibility of FCS when compared to NCS, we record the average time (in seconds) to generate one conditional simulation for varying number of observations, and divide it by the time needed to conditionally simulate with $|\mathring{\mathcal{S}}| = 1$. As shown through these time ratios in Figure~\ref{fig:brtime}, the average time to generate an FCS simulation when seven locations are observed is five times greater than the time needed when one location is observed. Since the Gibbs sampler underlying FCS involves evaluating multivariate Gaussian cumulative distribution functions, we expect FCS to scale exponentially with the number of observed locations $|\mathring{\mathcal{S}}|$ \citep{Hinrich2014,Balakrishnan2014,SpatialExtremes} and use this to show the likely scaling of FCS beyond seven observations---the maximum number of locations for which our implementation of the Gibbs sampler in the SpatialExtremes R package remained stable. We conclude that FCS is difficult to implement using available software, and its computational complexity quickly increases with the number of observed locations.

On the other hand, the time needed to generate an NCS simulation does not change with the number of observed locations. The average time to generate a single FCS and NCS simulation from seven observed locations was $28$ and $27$ seconds, respectively, with our setup. We currently evaluate the U-Net at every time step in the discretized conditional reverse SDE, but there are methods designed to reduce the number of steps needed to simulate from diffusion models \citep{Bond-Taylor2022}. Once implemented, these may further reduce the time needed for a single NCS simulation.

\subsection{Results: Amortizing with Respect to Observed Proportion and Process Parameters}
\label{sec:brvalidation}

In this section, we evaluate NCS's performance on the Brown--Resnick process with larger numbers of observed locations, and assess its adaptability to varying parameter values. Specifically, we consider a scenario where $\bm{\theta}$ is fixed but where $\rho$, the proportion of observed values, is allowed to vary, and another where $\rho$ is fixed and where the range parameter is allowed to vary. We trained the U-Net in the first scenario using simulations from a Brown--Resnick process with a fixed parameter vector $\bm{\theta} = (\lambda, \nu)' = (3, 1.5)$ and with $\rho$ sampled from a uniform distribution on $[0.01,0.5]$. This U-Net, which we refer to as the ``proportion U-Net'', is amortized with respect to the observed proportion of grid locations since it does not require re-training for different proportions $\rho\in [0.01,0.5]$. We trained the second U-Net using simulations from a Brown--Resnick process with fixed $\nu = 1.5$ and $\rho = 0.05$, but with $\lambda$ allowed to vary uniformly on $[1,5]$. This U-Net, which we refer to as the ``parameter U-Net'', is amortized with respect to the range parameter since it does not require re-training for different range values $\lambda \in [1,5]$. While in principle it is possible to train a single U-Net that amortizes with respect to both observed proportion and spatial-process parameters, we anticipate that larger U-Nets and computing capacity would be required for such capability.

When an element of $\bm{\theta}$, $\theta_1$ say, is allowed to vary, the score function estimate must also be allowed to vary with $\theta_1$. To accommodate this, in the parameter U-Net, we include $\theta_1$ as an extra input, i.e., we re-define the approximate score function to be $\bm{s}_{\bm{\phi}}(\bm{x}_{t},\mathring{\mathcal{S}},\theta_1,t)=\big(\tilde{\bm{s}}_{\bm{\phi}}(\bm{x}_{t},\mathring{\mathcal{S}},\theta_1,t)', \bm{0}_{\mathring{\mathcal{S}}}'\big)'$, and modify \eqref{eq:conditionallossfunction} accordingly to include an expectation with respect to a  distribution over $\theta_1$ (here, a uniform distribution on $[1,5]$). Through experimentation, we found that information about the parameter $\theta_1$ needs to be included at each step in the contracting and expanding paths of the U-Net to obtain good results. We implemented this by replacing the second input channel to the U-Net, $\bm{M}(\mathring{\mathcal{S}})$, with $\theta_1\bm{M}(\mathring{\mathcal{S}})$.

In this section, we validate the resulting NCS simulations with various metrics including empirical conditional univariate and bivariate densities as well as empirical unconditional densities of the minimum, maximum, and the absolute sum of the process over $\mathcal{S}$ via the approach described in Section~\ref{sec:validation}. See the Supplementary Material, Sections~\ref{supp:brtrain} and~\ref{supp:brvalidationdetails}, for training and validation details.

\begin{figure}[t!]
\centering
\includegraphics[scale=0.077]{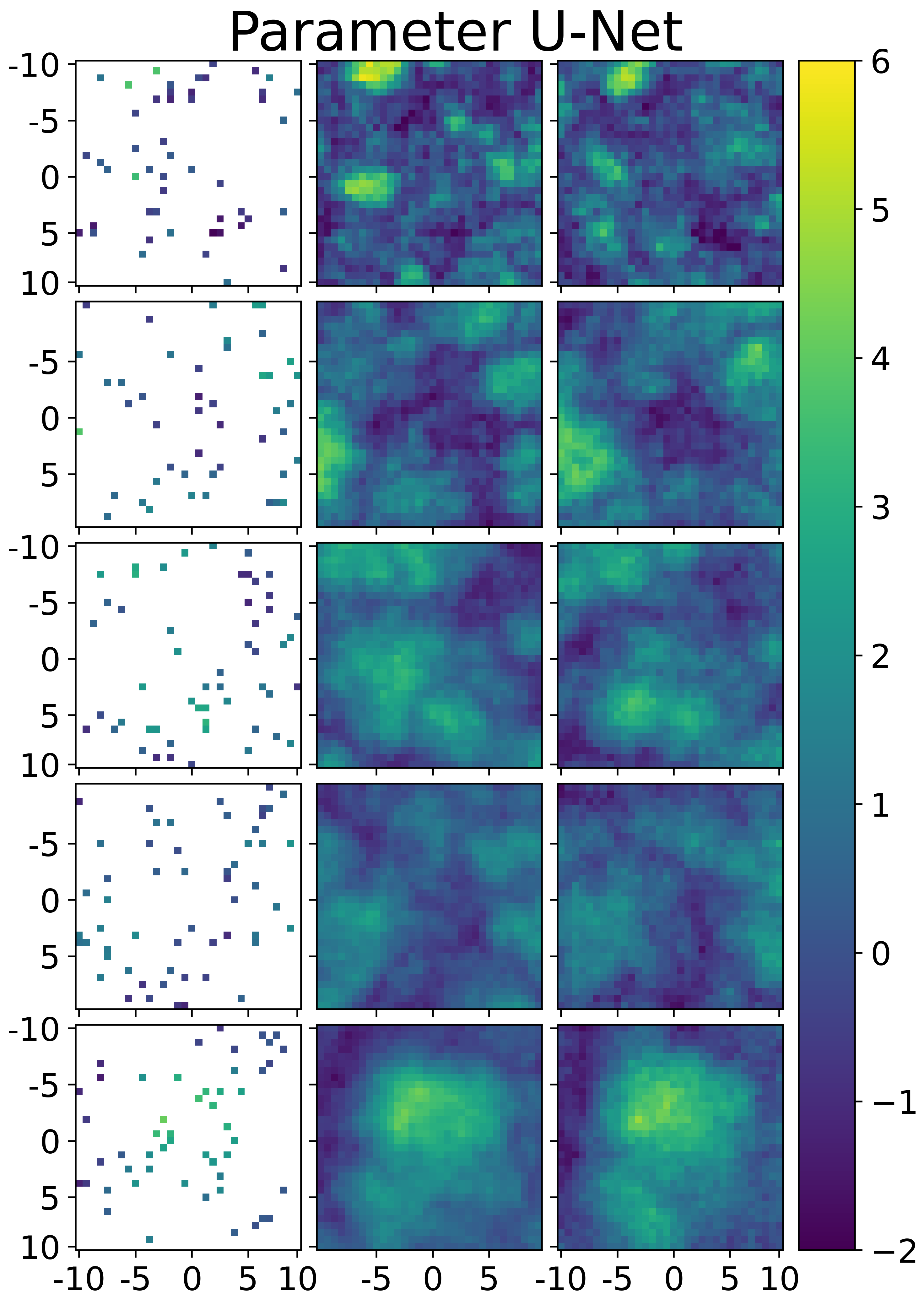}
\includegraphics[scale=0.077]{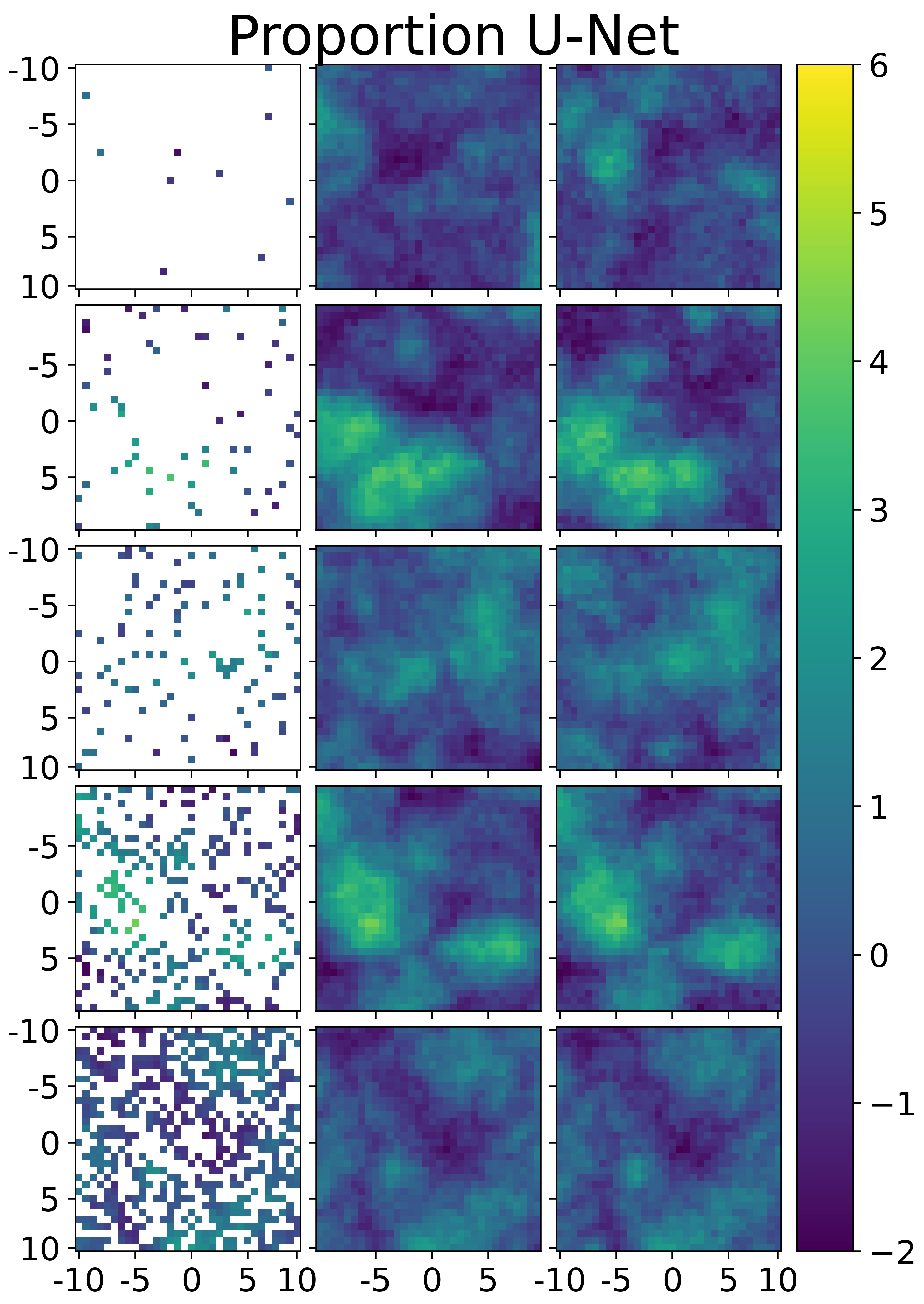}
\caption{Left panel: NCS with a Brown--Resnick process with smoothness  $\nu = 1.5$, range  $\lambda \in \{1,2,3,4,5\}$ (from top to bottom), and observed proportion $\rho = 0.05$. Right panel: NCS with a Brown--Resnick process with smoothness  $\nu = 1.5$,  range $\lambda = 3$, and observed proportion $\rho \in \{0.01,0.05,0.1,0.25,0.5\}$ (from top to bottom). For each panel: Left columns: Observations. Middle columns: Underlying spatial fields (generated using unconditional simulation). Right columns: Conditional simulations with NCS implemented with a parameter U-Net (left panel) or a proportion U-Net (right panel). All values are shown on the Gumbel scale (after a log transformation).}
\label{fig:brvisualization}
\end{figure}

In Figure \ref{fig:brvisualization} we show observations, underlying true fields, and conditional simulations obtained using NCS. In the left panel, we show results from fixing $\rho = 0.05$, varying the range parameter $\lambda$ between 1 and 5, and implementing NCS using the parameter U-Net. In the right panel, we show results from fixing $\lambda = 3$, varying the proportion parameter $\rho$ between $0.01$ and $0.5$, and implementing NCS using the proportion U-Net. We see that the NCS conditional simulations exhibit spatial patterns that are similar to those of the corresponding true (partially observed) fields, and that amortization with respect to either the range or the proportion of observed locations appears to be effective.

\begin{figure}[t!]
 \centering
    \includegraphics[scale = .096]{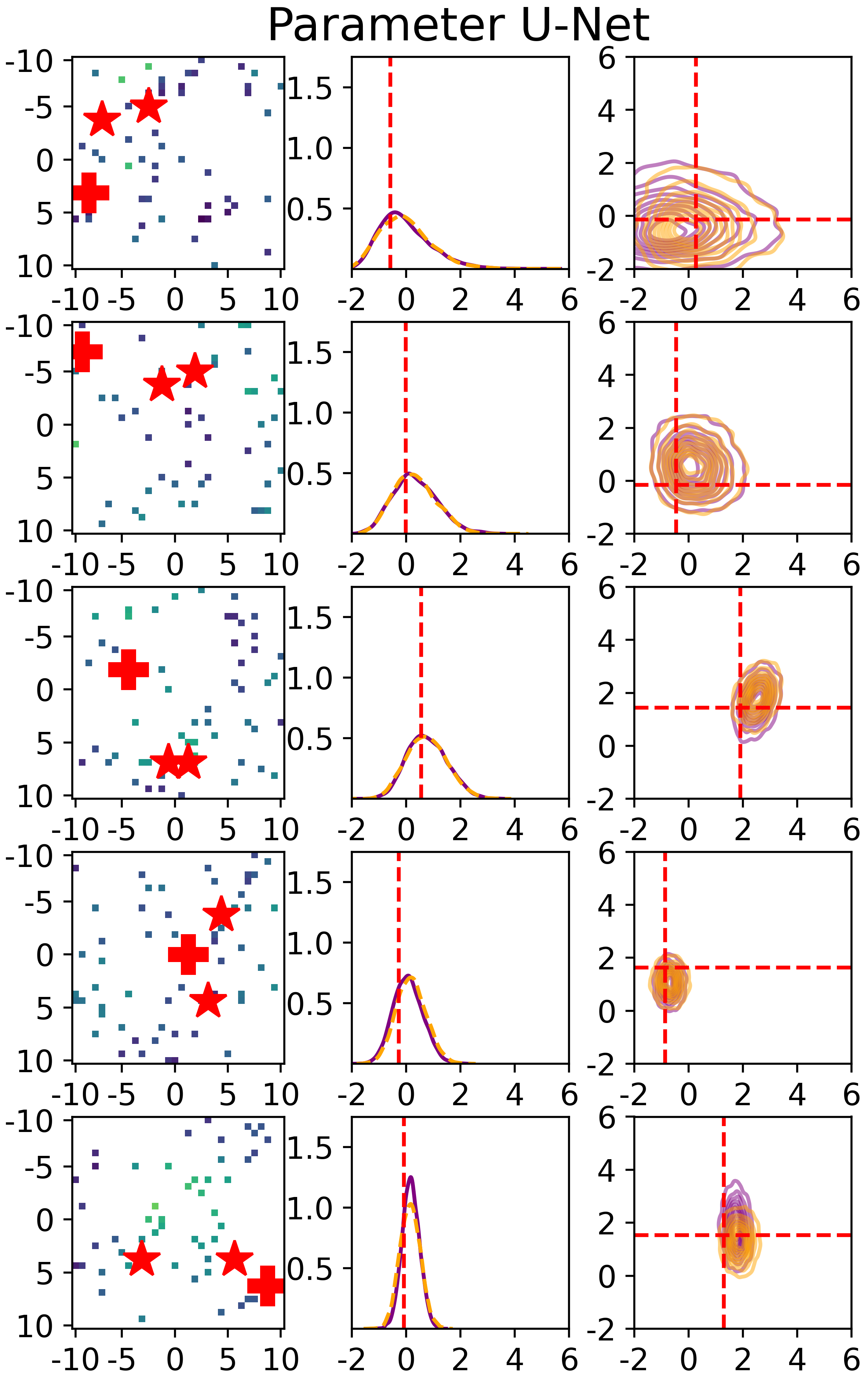}
    \includegraphics[scale = .096]{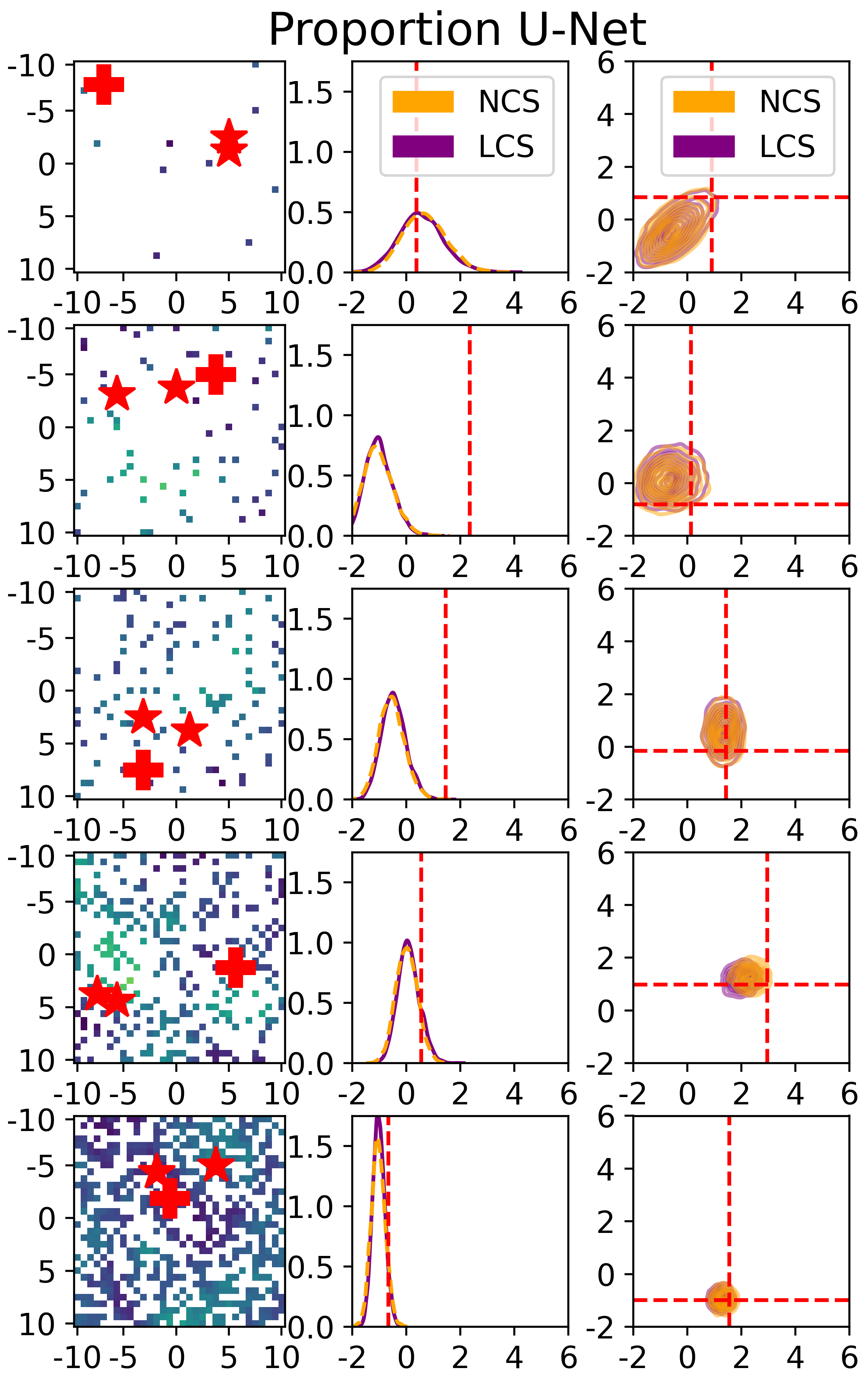}
    \caption{Left panel: Predictive densities of a Brown--Resnick process with smoothness $\nu = 1.5$, range $\lambda \in \{1,2,3,4,5\}$ (from top to bottom), and observed proportion $\rho = 0.05$. Right panel: As in the left panel, but with $\lambda = 3$, and with the observed proportion $\rho \in \{0.01,0.05,0.1,0.25,0.5\}$ (from top to bottom). For each panel: Left columns: Observations and prediction locations for which univariate densities (at the red crosses) and bivariate densities (across the red stars) are visualized. Middle columns: LCS (purple) and NCS (orange) empirical conditional univariate densities at the locations marked with crosses. Right columns:  LCS (purple) and NCS (orange) empirical conditional bivariate densities across the locations marked with stars. All shown process values are on the Gumbel scale and the red dashed lines indicate observed values.}
\label{fig:brdensity}
\end{figure}

Figure~\ref{fig:brdensity} shows the LCS (purple) and NCS (orange) conditional univariate and bivariate densities for select locations in $\mathcal{S}$. We do not expect the NCS and LCS densities to match exactly since LCS is an approximation that utilizes less information than NCS. Yet, we observe that the two predictive densities at these select locations are very similar. In particular, the densities correctly capture the true values shown by the dashed red lines as either the range $\lambda$ or the proportion of observed locations $\rho$ vary.

\begin{figure}[t!]
 \centering
    \includegraphics[width=0.48\textwidth]{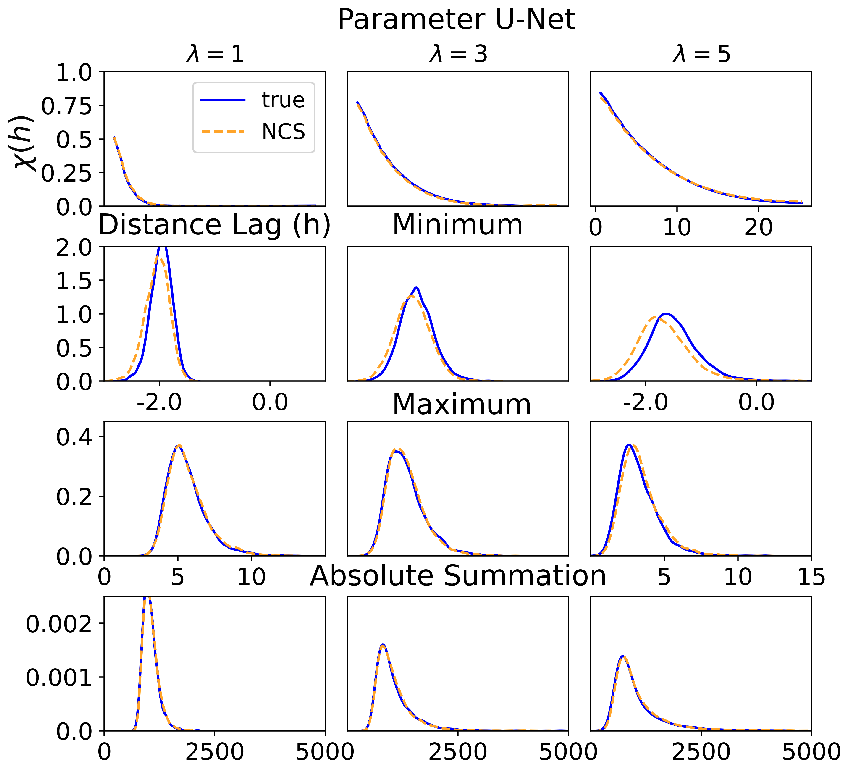}
    \includegraphics[width=0.48\textwidth]{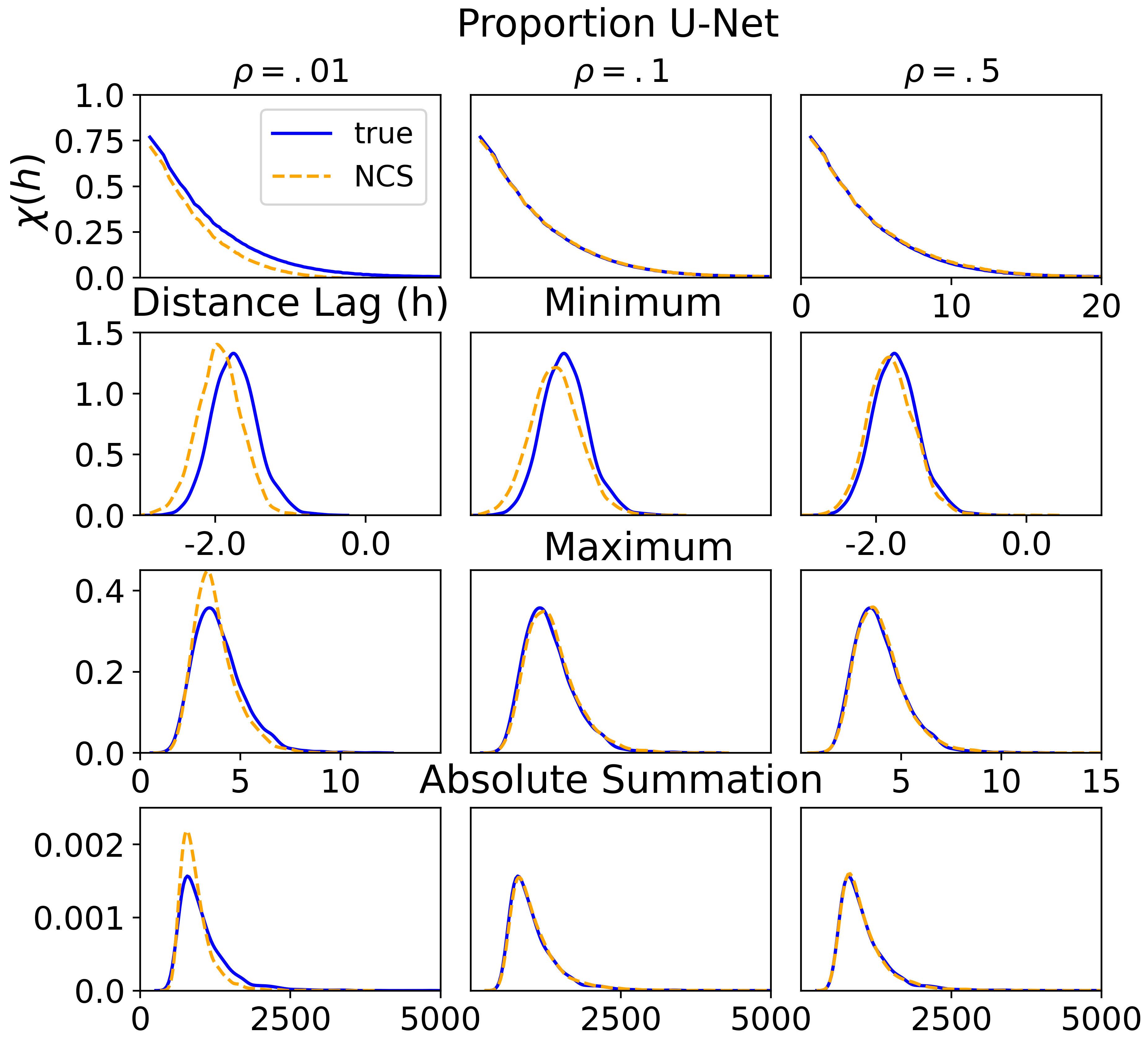}
    \caption{Extremal correlation $\chi(h)$ as a function of distance (top row), and the empirical distributions of the spatial minimum (second row), maximum (third row), and absolute sum (fourth row) for the true (blue) and the NCS-approximated (orange) unconditional distributions of a Brown--Resnick process with observed proportion $\rho = .05$, smoothness $\nu = 1.5$, and range $\lambda \in \{1,3,5\}$ (left panel, from left to right); and smoothness $\nu = 1.5$, range $\lambda = 3$, and observed proportion $\rho \in \{0.01, 0.1, 0.5\}$ (right panel, from left to right) on the Gumbel scale.}
\label{fig:brextremalcoefficient}
\end{figure}

Extremal correlation plots, as well as empirical distributions of the minimum, maximum, and absolute sum, are shown in Figure~\ref{fig:brextremalcoefficient}. The true and NCS extremal correlation functions match almost exactly, even as the range parameter $\lambda$ varies. The NCS-approximated unconditional distributions of the minimum, maximum, and absolute sum also match almost perfectly with the underlying true ones excepting the case when the proportion of observed locations is $\rho = 0.01$---corresponding to approximately $10$ observed locations. In Section~\ref{sec:results_small}, the metrics for the NCS-approximated unconditional distribution with only seven observed locations similarly deviated from the truth.

\subsection{Summary of Results}
In this case study, we found that NCS simulations are at least comparable to and often outperform the standard approximations (FCS and LCS) in generating simulations representative of the intractable true predictive distributions of a Brown--Resnick process. Importantly, NCS has fewer limitations: FCS becomes computationally intractable when the number of observed spatial locations exceeds seven, and LCS is inherently restricted in the degree of high-dimensional patterns it can capture. In addition, the high-dimensional patterns for the FCS-approximated unconditional distributions differ from those of the true unconditional distribution more significantly than the equivalent for NCS in many cases except when the range is large and roughly one percent of locations are observed. It is unclear whether this dip in performance is inherent to NCS when applied to Brown--Resnick processes or whether improved training in this case may address the decreased performance.

\section{Data Application}
\label{sec:da}
%In the previous section, we only applied NCS to simulated data. Since neural networks are notoriously brittle, it is not a foregone conclusion whether and to what degree NCS is robust to the inherent model misspecification in any real-world data. In this data application, we illustrate that NCS is robust with real world data in which slight model misspecification is at the very least present.

%In this section we evaluate the use of NCS to analyze annual maxima of sea surface temperature (SST) in the Red Sea with a Brown--Resnick process. 
In this section, we use NCS to model annual maxima of sea surface temperature (SST) in the Red Sea with a Brown--Resnick process to evaluate how well NCS performs with real data. Modeling SST annual maxima with Brown--Resnick processes is well-motivated as demonstrated in \citet{Huser2024}. We follow a similar setting as in \citet{Huser2024} with the additional assumption that $5$ percent of available data are observed; we then use the left-out data to validate NCS. %We set up an experiment where we assume that $5$ percent of available data are observed; we then use the left-out data to validate NCS.  %We modify slightly the the way in which we sample from the conditional reverse SDE To allow for the presence of coastal boundaries we adapt the sampling from the conditional reverse SDE; 
%Since the Red Sea has coastal boundaries, these coastal boundaries appear in the spatial fields produced from local windows intersecting the coast. As such, we slightly adapt sampling from the conditional reverse SDE to incorporate coastal boundaries into the NCS simulations. %The structure of this section is as follows: a brief description of the Red Sea SST data and the resulting annual SST maxima, a brief description of diffusion model training and NCS simulation details including how we incorporate coastal boundaries, and a selection of results focused on three locations in the Red Sea.
%To enable similar comparison between the simulated studies and this data application, 

Specifically, we consider three windows of size $2 \degree \times 2\degree$ centered at three different locations in the Red Sea, mask out $95\%$ of the residuals in these local windows, and apply NCS to generate $1000$ conditional simulations at the masked out locations. We then visualize select NCS simulations, together with select conditional marginal and bivariate densities and the conditional mean field constructed from the $1000$ NCS simulations. We refer the reader to \citet{Huser2024} for a full description of the Red Sea SST data, the processing steps we apply to extract annual SST maxima, and the removal of spatiotemporal trends to obtain detrended and standardized residuals of the annual maxima. The Supplementary Material, Section~\ref{supp:dasupp}, contains the parameter estimation procedure, our implementation of NCS in this setting, and the conditional mean field results.
%We then visualize the NCS simulations together with the conditional marginal and bivariate densities, and the conditional mean field, constructed from these NCS simulations. In the Supplementary Material, Section~\ref{supp:dasupp}, we give a full description of the Red Sea SST data, the processing steps we apply to extract annual SST maxima, the removal of spatiotemporal trends to get detrended and standardized residuals of the annual maxima, the parameter estimation procedure, and details specific to this application of NCS.

\subsection{Experimental Setup}
We consider SST from the Operational Sea Surface Temperature and Ice Analysis (OSTIA), a data product of daily observations on a fixed, non-equidistant grid. We extract $16703$ locations in the Red Sea at a resolution of approximately $0.05\degree \times 0.05 \degree$ for $31$ years, from 1985 to 2015. We re-grid the data to a fixed, equidistant grid of lower resolution ($4379$ locations on a $0.1\degree \times 0.1\degree$ grid) using nearest neighbor interpolation. Similar to \citet{Huser2024}, we then detrend the raw SST data to remove spatiotemporal trends, aggregate the resulting daily residuals into annual maxima residuals, and transform the annual maxima residuals to a common standard Gumbel scale in order to model spatial dependence.

Since the likelihood function of a Brown--Resnick process is computationally intractable, we perform parameter inference using a (composite) pairwise likelihood %--a standard approximation of the likelihood for which the summands in the logarithmic form are the bivariate log likelihoods between select pairs of spatial locations
\citep{Castruccio}. In order to avoid computing ${n(n-1)/2}$ bivariate log likelihoods, we only select pairs of locations that are within a distance $\delta$ of each other. The distance cut-off $\delta$ is a tuning parameter to which the pairwise likelihood parameter estimates are sensitive. See the Supplementary Material, Section~\ref{sec:comp_lik},
for details on how we choose $\delta$.

The $2\degree \times 2\degree$ local windows intersect with the coast to varying degrees. To incorporate coastal boundaries into the NCS simulations, we treat the land-based locations as unobserved during training and evaluation and simply remove the simulated values of the resulting NCS simulation at the land-based locations. See the Supplementary Material, Section~\ref{sec:data_NCS}, for further details.

\subsection{Results}

\begin{figure}[t!]
 \centering
  \begin{minipage}[c]{0.35\textwidth}
  \centering
    \includegraphics[width=\linewidth]{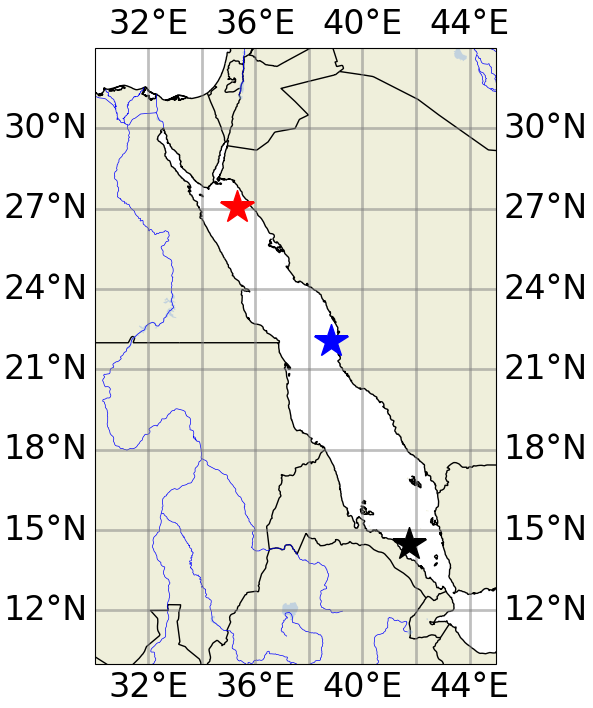}
  \end{minipage}
  \hfill
  \begin{minipage}[c]{0.6\textwidth}
  \centering
    \includegraphics[width=\linewidth]{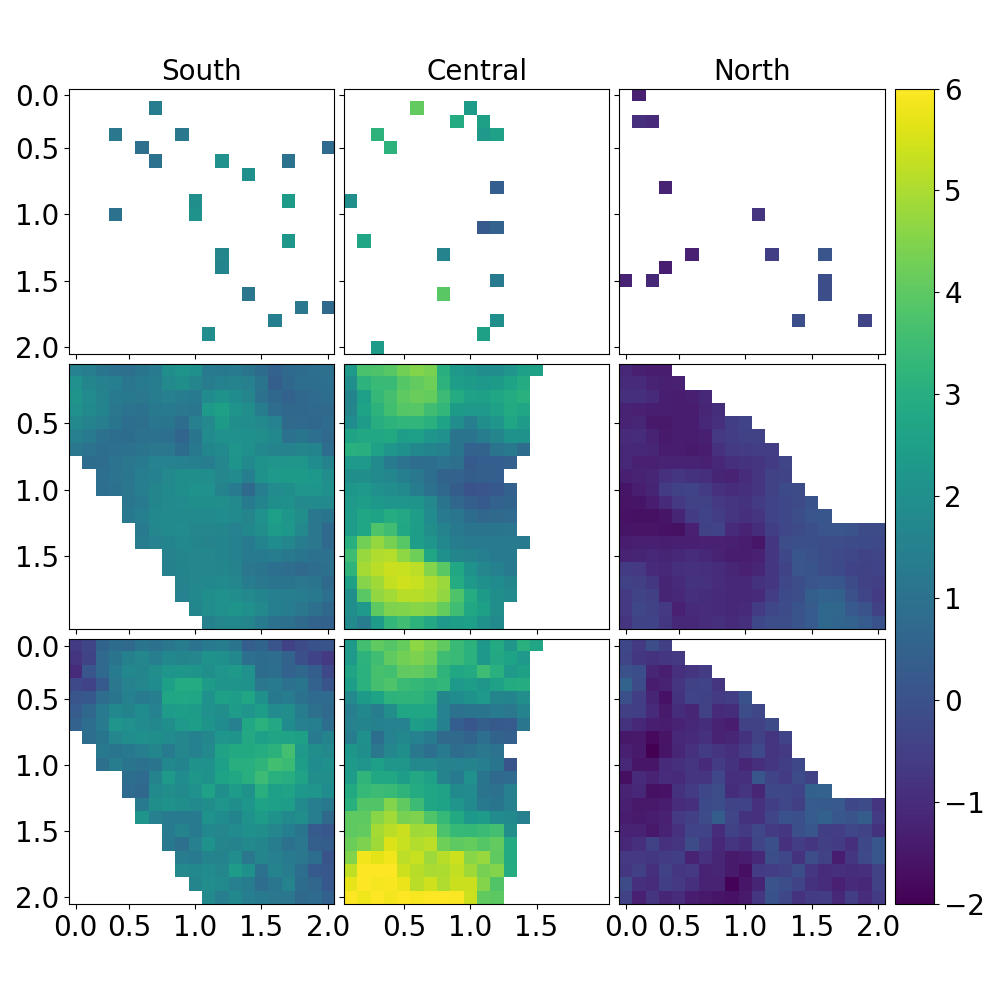}
  \end{minipage}
  \caption{Partially observed (right-top), fully observed (right-middle), and NCS simulations of (right-bottom) annual maxima residuals in the year 1985 on the Gumbel scale at three locations in the Red Sea (left panel). NCS is done using a Brown--Rensick process with parameter estimates $\hat{\bm{\theta}}$ estimated using a pairwise likelihood.}
  \label{fig:RSncsviz} 
\end{figure}

In this section, we provide low-dimensional validation metrics for the NCS simulations at the three chosen locations in the Red Sea. The left panel of Figure~\ref{fig:RSncsviz} gives a map of the Red Sea and shows the three locations considered in this study. The first and second rows in the right panel of Figure~\ref{fig:RSncsviz} show the partially observed and fully observed annual maxima standarized residuals at these three locations, while the third row contains NCS simulations conditioned on the partially observed spatial field and the pairwise likelihood estimates. The NCS simulations exhibit similar spatial patterns to those in the fully observed fields. %As in our simulation experiments, we observe some fine-scale ``graininess'' in our simulations. %The graininess of the NCS simulations is not unexpected as the NCS simulations from the simulation experiments in Section~\ref{sec:casestudies} follow the same pattern. %This graininess may not significantly impact the downstream products derived from NCS simulations.
%The remaining results in this section indicate that this does not impact low-dimensional aspects of the predictive distributions.

%Figure~\ref{fig:RSncsmeanfield} displays the partially and fully observed annual maxima residuals in the first two rows and the resulting empirical conditional mean fields computed from $n=1000$ NCS simulations in the bottom row. Notably, the empirical conditional mean fields smoothly integrate the large-scale spatial patterns contained in the fully observed annual maxima residuals in the middle row.

% \begin{figure}
%  \centering
%   \begin{minipage}[c]{0.25\textwidth}
%   \centering
%     \includegraphics[width=\linewidth]{Figures/RS/results/locations_map}
%   \end{minipage}
%   \hfill
%   \begin{minipage}[c]{0.7\textwidth}
%   \centering
%     \includegraphics[width=\linewidth]{Figures/RS/results/NCS/ncs_conditional_mean_field_year_0}
%   \end{minipage}
% \caption{Partially (top) and fully (middle) observed annual maxima residuals and NCS (bottom) empirical conditional mean fields with pairwise parameter estimates $\hat{\bm{\theta}}$ on the Gumbel scale (right subfigure) for $1985$ and locations (red star) in the left subfigure}
% \label{fig:RSncsmeanfield}
% \end{figure}

\begin{figure}[t!]
 \centering
  % \begin{minipage}[c]{0.35\textwidth}
  %   \centering
  %   \includegraphics[width=\linewidth]{Figures/RS/results/locations_map}
  % \end{minipage}
  % \hfill
  \begin{minipage}[c]{0.8\textwidth}
    \centering
    \includegraphics[width=\linewidth]{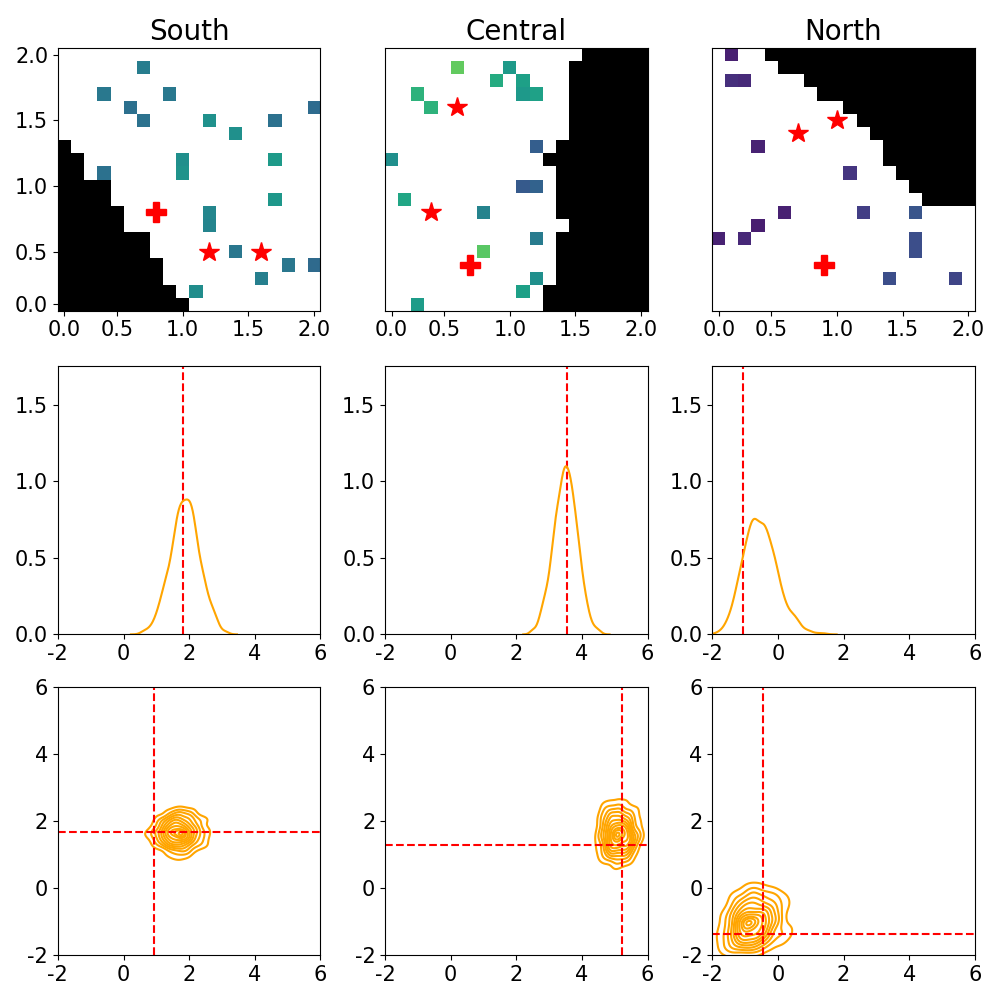}
  \end{minipage}
  \caption{Partially observed annual maxima residuals (top row), NCS empirical conditional marginal densities (middle row), and empirical conditional bivariate densities (bottom row) at spatial locations marked by the red crosses for the marginal distributions, and the red stars for the bivariate distributions. Results are shown on the Gumbel scale. NCS is done using a Brown--Resnick process with parameter estimates $\hat{\bm{\theta}}$ estimated using composite likelihood.}
  \label{fig:RSncsmarginalbivariate}
\end{figure}

In Figure~\ref{fig:RSncsmarginalbivariate}, we show the marginal (middle row) and bivariate (bottom row) distributions for single and pairs of locations marked with red crosses and stars (top row), as generated by NCS conditional on the observed data (top row). Note that the land-based locations are marked as black pixels. In all marginal and bivariate cases, the NCS conditional distributions contain the observed values in the bulk of their probability mass. Note how the conditional distributions are more concentrated at locations with nearby observations, as expected.

\section{Discussion and Conclusions}
\label{sec:discuss}

In this paper, we introduced neural conditional simulation (NCS) as a method to sample from predictive distributions of a spatial process using a conditional score-based diffusion model within an SDE framework. For each partially observed field and spatial process parameters, there is an individual conditional forward SDE which diffuses the values of the unobserved locations into white noise and a corresponding conditional reverse SDE which evolves white noise to simulations from the true predictive distribution. While the conditional forward SDEs are known, the corresponding conditional reverse SDEs are unknown due to an intractable conditional score function. To sample from the conditional reverse SDEs, we first train a neural network that depends on the values at the observed locations and the locations themselves to approximate the score function. We also show how the score-function neural estimator can be amortized with respect to the proportion of observed locations, and the parameters of the underlying spatial process. Importantly, training the neural network to approximate the conditional score function only requires simulations from the unconditional distribution. NCS thus implicitly learns the true predictive distributions of spatial processes for which unconditional simulation is fast and efficient yet conditional simulation is not. With NCS, fast and accurate spatial prediction and uncertainty quantification using complex, non-Gaussian spatial processes and partially observed data becomes possible.

To demonstrate the accuracy of NCS, we evaluated how representative NCS-generated simulations are of the true predictive distributions for Gaussian processes (Supplementary Material, Section~\ref{supp:gpvalidation}) and Brown--Resnick processes (Section~\ref{sec:casestudies}), which have tractable and intractable predictive distributions, respectively. Through the use of diagnostics that depend on the joint properties of the predictive distribution, such as the minimum and maximum, we gave empirical evidence that the NCS-approximated conditional simulations are accurate. Finally, we showed that NCS remains computationally efficient regardless of the number of observed locations. We also demonstrated that NCS can be more accurate and comes with fewer limitations than MCMC in practical settings. NCS therefore appears to be ideally suited for efficiently simulating from intractable predictive distributions for a wide class of complex spatial processes.

In this paper, we have presented NCS in the context of spatial process models. However, the methodology presented in Section~\ref{sec:conditionalsde} is sufficiently flexible to also apply to spatial hierarchical models. Specifically, in the case where one has a data model and a latent process model, \eqref{eq:conditionalforwardsde} and \eqref{eq:conditionalreversesde} can be modified slightly so that the target distribution is that of the latent process given the data. Furthermore, we envision that NCS can enable accurate prediction and uncertainty quantification for a broad class of statistical and physical models beyond spatial statistics for which conditional simulation currently necessitates approximations.

\section*{Data Availability Statement}

The code used to produce these results is publicly available at~\url{https://github.com/jmwalchessen/ncs}. The simulation experiments in this paper can be reproduced with the given code.

\section*{Acknowledgments}
Part of this work was done when J.W. visited A.Z.-M. at University of Wollongong and R.H. at the King Abdullah University of Science and Technology (KAUST). We would like to thank Jordan Richards and Brian Trippe for helpful discussions about masking approaches for partially observed spatial data and conditional diffusion, respectively. We would like to acknowledge Microsoft for providing Azure computing resources for this work. J.W., A.Z.-M. and R.H.\ were supported by KAUST grant ORFS-2023-OFP-55020. J.W. and M.K. were supported in part by National Science Foundation grant DMS-2053804, NOAA grant NA21OAR4310258 and a grant from C3.ai Digital Transformation Institute.

\section*{Declaration of Interest}
No competing interests are declared.

\spacingset{1.5}
\bibliographystyle{Chicago}
\bibliography{references}

\clearpage

\setcounter{page}{1}

\beginsupplement
\spacingset{2}
\section{Supplementary Material}

\subsection{Methodology Details}
\label{supp:methods}

This section provides detailed justification for the loss function \eqref{eqn:unconditionalsdelossfunction} in Section~\ref{sec:unconditionalsde} and the discretized forward \eqref{eq:forwardconditionalsdediscretized} and reverse SDEs~\eqref{eq:reverseconditionalsdediscretized} in Section~\ref{sec:discrete}. All the derivations presented here are for the unconditional case; they are applicable to the conditional case with minor adjustments.

\subsubsection{Loss Function for Unconditional Score-Based Diffusion}
\label{supp:supplossfunction}
Here, we demonstrate how the seemingly intractable loss function in the first line of \eqref{eqn:unconditionalsdelossfunction} can be rewritten as the tractable expression in the second line via a derivation from \citet{Vincent2011}. First, expand the norm difference in the intractable loss function via matrix multiplication:
\begin{equation}
\begin{aligned}
\ell(\bm{\phi}) & = \mathbb{E}_{t\sim U(0,T)}\Big(\lambda(t)\mathbb{E}_{\bm{x}_{t}} \big(\norm{\bm{s}_{\bm{\phi}}(\bm{x}_{t},t)-\nabla_{\bm{x}_{t}} \log(p(\bm{x}_{t}))}^{2}\big)\Big)\\
& = \mathbb{E}_{t\sim U(0,T)}\Big(\lambda(t)\mathbb{E}_{\bm{x}_{t}}\big(\langle \bm{s}_{\bm{\phi}}(\bm{x}_{t},t)-\nabla_{\bm{x}_{t}} \log(p(\bm{x}_{t})),  \bm{s}_{\bm{\phi}}(\bm{x}_{t},t)-\nabla_{\bm{x}_{t}} \log(p(\bm{x}_{t})) \rangle\big)\Big)\\
& = \mathbb{E}_{t\sim U(0,T)}\Big(\lambda(t)\big(\mathbb{E}_{\bm{x}_{t}} (\norm{\bm{s}_{\bm{\phi}}(\bm{x}_{t},t)}^{2})-2\mathbb{E}_{\bm{x}_{t}} (\la \bm{s}_{\bm{\phi}}(\bm{x}_{t},t), \nabla_{\bm{x}_{t}} \log(p(\bm{x}_{t})) \ra)\\
&\hspace{2cm} +\mathbb{E}_{\bm{x}_{t}}(\norm{\nabla_{\bm{x}_{t}}\log(p(\bm{x}_{t}))}^{2})\big)\Big),
\end{aligned}
\label{eqn:expandedlossfunction}
\end{equation} 
where $\la \cdot ,\cdot \ra$ is the inner product operator. Second, rewrite the second term in the expansion as follows,
\begin{subequations}
\begin{align}
\mathbb{E}_{\bm{x}_{t}} \big(\la \bm{s}_{\bm{\phi}}(\bm{x}_{t},t), \nabla_{\bm{x}_{t}} \log(p(\bm{x}_{t})) \ra\big)&=
\int p(\bm{x}_{t})  \la \bm{s}_{\bm{\phi}}(\bm{x}_{t},t), \nabla_{\bm{x}_{t}} \log(p(\bm{x}_{t})) \ra \mathrm{d} \bm{x}_{t} \nonumber\\
& \hspace{-8em}= \int p(\bm{x}_{t}) \la \bm{s}_{\bm{\phi}}(\bm{x}_{t},t),\frac{\nabla_{\bm{x}_{t}} p(\bm{x}_{t})}{p(\bm{x}_{t})} \ra \mathrm{d} \bm{x}_{t} \nonumber\\
& \hspace{-8em} = \int \la \bm{s}_{\bm{\phi}}(\bm{x}_{t},t), \nabla_{\bm{x}_{t}} p(\bm{x}_{t}) \ra \mathrm{d} \bm{x}_{t} \nonumber\\
& \hspace{-8em}= \int \la \bm{s}_{\bm{\phi}}(\bm{x}_{t},t), \nabla_{\bm{x}_{t}} \int p(\bm{x}_{0}) p(\bm{x}_{t} \mid \bm{x}_{0}) \mathrm{d} \bm{x}_{0} \ra \mathrm{d} \bm{x}_{t} \nonumber\\
& \hspace{-8em} =  \int \la \bm{s}_{\bm{\phi}}(\bm{x}_{t},t), \int p(\bm{x}_{0}) \nabla_{\bm{x}_{t}} p(\bm{x}_{t} \mid \bm{x}_{0}) \mathrm{d} \bm{x}_{0} \ra \mathrm{d} \bm{x}_{t} \nonumber\\
& \hspace{-8em} = \int \la \bm{s}_{\bm{\phi}}(\bm{x}_{t},t), \int p(\bm{x}_{0}) p(\bm{x}_{t} \mid \bm{x}_{0}) \nabla_{\bm{x}_{t}} \log(p(\bm{x}_{t} \mid \bm{x}_{0})) \mathrm{d} \bm{x}_{0} \ra \mathrm{d} \bm{x}_{t} \label{eqn:secondterm1}\\
& \hspace{-8em} = \iint p(\bm{x}_{0})p(\bm{x}_{t} \mid \bm{x}_{0}) \la \bm{s}_{\bm{\phi}}(\bm{x}_{t},t), \nabla_{\bm{x}_{t}} \log(p(\bm{x}_{t} \mid \bm{x}_{0})) \ra \mathrm{d} \bm{x}_{0} \mathrm{d} \bm{x}_{t} \nonumber \\
& \hspace{-8em} = \iint p(\bm{x}_{0})p(\bm{x}_{t} \mid \bm{x}_{0}) \la \bm{s}_{\bm{\phi}}(\bm{x}_{t},t), \nabla_{\bm{x}_{t}} \log(p(\bm{x}_{t} \mid \bm{x}_{0})) \ra \mathrm{d} \bm{x}_{t} \mathrm{d} \bm{x}_{0} \label{eqn:secondterm2}\\
& \hspace{-8em} = \mathbb{E}_{\bm{x}_{0}\sim p(\bm{x}_{0})} \mathbb{E}_{\bm{x}_{t} \sim p(\bm{x}_{t} \mid \bm{x}_{0})} \la \bm{s}_{\bm{\phi}}(\bm{x}_{t},t), \nabla_{\bm{x}_{t}} \log(p(\bm{x}_{t} \mid \bm{x}_{0})) \ra \nonumber,
\end{align}
\end{subequations} 
where \eqref{eqn:secondterm1} is a rearrangement of the gradient $\nabla_{\bm{x}_{t}} \log(p(\bm{x}_{t} \mid \bm{x}_{0}))$ and \eqref{eqn:secondterm2} follows from Fubini's Theorem. Next, the first term in the expansion of $\ell(\bm{\phi})$ can be reworked via the Law of Total Expectation,
\begin{equation}
\mathbb{E}_{\bm{x}_{t}\sim p(\bm{x}_{t})}( \norm{\bm{s}_{\bm{\phi}}(\bm{x}_{t},t)}^{2}) = \mathbb{E}_{\bm{x}_{0}\sim p(\bm{x}_{0})} \mathbb{E}_{\bm{x}_{t} \sim p(\bm{x}_{t} \mid \bm{x}_{0})}(\norm{\bm{s}_{\bm{\phi}}(\bm{x}_{t},t)}^{2}).
\end{equation}

Finally, substitute the reworked first and second terms into the expanded loss function in \eqref{eqn:expandedlossfunction} and minimize $\ell(\bm{\phi})$ with respect to the parameters $\bm{\phi}$ according to
\begin{subequations}
\small
\begin{align}
\bm{\phi}^{*}& = \argmin_{\bm{\phi}} \mathbb{E}_{t\sim U(0,T)}\Bigg(\lambda(t)\Big(\mathbb{E}_{\bm{x}_{t}} \big((\norm{\bm{s}_{\bm{\phi}}(\bm{x}_{t},t)}^{2})-2\mathbb{E}_{\bm{x}_{t}} (\la \bm{s}_{\bm{\phi}}(\bm{x}_{t},t), \nabla_{\bm{x}_{t}} \log(p(\bm{x}_{t})) \ra)\\
& \hspace{9em} +\mathbb{E}_{\bm{x}_{t}}(\norm{\nabla_{\bm{x}_{t}}\log(p(\bm{x}_{t}))}^{2})\big)\Big)\Bigg) \nonumber \\
& = \argmin_{\bm{\phi}} \mathbb{E}_{t\sim U(0,T)}\Bigg( \lambda(t)\Big(\mathbb{E}_{\bm{x}_{0}\sim p(\bm{x}_{0})} \mathbb{E}_{\bm{x}_{t} \sim p(\bm{x}_{t} \mid \bm{x}_{0})}(\norm{\bm{s}_{\bm{\phi}}(\bm{x}_{t},t)}^{2}) \nonumber \\
&\hspace{9em} -2 \mathbb{E}_{\bm{x}_{0}\sim p(\bm{x}_{0})} \mathbb{E}_{\bm{x}_{t} \sim p(\bm{x}_{t} \mid \bm{x}_{0})} \la \bm{s}_{\bm{\phi}}(\bm{x}_{t},t), \nabla_{\bm{x}_{t}} \log(p(\bm{x}_{t} \mid \bm{x}_{0})) \ra + \mathbb{E}_{\bm{x}_{t}}(\norm{\nabla_{\bm{x}_{t}}\log(p(\bm{x}_{t}))}^{2})\Big)\Bigg) \nonumber \\
& = \argmin_{\bm{\phi}} \mathbb{E}_{t\sim U(0,T)}\Big(\lambda(t)\mathbb{E}_{\bm{x}_{0}\sim p(\bm{x}_{0})} \mathbb{E}_{\bm{x}_{t}\sim p(\bm{x}_{t} \mid \bm{x}_{0})} \big(\norm{\bm{s}_{\bm{\phi}}(\bm{x}_{t},t)}^{2} - 2 \la \bm{s}_{\bm{\phi}}(\bm{x}_{t},t), \nabla_{\bm{x}_{t}} \log(p(\bm{x}_{t} \mid \bm{x}_{0})) \ra \big)\Big) \nonumber \\
& = \argmin_{\bm{\phi}} \mathbb{E}_{t\sim U(0,T)}\Big(\lambda(t) \mathbb{E}_{\bm{x}_{0}\sim p(\bm{x}_{0})} \mathbb{E}_{\bm{x}_{t}\sim p(\bm{x}_{t} \mid \bm{x}_{0})} \big(\norm{\bm{s}_{\bm{\phi}}(\bm{x}_{t},t)}^{2} - 2 \la \bm{s}_{\bm{\phi}}(\bm{x}_{t},t), \nabla_{\bm{x}_{t}} \log(p(\bm{x}_{t} \mid \bm{x}_{0})) \ra \nonumber \\
& \hspace{10em} + \norm{\nabla_{\bm{x}_{t}} \log(p(\bm{x}_{t} \mid \bm{x}_{0}))}^{2}\big)\Big) \nonumber \\
& = \argmin_{\bm{\phi}} \mathbb{E}_{t\sim U(0,T)}\Big(\lambda(t)\mathbb{E}_{\bm{x}_{0}\sim p(\bm{x}_{0})} \mathbb{E}_{\bm{x}_{t}\sim p(\bm{x}_{t} \mid \bm{x}_{0})} \big(\norm{\bm{s}_{\bm{\phi}}(\bm{x}_{t},t) -\nabla_{\bm{x}_{t}} \log(p(\bm{x}_{t} \mid \bm{x}_{0}))}^{2}\big)\Big). \label{eqn:finallossfunction}
\end{align}
\end{subequations}
\normalsize

\subsubsection{Discretization of the Forward and Reverse SDEs}
\label{supp:suppdiscretization}
Here, we provide the details on the time-discretized forward and reverse SDEs for the unconditional case. Note that the derivation is applicable to the conditional case, as well, with minor adjustments. For simplicity, we assume $T$ is sufficiently large such that ${{\Delta{t}/T} \approx 0}$ for $\Delta t = 1$.

To emphasize the connection between denoising diffusion probabilistic models (DDPM) and the VPSDE framework for score-based diffusion, we start by writing down our derivations from the forward and reverse processes of a DDPM \citep{Ho2020,Chan2025}.  The discretized forward process of a DDPM is
\begin{equation}
\begin{aligned}
\bm{x}_{t+1}& = \sqrt{1-\beta_{t}} \bm{x}_{t} + \sqrt{\beta_{t}}\bm{\epsilon}, \quad \bm{\epsilon}\sim \mathcal{N}(\bm{0}, \bm{I}),\\
&  = \big(1-{\beta_{t}/2} + \smallO(\beta_{t}^{2})\big) \bm{x}_{t} + \sqrt{\beta_{t}} \bm{\epsilon}\\
& \approx \big(1-{\beta_{t}/2}\big) \bm{x}_{t} + \sqrt{\beta_{t}} \bm{\epsilon},
\end{aligned}
\end{equation}
and therefore,
\begin{equation}
\begin{aligned}
\bm{x}_{t+1} - \bm{x}_{t}& \approx -{\beta_{t} \bm{x}_{t}/2} + \sqrt{\beta_{t}} \bm{\epsilon}\\
& = \bm{f}(\bm{x}_{t},t) + g(t)\bm{\epsilon},
\end{aligned}
\end{equation}
where $\bm{f}(\bm{x}_{t},t) = -{\beta_{t}\bm{x}_{t}}/2$, $g(t)=\sqrt{\beta_{t}}$, and the first approximation follows due to a MacLaurin series expansion of $h(\beta_{t})=\sqrt{1-\beta_{t}}$ for $\beta_{t}$ chosen such that $\beta_{t}\approx 0$ for $t\in [0,T]$. The discretized reverse process of a DDPM is
\begin{equation}
\begin{aligned}
\bm{x}_{t-1}& = (1-\beta_{t})^{-\frac{1}{2}}\big(\bm{x}_{t} + \beta_{t}\bm{s}_{\bm{\phi}}(\bm{x}_{t},t)\big)+\sqrt{\beta_{t}} \bm{\epsilon}, \quad \bm{\epsilon}\sim \mathcal{N}(\bm{0}, \bm{I}),\\
& = \big(1+{\beta_{t}/2}+\smallO(\beta_{t}^{2})\big)\big(\bm{x}_{t} + \beta_{t}\bm{s}_{\bm{\phi}}(\bm{x}_{t},t)\big)+\sqrt{\beta_{t}} \bm{\epsilon}\\
& \approx \big(1+{\beta_{t}/2}\big)\big(\bm{x}_{t} + \beta_{t}\bm{s}_{\bm{\phi}}(\bm{x}_{t},t)\big)+\sqrt{\beta_{t}} \bm{\epsilon}\\
& = \big(1+{\beta_{t}/2}\big)\bm{x}_{t} + \beta_{t} \big(1+{\beta_{t}/2}\big)\bm{s}_{\bm{\phi}}(\bm{x}_{t},t) + \sqrt{\beta_{t}} \bm{\epsilon}\\
& = \big(1+{\beta_{t}/2}\big)\bm{x}_{t} + \beta_{t} \bm{s}_{\bm{\phi}}(\bm{x}_{t},t) + \frac{\beta_{t}^{2}}{2} \bm{s}_{\bm{\phi}}(\bm{x}_{t},t) + \sqrt{\beta_{t}} \bm{\epsilon}\\
& \approx \big(1+{\beta_{t}/2}\big) \bm{x}_{t} + \beta_{t} \bm{s}_{\bm{\phi}}(\bm{x}_{t},t) + \sqrt{\beta_{t}} \bm{\epsilon}, \quad  \textrm{since }\beta_{t}^{2} \approx 0,
\end{aligned}
\end{equation}
and therefore,
\begin{equation}
\begin{aligned}
\bm{x}_{t-1} - \bm{x}_{t}& \approx {\beta_{t}\bm{x}_{t}}/2 + \beta_{t} \bm{s}_{\bm{\phi}}(\bm{x}_{t},t)+\sqrt{\beta_{t}} \bm{\epsilon}\\
& \approx -\bm{f}(\bm{x}_{t},t)+g(t)^{2}\bm{s}_{\bm{\phi}}(\bm{x}_{t},t)+g(t)\bm{\epsilon},\\
\end{aligned}
\end{equation}
where $\bm{f}(\bm{x}_{t},t) = -{\beta_{t}\bm{x}_{t}}/2$, $g(t)=\sqrt{\beta_{t}}$, and the first approximation follows due to a MacLaurin series expansion of $h(\beta_{t})=(1-\beta_{t})^{-\frac{1}{2}}$ for $\beta_{t}$ chosen such that $\beta_{t}\approx 0$ for $t\in [0,T]$.

\subsection{Case Study Methodology}
\label{supp:casestudymethods}
\subsubsection{Brief Description of Spatial Processes}
\label{supp:spatialprocessdescription}
\paragraph*{Gaussian Process}
A Gaussian process is a stochastic process whose finite-dimensional distributions are Gaussian. Its predictive distributions are analytically available, and therefore we can use them to validate NCS. Results from applying NCS to Gaussian processes are shown in Supplementary Material, Section~\ref{supp:gpvalidation}.

\paragraph*{Brown--Resnick Process}

Here, we give more details on the Brown--Resnick process employed in Section~\ref{sec:casestudymethodology}. Max-stable processes, such as Brown--Resnick processes, can be expressed as the pointwise maxima of an infinite collection of random spatial processes \citep{Davison, Huser2019, Davison2019}. More specifically, a max-stable process has the following construction \citep{deHaan1985}
\begin{equation}
    Z(\bm{s}) = \max_{i=1,2,\ldots} \eta_{i} W_{i}(\bm{s}), \textrm{ for } \bm{s} \textrm{ on a domain }\mathcal{D}\subset \mathbb{R}^{d},
\label{eqn:maxstable}
\end{equation}  
in which $\{\eta_{i}\}_{i=1}^{\infty}$ is a Poisson point process on $(0,\infty)$ with intensity function $\d\Lambda(\eta) = \eta^{-2} \d \eta$ and $\{W_{i}(\bm{s})\}$ are independent copies of a non-negative stochastic process with mean one \citep{Kabluchko2009}. Different types of max-stable processes arise from the definition of $W_{i}(\bm{s})$. For a Brown--Resnick model, the process $W_{i}(\bm{s})$ is log-normal and has the form:
\begin{equation}
W_i(\bm{s})=\exp(\epsilon_i(\bm{s}) - \gamma(\bm{s})), \quad i \in \mathbb{N}^{+}
\end{equation}
where $\{\epsilon_i(\bm{s})\}_{i \in \mathbb{N}^{+}}$ are realizations of an intrinsically stationary Gaussian process, conventionally with $\epsilon_{i}(\bm{s})=0$ almost surely, and with semivariogram $\gamma(\bm{h})$, $i\in \mathbb{N}^{+}$, where $\mathbb{N}^{+}$ denotes the positive integers. A common choice, which we also adopt in this work, is to set $\gamma(\bm{h})=(\slfrac{\norm{\bm{h}}}{\lambda})^{\nu}$, with $\bm{h}$ the spatial separation between two locations in spatial domain $\mathcal{D}$ , $\lambda \in \mathbb{R}^{+}$ a range parameter, and $\nu\in (0,2]$ a smoothness parameter \citep{Castruccio}.

\subsubsection{Training Details Common to the Gaussian Process and Brown--Resnick Process Case Studies}
\label{supp:train}

\paragraph*{Weighting Function}
In our loss function \eqref{eq:conditionallossfunction}, we use the same weighting function $\lambda(\cdot)$ as in \citet{Song2021}, which is proportional to the inverse of the expectation of the conditional score function. This is equivalent to setting $\lambda(t)=\underline{\sigma}_{t}^{2}$ for a variance preserving SDE (VPSDE), since, from \eqref{eq:transitionkernel},
\begin{equation}
\begin{aligned}
\lambda(t) & = \Big(\mathbb{E}\big(\norm{\nabla_{\tilde{\bm{x}}_{t}} \log(p (\tilde{\bm{x}}_{t}\mid \bm{x}_{0}))}_{2}^{2}\big)\Big)^{-1} \\
& = \Big(\frac{1}{\underline{\sigma}_{t}^{2}}\mathbb{E}(\norm{\tilde{\bm{\epsilon}}}_{2}^{2})\Big)^{-1} \textrm{ such that } \tilde{\bm{\epsilon}} \sim \mathcal{N}(\tilde{\bm{0}},\tilde{\bm{I}})\\
& = \underline{\sigma}_{t}^{2}
\end{aligned}
\label{eqn:weightingfunction}
\end{equation}

\paragraph*{Spatial Process Parameters}
In the main text, we consider two differently amortized U-Nets, one amortized with respect to a spatial process parameter referred to as a parameter U-Net and the other amortized with respect to the proportion of missing observations referred to as a proportion U-Net. For the parameter U-Net, we consider the parameter space $\Theta = [1,5] \times \{1.5\}$, where $\nu = 1.5$ is fixed for both processes described in Supplementary Material, Section~\ref{supp:spatialprocessdescription}, and where the length scale $\ell$ and range $\lambda$ vary from one to five for the Gaussian process and Brown--Resnick process, respectively. We define the training parameter space $\tilde{\Theta}$---the space from which we draw unconditional simulations during training---to extend beyond $\Theta = [1,5]\times \{1.5\}$ to avoid suboptimal behavior of the U-Net at the boundary. On the other hand, when training the proportion U-Net, we do not vary the spatial process parameters. Specifically, we fix the length scale $\ell = 3$ and variance $\tau^{2} = 1.5$ for the Gaussian process, and the range $\lambda = 3$ and smoothness $\nu = 1.5$ for the Brown--Resnick process. Therefore, for the proportion U-Net, $\Theta = \tilde{\Theta} = \{3\} \times \{1.5\}$.

\paragraph*{Simulating Training and Validation Data}

We employ data generation ``on the fly." That is, we simulate a dataset, use it to train and evaluate the diffusion model for a fixed number of epochs, and repeat the process with a new simulated dataset each time. This approach avoids overfitting and minimizes data storage \citep{Chan2018}.

%For each data draw, we simulate a new dataset and use it to train and evaluate the diffusion model for a fixed number of epochs.

We simulated training and validation data for the parameter and proportion U-Nets according to the following procedure: For each data draw, we first simulate $p$ parameters from the training parameter space $\Tilde{\Theta}$, and for each of those $p$ parameters, we simulate $s$ full spatial field realizations via the unconditional simulator. Note that for the proportion U-Net, these $p$ parameter vectors are identical. Finally, we simulate $m$ masks and $m$ discrete timesteps uniformly from $\{0,1,\dots,T\}\subset \mathbb{Z}$ per each full spatial field realization. Each mask is simulated using the Bernoulli distribution with $\rho$---the probability of observing a single location---for all $n=32^{2}$ locations, and our convention is that $0$ indicates missingness. For the parameter U-Net, $\rho = 0.05$ is kept fixed. For the proportion U-Net, we sample $\rho$ uniformly from $[0.01,0.525]$ during training to avoid boundary issues during evaluation, where we instead restrict the space of proportions to $[0.01,0.5]$. Algorithm~\ref{algo:trainingdata} illustrates this data generation process. Once finished with simulation, we have training data of the form:
\begin{equation}
\{(\bm{\mathring{x}}_{0}',\tilde{\bm{x}}_{t}')', \bm{M}(\mathring{\mathcal{S}}),\bm{\theta},t\}
\label{eqn:dataform}
\end{equation}
where 
$\bm{\mathring{x}}_{0} = (x_{0,j}: M_{j}(\mathring{\mathcal{S}})=1)', \quad \bm{\tilde{x}}_{0} = (x_{0,j}: M_{j}(\mathring{\mathcal{S}})=0)', \quad \tilde{\bm{x}}_{t}=\sqrt{\underline{\alpha}_{t}}\cdot \tilde{\bm{x}}_{0}+\underline{\sigma}_{t}\tilde{\bm{\epsilon}}, \quad \tilde{\bm{\epsilon}}\sim \mathcal{N}(\tilde{\bm{0}},\tilde{\bm{I}})$, and where $M_{j}(\mathring{\mathcal{S}})$ is defined in \eqref{eqn:mask}.

\spacingset{1}
\begin{algorithm}[t!]
\caption{Simulating Training and Validation Data for the Parameter and Proportion U-Nets}
\begin{algorithmic}
\STATE Set $\mathcal{D}=\emptyset$
\STATE Parameter U-Net: Set $q_{L}=q_{U}=0.05$, $\tilde{\Theta}=[.5,5.5]\times \{1.5\}$, $\Theta=[1,5]\times \{1.5\}$.
\STATE Proportion U-Net: Set $q_{L}=0.01$, $\Theta = \{3\}\times \{1.5\}$, and $q_{U}=0.525$ for the training data and $q_{U}=0.5$ for the validation data. 
\STATE Sample proportions of observed locations $q_{1},\dots,q_{r}\sim U([q_{L},q_{U}])$.
\FOR{$q_{i}$ in $\{q_{i}\}_{i\in [r]}$}
\STATE Sample spatial process parameters $\bm{\theta}_{1},\dots,\bm{\theta}_{p} \sim U(\tilde{\Theta})$.
\FOR{$\bm{\theta}_{j}$ in $\{\bm{\theta}_{j}\}_{j\in [p]}$}
\STATE Simulate unconditionally $\{\bm{x}_{0,k}\}_{k \in [s]} \sim p(\cdot \mid \bm{\theta}_{j})$.
\FOR{$\bm{x}_{0,k}\in \{\bm{x}_{0,k}\}_{k \in [s]}$}
\STATE $\bm{M}(\mathring{\mathcal{S}}_{l}) \sim \textrm{Bin}(n = 32^{2},p = q_{i})$ \textrm{ for } $l\in [m]$
\STATE $t_{l}\sim U(\{0,1,\dots,T\}) \textrm{ for } l\in [m]$
\STATE $\bm{\mathring{x}}_{0,l}=(x_{0,k,j}: M_{j}(\mathring{\mathcal{S}}_{l})=1)'\textrm{ for } l \in [m]$
\STATE $\tilde{\bm{x}}_{0,l}=(x_{0,k,j}: M_{j}(\mathring{\mathcal{S}}_{l})=0)'\textrm{ for } l \in [m]$
\STATE $\tilde{\bm{x}}_{t_{l}}\sim p(\cdot \mid \bm{\mathring{x}}_{0,l}, \tilde{\bm{x}}_{0,l}):=\mathcal{N}(\sqrt{\underline{\alpha}_{t}}\tilde{\bm{x}}_{0,l}, \underline{\sigma}_{t}^{2}\tilde{\bm{I}}) \textrm{ for } l \in [m]$
\STATE $\mathcal{D} \leftarrow \mathcal{D} \cup \{ \{(\bm{\mathring{x}}_{0,l}',\tilde{\bm{x}}_{t_{l}}')',\bm{M}(\mathring{\mathcal{S}}_{l}),\bm{\theta}_{j},t_{l}\big\} : l \in [m]\}$
\ENDFOR
\ENDFOR
\ENDFOR
\STATE Return $\mathcal{D}$
\end{algorithmic}
\label{algo:trainingdata}
\end{algorithm}
\spacingset{2}

\subsubsection{Validation Details Common to Both the Gaussian Process and Brown--Resnick Process Case Studies}
\paragraph{Generation of Validation Datasets}
\label{supp:validation}
In this section, we give details on how we generate test data for the simulation experiments conducted in Section~\ref{sec:casestudies} and the Supplementary Material, Sections~\ref{supp:GPcasestudy} and~\ref{supp:br}.

\subparagraph*{Conditional Validation Dataset (Proportion U-Net)} For each of the five observed proportions $\rho \in \{0.01,0.05,0.1,0.25,0.5\}$, we generated a conditioning set $\mathring{\mathcal{S}}$ and mask $\bm{M}(\mathring{\mathcal{S}})$ using the Bernoulli distribution with probability $\rho$, and a reference spatial field $\bm{x}_{0}$ from its unconditional distribution with either variance $\tau^{2} = 1.5$ and length scale $\ell = 3$ (Gaussian process), or smoothness $\nu = 1.5$ and range $\lambda = 3$ (Brown--Resnick process). For each of the resulting five partially observed spatial fields $\bm{\mathring{x}}_{0}=(x_{0,j}: M_{j}(\mathring{\mathcal{S}})=1)'$, we generated $m = 4000$ replicates of the unobserved spatial field $\{\tilde{\bm{x}}_{0,i}\}_{i\in [m]}$ with NCS and either the true predictive distribution (Gaussian process) or an MCMC approximation (Brown--Resnick process). We use this dataset in Section~\ref{sec:brvalidation} to validate NCS when the percentage of observed locations varies.

\subparagraph*{Conditional Validation Dataset (Parameter U-Net)} In the Gaussian process case, we unconditionally simulated a reference spatial field $\bm{x}_{0}$ for each parameter length scale $\ell\in\{1,\ldots,5\}$ and variance $\tau^2=1.5$. In the Brown--Resnick case, we unconditionally simulated a reference spatial field $\bm{x}_{0}$ for each range parameter $\lambda\in\{1,\ldots,5\}$ and smoothness $\nu=1.5$. Then, for each reference spatial field, we generated a conditioning set $\mathring{\mathcal{S}}$ and corresponding mask $\bm{M}(\mathring{\mathcal{S}})$ using the Bernoulli distribution with probability $\rho=0.05$ of observing each location. In each case, using each of the five partially observed fields $\bm{\mathring{x}}_{0}=(x_{0,j}: M_{j}(\mathring{\mathcal{S}})=1)'$, we generated $m=4000$ unobserved spatial field replicates $\{\tilde{\bm{x}}_{0,i}\}_{i\in [m]}$ with NCS and either the true predictive distribution (Gaussian process) or an MCMC approximation (Brown--Resnick process). We use this dataset in Section~\ref{sec:brvalidation} to validate NCS when the spatial process parameter varies.

\paragraph{Diagnostic and Validation Metrics}
\label{supp:validationmetricsboth}
In this section, we describe diagnostics and metrics used to assess the quality of conditional simulation in Sections~\ref{sec:results_small},~\ref{sec:brvalidation},~\ref{sec:da},~\ref{supp:gpvalidationmetric} and~\ref{supp:brvalidation}.

\subparagraph*{Conditional Univariate and Bivariate Densities}
Conditional univariate and bivariate densities are estimated using the conditional validation datasets and kernel density estimation with a Gaussian kernel and bandwidth selected according to Scott's Rule \citep{Scott1992} on the conditional validation datasets. Specifically, in our results (e.g., Figure~\ref{fig:brdensity}), we show a $5\times 3$ grid of plots in which the first column contains the partially observed spatial fields and the second and third columns contain the conditional univariate and bivariate kernel density estimates respectively for the selected unobserved locations.

\subparagraph*{Conditional Mean Fields}
The empirical conditional mean fields are computed by averaging the $m=4000$ replicates at each of the unobserved locations in the conditional validation dataset. In our results (e.g., Figures~\ref{fig:gpconditionalmean} and~\ref{fig:brconditionalmean}), we display a $5\times 3$ grid of plots that contains the partially observed spatial fields in the first column, the true empirical (Gaussian) or an MCMC-approximated (Brown--Resnick) conditional mean field in the second column, and the empirical NCS conditional mean field in the third column.

\subsection{Gaussian Process: Tractable Case Study}
\label{supp:GPcasestudy}
We applied NCS to a Gaussian process with zero mean and exponential covariance function parametrized by the length scale $\ell>0$ and the variance $\tau^{2}>0$. Hence, the model is
\begin{equation}
\begin{aligned}
h&\sim \textrm{GP}(0,k(\bm{s}_{1},\bm{s}_{2})), \quad\textrm{where}\quad k(\bm{s}_{1},\bm{s}_{2})=\tau^{2}\exp(-\slfrac{\norm{\bm{s}_{1}-\bm{s}_{2}}_{2}}{\ell}),\textrm{ }\bm{s}_{1},\bm{s}_{2}\in \mathcal{S}.
\end{aligned}
\label{eqn:gpkernel}
\end{equation}
%where $\bm{s}_{1},\bm{s}_{2}\in \mathcal{S}$.

\subsubsection{Training Details}
\label{supp:gptrain}

\paragraph*{Parameter U-Net} For the parameter U-Net in the Gaussian process case, we fixed the variance to $\tau^2 = 1.5$ and the proportion of observed locations to $\rho = 0.05$, but let the length scale $\ell \in [1.0,5.0]$ vary. To simulate training data, we follow the process described in Section~\ref{supp:train}. Specifically, for training, we simulated $p=1000$ parameters samples from $\tilde{\Theta}=[0.5,5.5]\times \{1.5\}$, $s$ = 1 full spatial field realizations per parameter sample, and $m = 100$ masks and timesteps per full spatial field realization.

\paragraph*{Proportion U-Net} For the proportion U-Net in the Gaussian process case, we fixed the variance to $\tau^{2} = 1.5$, the length scale to $\ell = 3$, while we allowed the proportion of observed locations $\rho \in [0.01,0.5]$ to vary. We follow the process in Section~\ref{supp:train} for simulating training data. Specifically, for training, we sampled $r = 50$ observed proportions $\rho$ uniformly from $[0.01,0.525]$, $p=1$ parameters uniformly from $\Theta = \tilde{\Theta}=\{3\}\times \{1.5\}$ for each $\rho$, $s = 25$ full spatial field realizations for each $\rho$, and used $m = 100$ masks and timesteps for each full spatial field realization.

\subsubsection{Validation Metrics Particular to the Gaussian Process Case Study}
\label{supp:gpvalidationmetric}
\paragraph*{Conditional Correlation Metrics}
To assess how well the NCS simulations capture the spatial correlations induced by the true predictive distributions in the Gaussian case study, we produce empirical conditional correlation heatmaps (Figure~\ref{fig:gpcorrelationheatmap}). For a fixed unobserved spatial location in a partially observed spatial field from the conditional validation dataset, we compute the Pearson correlation coefficient (PCC) between
the unobserved spatial location and all other unobserved spatial locations using both NCS simulations and conditional simulations from the true predictive distribution. PCC measures the linear correlation between the process evaluated at two unobserved spatial locations $\bm{s}_{1}, \bm{s}_{2}$ given the partially observed spatial field. If the PCC is $-1$ or $1$, there is perfect linear correlation given the partially observed spatial field, while a PCC value of $0$ indicates no linear correlation given the partially observed field. Specifically, we display a $5\times 2$ grid of plots for both the parameter and proportion U-Nets, in which the first column contains the true empirical correlation heat maps and the second column contains the heatmaps derived from NCS.

\subsubsection{Results}
\label{supp:gpvalidation}
\begin{figure}[!t]
    \centering
    \includegraphics[scale = .077]{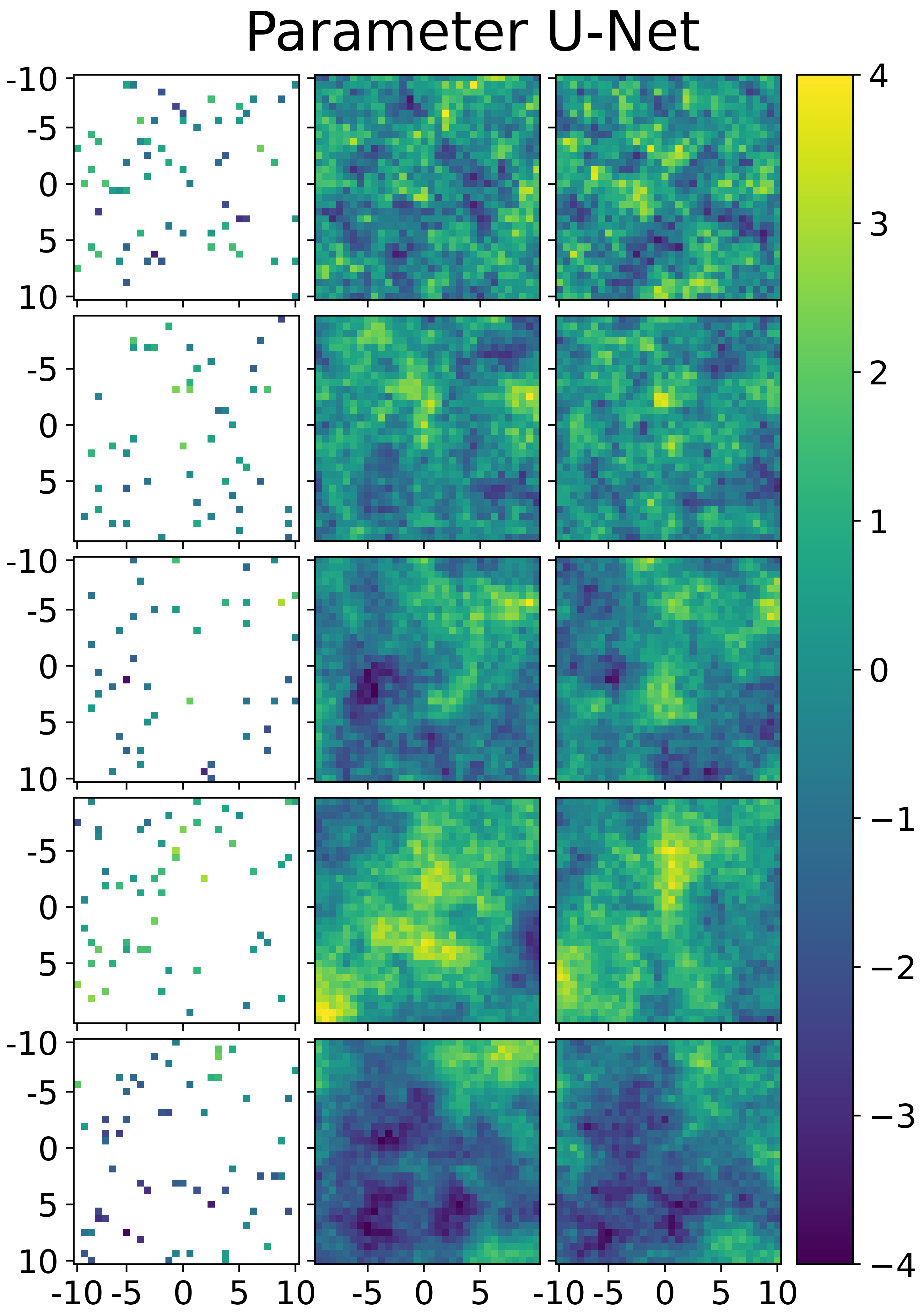}
    \includegraphics[scale = .077]{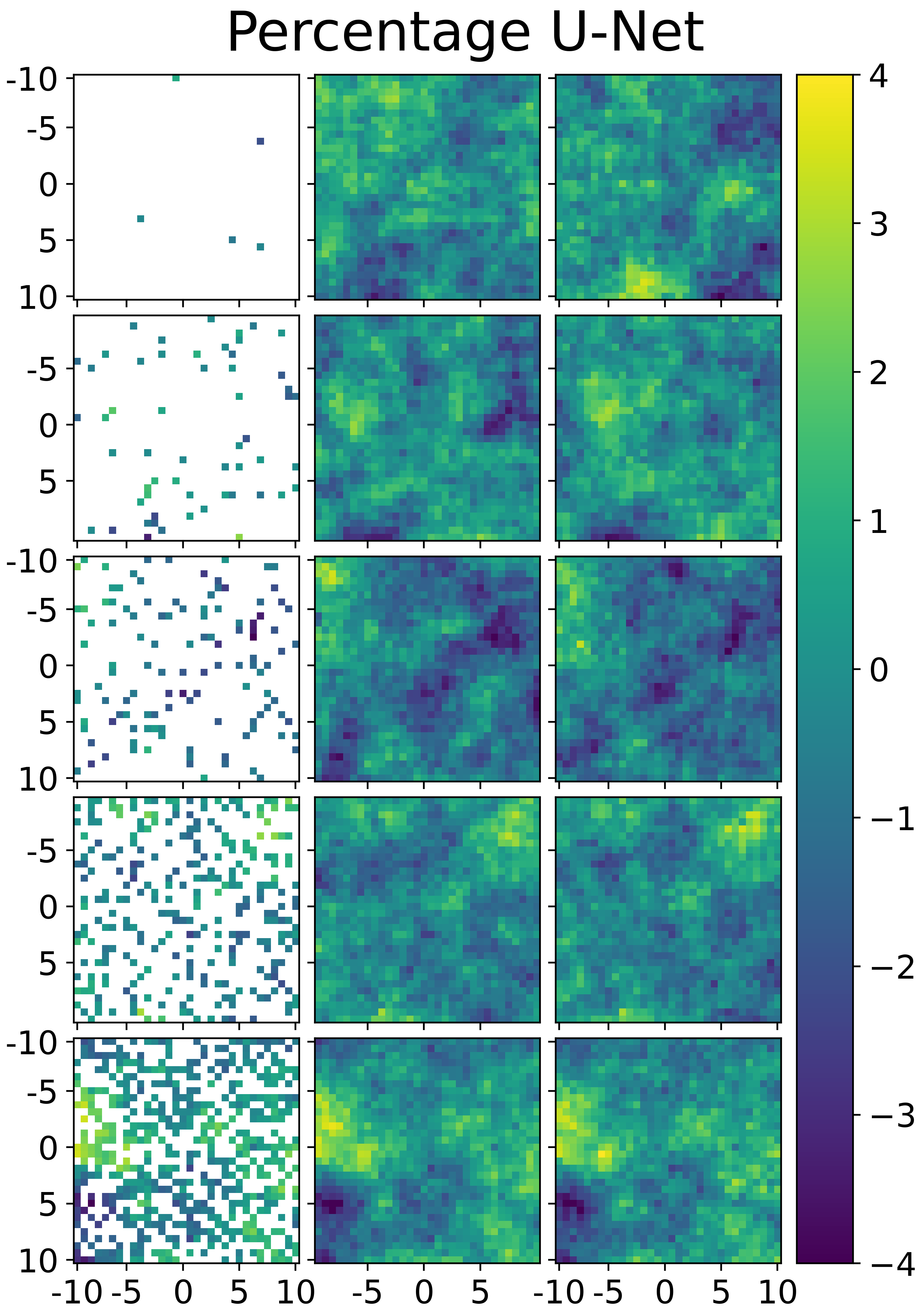}
    \caption{Left panel: NCS with a Gaussian process with variance $\tau^{2} = 1.5$, length scale  $\ell \in \{1,2,3,4,5\}$ (from top to bottom), and proportion of observed locations $\rho = 0.05$. Right panel: NCS with a Gaussian process with variance $\tau^{2} = 1.5$, length scale $\ell = 3$, and observed proportions $\rho \in \{0.01,0.05,0.1,0.25,0.5\}$ (from top to bottom). For each panel: Left columns: Observations. Middle columns: Underlying spatial fields (generated using unconditional simulation, only partially observed). Right columns: Conditional simulations with NCS implemented with a parameter U-Net (left panel) or a proportion U-Net (right panel).}
\label{fig:gpvisualization}
\end{figure}

\paragraph*{Visualizations} In Figure~\ref{fig:gpvisualization}, we show simulations from the true predictive distributions and NCS. These simulations display similar spatial patterns and seamlessly incorporate the partially observed field as intended. Additionally, amortization appears to be highly effective: as the length scale increases in the left panel, the spatial correlations increase in the NCS simulations, mirroring the true conditional simulations. Further, no matter the number of observed spatial locations, NCS seamlessly integrates the simulated unobserved values with the partially observed field as shown in the right panel. From these results, we conclude that NCS simulations are reasonable as one cannot discriminate between the truth and NCS by the naked eye.

\begin{figure}[!t]
    \centering
    \includegraphics[scale = .085]{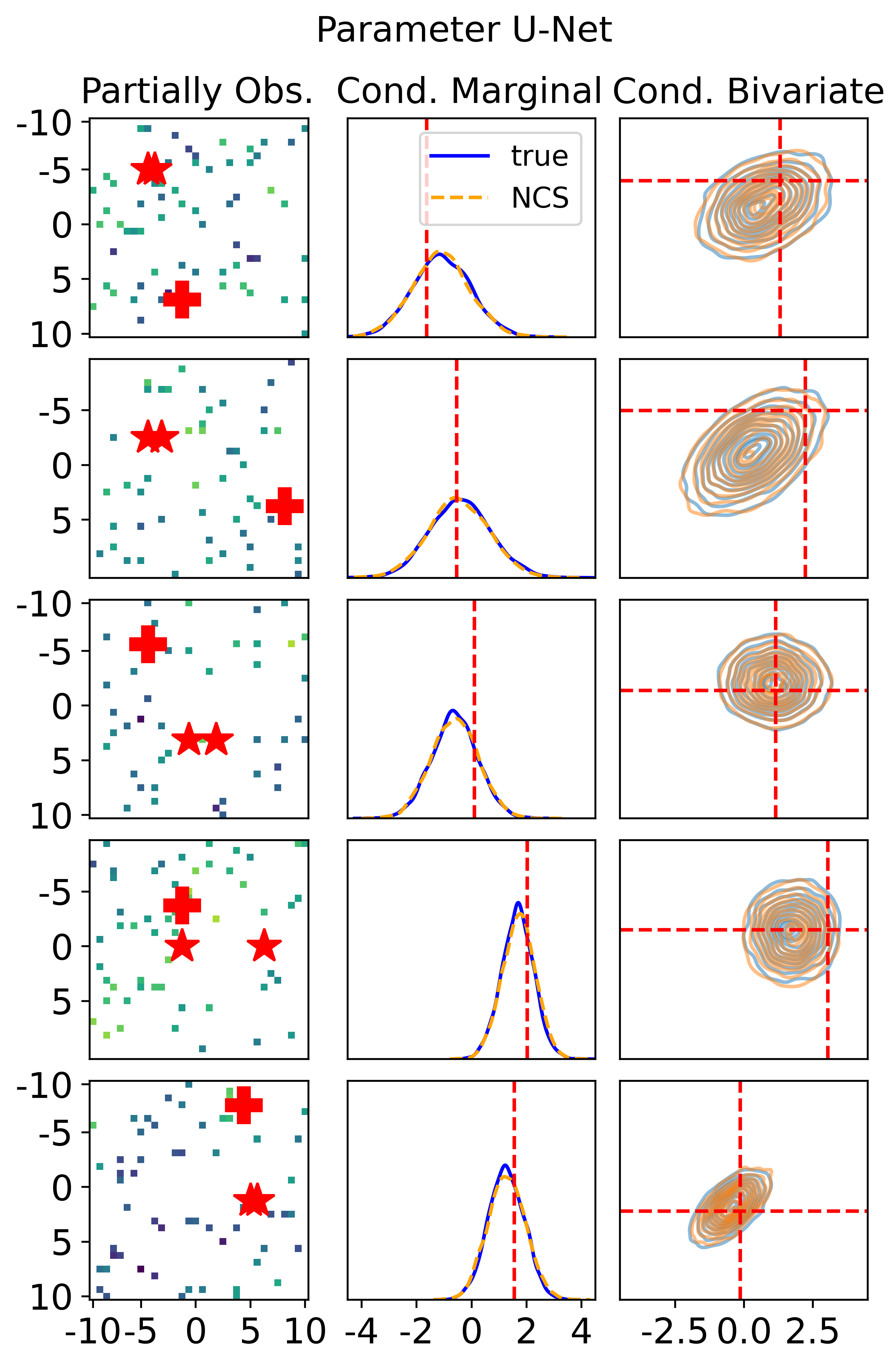}
    \includegraphics[scale = .085]{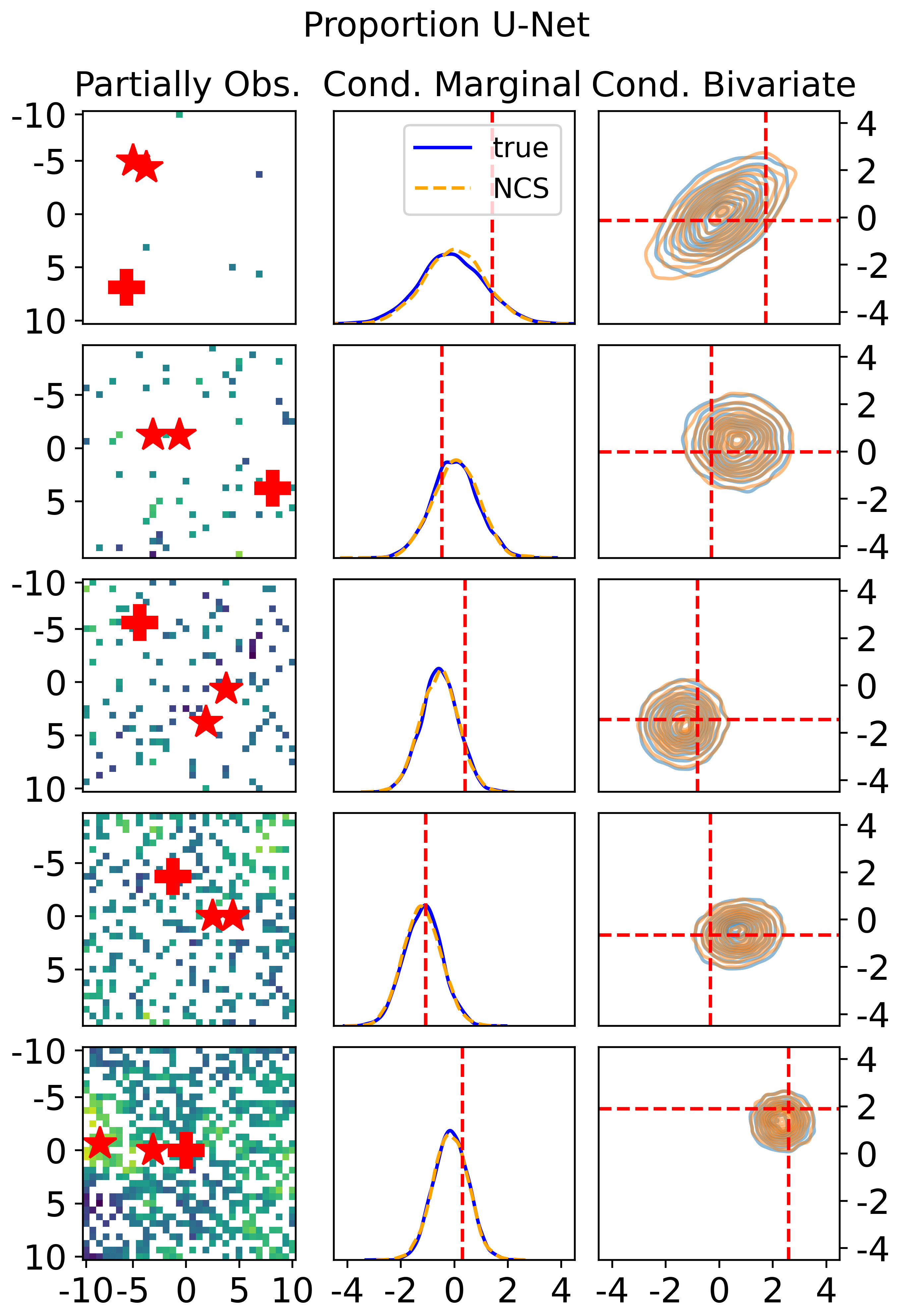}
    \caption{Left panel: Predictive densities of a Gaussian process with variance $\tau^2 = 1.5$, length scale $\ell \in \{1,2,3,4,5\}$ (from top to bottom), and observed proportion $\rho = 0.05$. Right panel: As in left panel, but with $\tau^{2} = 3$, and with observed proportions $\rho \in \{0.01,0.05,0.1,0.25,0.5\}$ (from top to bottom). For each panel: Left columns: Observations and prediction locations for which univariate densities (at the red crosses) and bivariate densities (across the red stars) are visualized. Middle columns: Exact (blue) and NCS (orange) empirical conditional univariate densities at the locations marked with crosses. Right columns:  Exact (blue) and NCS (orange) empirical conditional bivariate densities across the locations marked with stars.}
\label{fig:gpdensity}
\end{figure}

\paragraph*{Conditional Marginal and Bivariate Densities}
The true predictive distributions of a Gaussian process are fully determined by its first and second moments. As such, if all the true and NCS conditional univariate and bivariate densities are equivalent up to some small approximation error due to finite sample estimation, then the NCS simulations are representative of the true predictive distributions. In Figure~\ref{fig:gpdensity}, the true and NCS conditional univariate and bivariate densities appear equivalent up to approximation error. While Figure~\ref{fig:gpdensity} only shows densities for a select number of spatial locations and partially observed spatial fields, we compared conditional univariate and bivariate densities for several other spatial locations and observed similar behavior. Thus, we conclude that the first and second moments of the NCS and true predictive distributions are approximately equivalent, and that the NCS simulations are representative of the true predictive distributions.

\begin{figure}[t!]
    \centering
    \includegraphics[scale = .077]{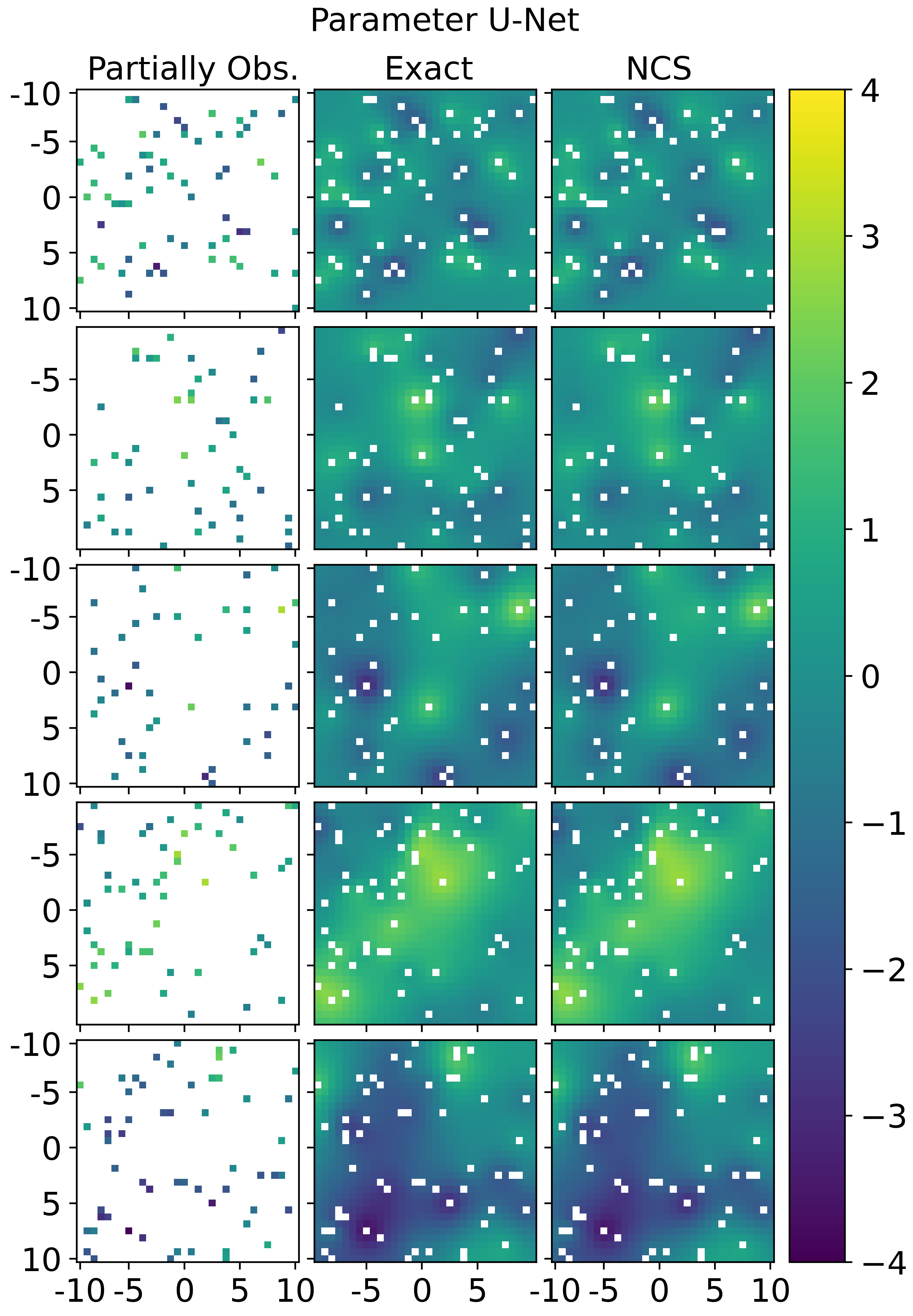}
    \includegraphics[scale = .077]{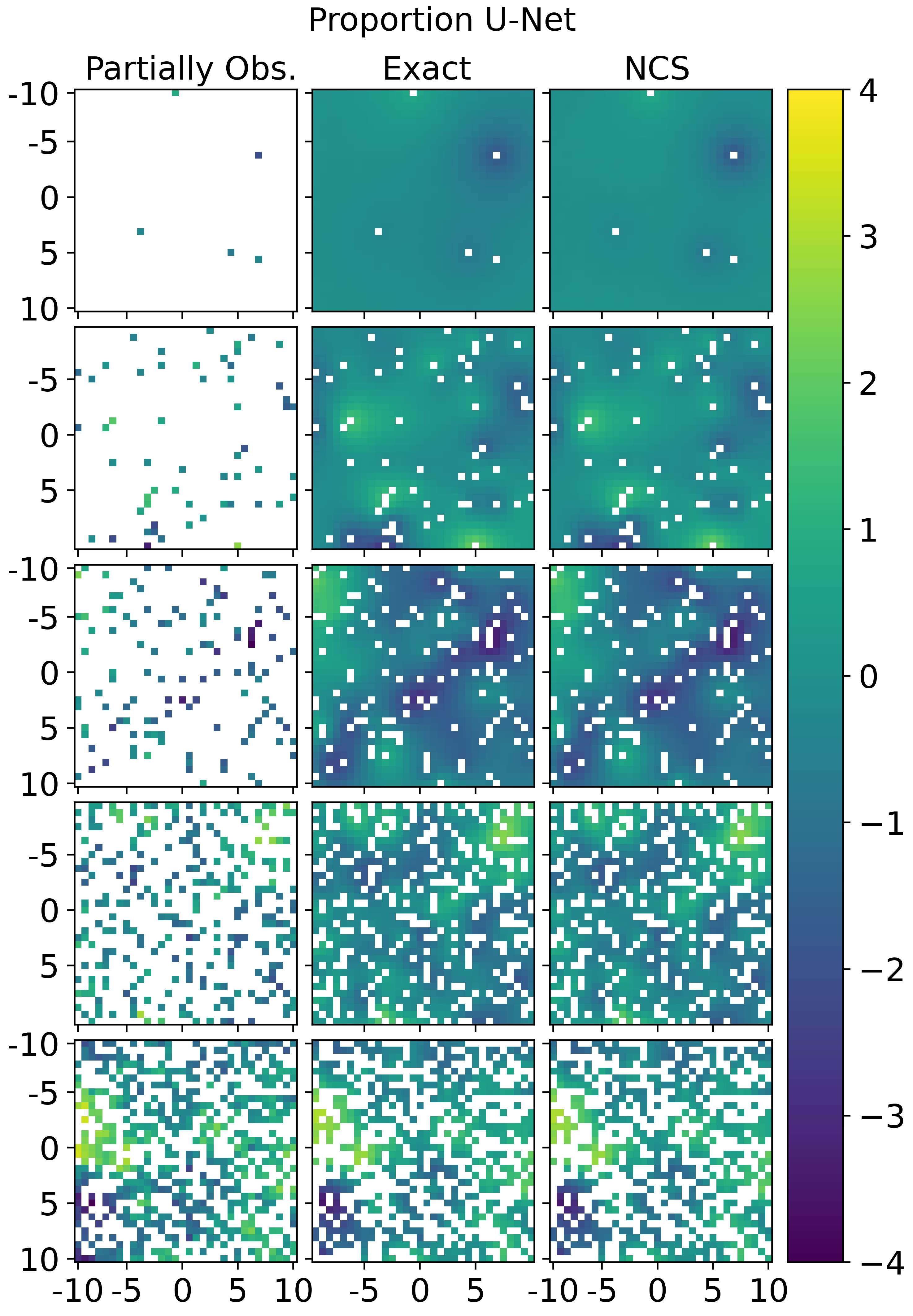}
    \caption{Left panel: NCS with a Gaussian process with variance $\tau^{2} = 1.5$, length scale  ${\ell \in \{1,2,3,4,5\}}$ (from top to bottom), and observed proportions $\rho = 0.05$. Right panel: NCS with a Gaussian process with variance $\tau^{2} = 1.5$, length scale $\ell = 3$, and observed proportions ${\rho \in \{0.01,0.05,0.1,0.25,0.5\}}$ (from top to bottom). For each panel: Left columns: Observations. Middle columns: True empirical conditional mean field. Right columns: empirical NCS conditional mean field using a parameter U-Net (left panel) or a proportion U-Net (right panel).}
    \label{fig:gpconditionalmean}
\end{figure}

\begin{figure}[t!]
    \centering
    \includegraphics[scale = .11]{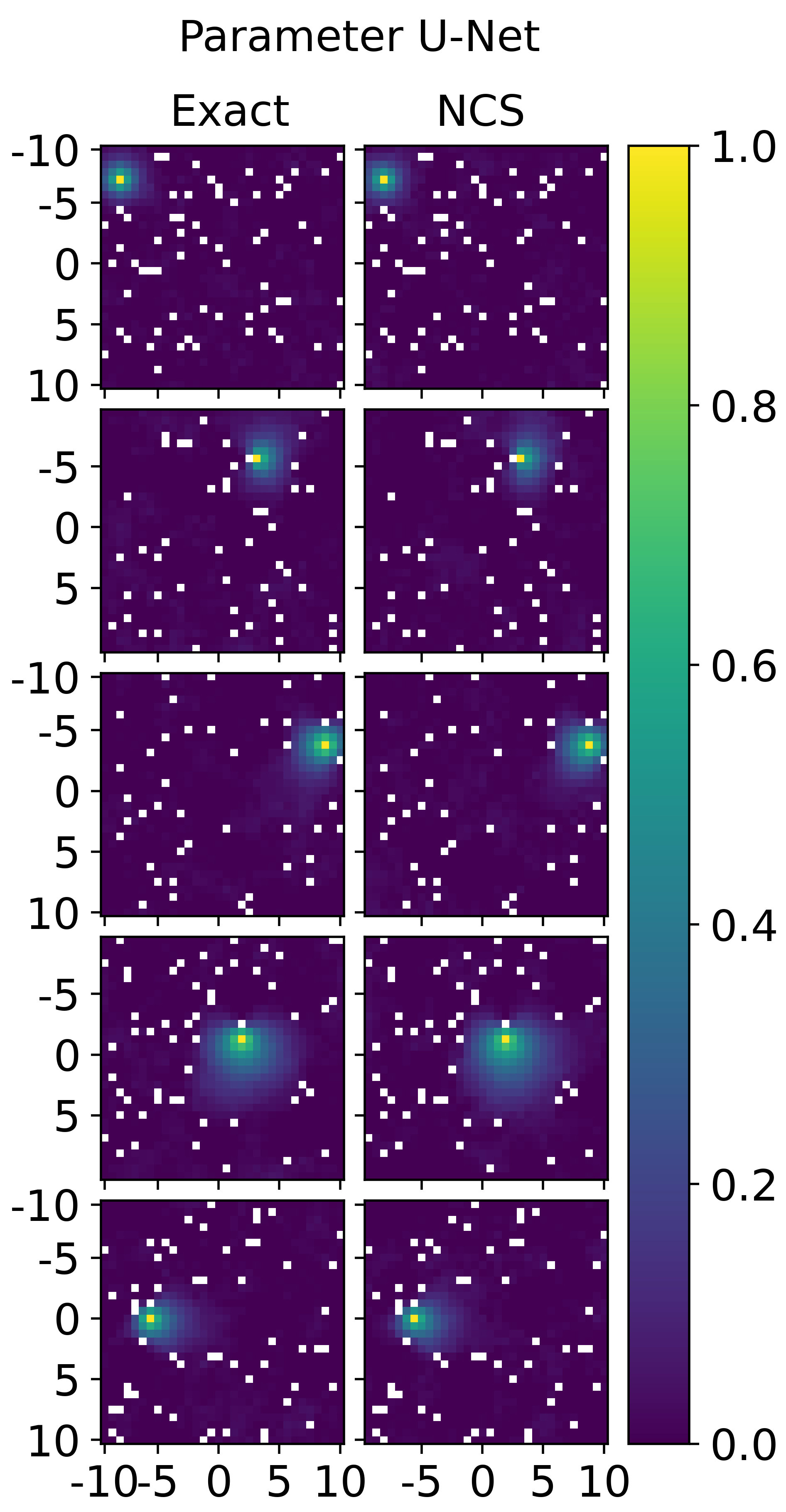}
    \includegraphics[scale = .11]{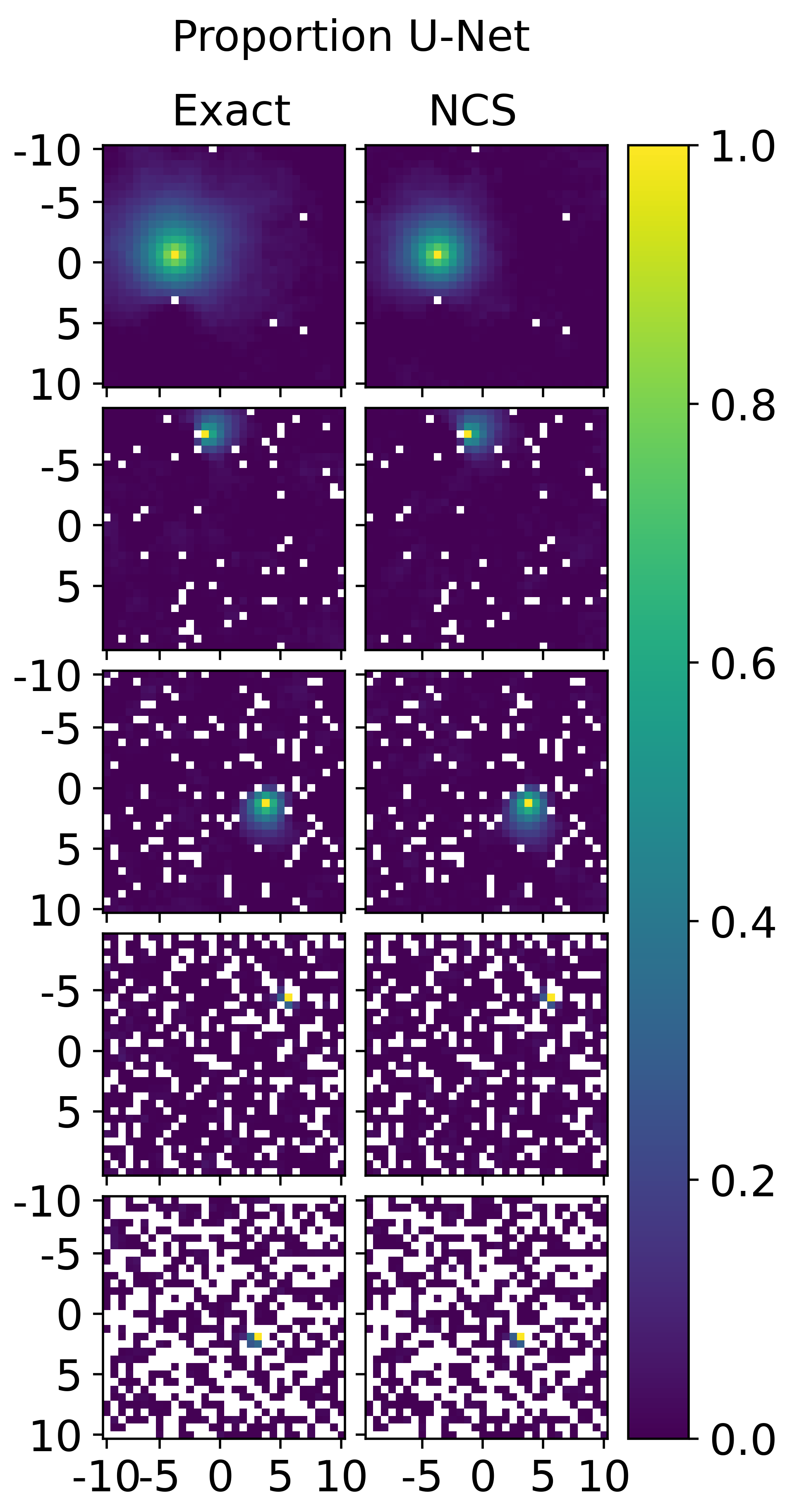}
    \caption{True (left) and NCS (right) empirical conditional correlation heatmaps for location of interest (in yellow) for a Gaussian process with $\tau^{2} = 1.5$ and $\ell = 1,2,3,4,5$ (top to bottom, left panel) and $\ell = 3$ and observed proportions $\rho = .01,.05,.1,.25,.5$ (top to bottom, right panel).}
    \label{fig:gpcorrelationheatmap}
\end{figure}

\paragraph*{Conditional Mean and Correlation Metrics}
Figure~\ref{fig:gpconditionalmean} and Figure~\ref{fig:gpcorrelationheatmap} contain the true and NCS conditional mean fields and correlation heatmaps, respectively. These serve as additional diagnostics to assess whether---and, if so, how---the first and second moments of the NCS and true predictive distributions differ. In both figures, there are little to no visual differences between the true and NCS empirical conditional mean fields and correlation heatmaps. For both the NCS and true predictive distributions in Figure~\ref{fig:gpcorrelationheatmap}, spatial (conditional) correlations generally increase with the length scale, unless there are observed locations (marked in white) near the spatial location of interest (marked in yellow). This behavior is best illustrated in the fourth row of the left panel; the conditional correlations sharply decrease above the spatial location of interest (in yellow) due to the presence of an observation two pixels above (in white). Yet, the conditional correlations are unaffected below the spatial location of interest due to the absence of observed data in that area. Such behavior is expected for the true predictive distributions; it is yet a striking indication how effective NCS is at correctly conditioning on the given partially observed spatial fields.

\begin{comment}
\paragraph{Summary of Results}
From this first case study, we demonstrated that NCS is effective at generating simulations which are representative of the true predictive distributions of a Gaussian process by simply examining metrics which evaluate the first and second moments of predictive distributions. For a Gaussian process, examining such metrics is sufficient because the true predictive distributions are tractable and fully determined by their first and second moments. Thus, we have validated our method for a spatial process with tractable predictive distributions and instilled greater confidence in our method when considering spatial processes with intractable predictive distributions.
\end{comment}

\subsection{Brown--Resnick: Intractable Case Study}
\label{supp:br}
\subsubsection{Details on Standard Approximation for the Brown--Resnick Process}
\label{supp:brapprox}
\paragraph*{Further Description}

For both FCS and LCS, we utilize the function \texttt{condrmaxstab} in the SpatialExtremes package to approximate the predictive distributions \citep{SpatialExtremes}. This function applies a three-step procedure in which the first step involves identifying feasible partitions of the $k$ observed locations based on whether they share common extremal functions \citep{Dombry2012}. The extremal functions $\varphi_{\bm{s}_{i}}^{+}$ of a max-stable realization $\bm{z}$ are functions $\eta_{i} W_{i}(\bm{s})$ from the stochastic representation of \citet{deHaan1985} in \eqref{eqn:maxstable} over the set of spatial locations $\mathcal{S}$ for which the realization $\bm{z}$ and the extremal function $\varphi_{\bm{s}_{i}}^{+}$ evaluated at location $\bm{s}_{i}$ are equal. Extremal functions are not necessarily unique for all spatial locations $\bm{s}_{i}\in \mathcal{S}$ because an extremal function $\varphi_{\bm{s}_{i}}^{+}$ evaluated at another location $\bm{s}_{j}$ may equal the realization $\bm{z}$ at location $\bm{s}_{j}$ as well. As such, this first step requires an approximation via a MCMC algorithm, specifically a Gibbs sampler, to achieve computational tractability because the number of all possible partitions for $k$ locations is the Bell number $B_{k}$ which grows more than exponentially with respect to $k$ \citep{Castruccio}. The second and third steps involve simulating extremal and sub-extremal functions respectively from tractable, closed-form distributions and as such, are less computationally intensive. See \citet{Dombry2012} for full mathematical details and \citet{Huser2019} for another application and assessment of the above Gibbs sampler.

As shown in Figure~\ref{fig:brfcsvisualization} below, FCS suffers from negative bias for unobserved locations far away from the spatial domain center when the range parameter is small. This is probably due to the insufficient number of subextremal functions drawn by the \texttt{condrmaxstab} function, which affects mostly locations far away from the observed locations when the variogram increases to infinity.  

\paragraph*{Implementation}
To implement FCS and LCS, we use the \texttt{condrmaxstab} function with the given set of observed locations (FCS) or the nearest seven neighbors to the select unobserved location or locations (LCS), the nugget set to $0.00001$, the burn-in period set to 1000, and the thinning factor set to $100$. Here, a nugget of $0.00001$ adds an insignificant amount of noise to the Brown--Resnick unconditional simulations used during the third step, but helps address computational problems which occur in the absence of a nugget.

We use univariate and bivariate LCS to produce conditional univariate and bivariate densities and conditional mean fields. In the univariate case, for LCS, we condition on the seven observations closest to the location of interest. In the bivariate case, we first identify the three nearest neighbors for each of the two locations of interest, and then randomly sample from the remaining unique seven nearest neighbors for each of the two select locations until a conditioning set of size seven is obtained. This process ensures there are neighbors for each of the two locations of interest in the conditioning set for bivariate LCS.

\subsubsection{Training Details}
\label{supp:brtrain}
The major difference between the Gaussian and Brown--Resnick case studies is the log-transformation of the training data simulated from the Brown--Resnick process. As such, NCS produces conditional simulations on the (log-transformed) Gumbel scale, and all results are on the Gumbel scale.

\paragraph*{Parameter U-Net} The details for simulating training are the same as for the Parameter U-Net in the Gaussian process case study (Section~\ref{supp:gptrain}) except that the number of proportion values was set to $r = 200$ for the training data.

\paragraph*{Proportion U-Net} The details for simulating training data are the same as for the Proportion U-Net in the Gaussian process case study.

\paragraph*{Small Conditioning Set U-Net}

We train U-Nets for the small conditioning sets via the following data generation process: We first fix the parameter $\bm{\theta}=(\lambda,\nu)'$ and select a minimum $o_{L}=1$ and maximum $o_{U}=10$ number of observations for the conditioning sets according to the experimental settings. For each data draw, we iterate through each number of observed locations $o_{j}$ starting from the minimum $o_{L}$ and ending with the maximum $o_{U}$. For each observed number $o_{j}$, we simulate $s$ full spatial field realizations via the unconditional simulator with fixed parameter $\bm{\theta}$. For each of these $s$ full spatial fields, we construct $m$ masks and simulate $m$ discrete timesteps uniformly from $\{0,1,\dots,T\}$. Each mask is constructed from a conditioning set $\mathring{\mathcal{S}}$ containing $o_j$ locations uniformly sampled without replacement from $\mathcal{S}$. Algorithm~\ref{algo:trainingdatasmall} illustrates this data process, and the resulting data have the form shown in \eqref{eqn:dataform}.

\spacingset{1}
\begin{algorithm}[t!]
\caption{Training and Validation Data for Small Conditioning Set U-Net}
\begin{algorithmic}
\STATE $\mathcal{D}=\emptyset$
\STATE Fix parameter $\bm{\theta}=(\lambda, \nu)'$.
\STATE Select a minimum and maximum number of observed locations, $o_{L}<o_{U}\in \mathbb{N}^{+}$.
\FOR{$o_{j}$ in $o_{L},\ldots,o_{U}$}
\STATE Simulate unconditionally $\{\bm{x}_{0,k}\}_{k \in [s]} \sim p(\cdot \mid \bm{\theta})$
\FOR{$\bm{x}_{0,k}\in \{\bm{x}_{0,k}\}_{k \in [s]}$}
\STATE $\mathring{\mathcal{S}}_{l}=\{\bm{s}_{l_{i}}\in \mathcal{S} \mid \bm{s}_{l_{i}}\sim \textrm{U}(\mathcal{S}), i \in [o_{j}]\}$
\STATE $t_{l}\sim U(\{0,1,\dots,T\}) \textrm{ for } l\in [m]$
\STATE $\bm{\mathring{x}}_{0,l}=(x_{0,k,j}: M_{j}(\mathring{\mathcal{S}}_{l})=1)' \textrm{ for } l \in [m]$
\STATE $\tilde{\bm{x}}_{0,l}=(x_{0,k,j}: M_{j}(\mathring{\mathcal{S}}_{l})=0)' \textrm{ for } l \in [m]$
\STATE $\tilde{\bm{x}}_{t_{l}}\sim p(\cdot \mid \bm{\mathring{x}}_{0,l}, \tilde{\bm{x}}_{0,l}):=\mathcal{N}(\sqrt{\underline{\alpha}_{t}}\tilde{\bm{x}}_{0,l}, \underline{\sigma}_{t}^{2}\tilde{\bm{I}}) \textrm{ for } l\in [m]$
\STATE $\mathcal{D} \leftarrow \mathcal{D} \cup \{\{ ({\bm{\mathring{x}}_{0,l}}',{\tilde{\bm{x}}_{t_{l}}}')',\bm{M}(\mathring{\mathcal{S}}_{l}),\bm{\theta}_{j},t_{l}\} : l \in [m]\}$
\ENDFOR
\ENDFOR
\STATE Return $\mathcal{D}$
\end{algorithmic}
\label{algo:trainingdatasmall}
\end{algorithm}
\spacingset{2}

Specifically, we train one U-Net for each specific range parameter ${\lambda \in \{1,2,3,4,5\}}$. However, we fix the smoothness to $\nu = 1.5$ and the minimum and maximum number of observations to one and ten respectively for all U-Nets. For validation, the number of observed locations does not exceed seven because FCS as implemented in the SpatialExtremes R package is fragile beyond seven. To simulate training and validation data, we follow the process described in Section~\ref{supp:train}. For simulating training data, the minimum and maximum number of observations were set to $o_{L}=1$ and $o_{U}=10$, $s=32$ full spatial field realizations were generated per number of observations between $o_{L}$ and $o_{U}$, and $m=100$ conditioning sets, masks and timesteps were simulated for each full spatial field realization.

\subsubsection{Validation Details}
\label{supp:brvalidationdetails}

\subparagraph*{Conditional Validation Dataset (Small Conditioning Set)}

For each range value $\lambda \in \{1,2,3,4,5\}$ and each conditioning set size $m$ ranging from one to seven, we unconditionally simulated a reference spatial field $\bm{x}_{0}$ from a Brown--Resnick process with the given range $\lambda$ and fixed smoothness $\nu = 1.5$. For each reference spatial field, we generated a conditioning set $\mathring{\mathcal{S}}$ and corresponding mask $\bm{M}(\mathring{\mathcal{S}})$ by uniformly sampling from $\mathcal{S}$ without replacement, such that $\abs{\mathring{\mathcal{S}}}=m$. Using each of the $35$ partially observed fields $\bm{\mathring{x}}_{0}=(x_{0,j}: M_{j}(\mathring{\mathcal{S}})=1)'$, we generated $m=4000$ full spatial field replicates $\{(\bm{\mathring{x}}_{0}',\tilde{\bm{x}}_{0,i}')'\}_{i\in [m]}$ with NCS and FCS (an MCMC approximation described in Section~\ref{sec:casestudymethodology}). We use this dataset in Section~\ref{sec:results_small} to validate NCS when the number of observed locations does not exceed seven.

\paragraph*{Unconditional Validation Dataset (Small Conditioning Set U-Net)}
For the validation method described in Section~\ref{sec:validation}, we need to generate unconditional samples from the conditional ones. As discussed in Section~\ref{sec:validation}, we do this by generating unconditionally at observed locations and then conditionally simulating at unobserved locations; the resulting concatenated sample is then one from the unconditional distribution. Algorithm~\ref{algo:linhartmethodsmall} summarizes how we do this, using $m=4000$ NCS-approximated or FCS-approximated unconditional simulations using varying range ${\lambda \in \{1,2,3,4,5\}}$, fixed smoothness $\nu = 1.5$, and conditioning sets with varying sizes ranging from one to seven.

\spacingset{1}
\begin{algorithm}[t!]
\caption{Generating NCS-approximated Unconditional Validation Dataset (Small Conditioning set)}
\begin{algorithmic}
\STATE Fix parameter $\bm{\theta}=(\lambda, \nu)'$ and number of observed locations $o_{j}$ and set $\mathcal{D}_{\textrm{val}}=\emptyset$
\FOR{$i$ in $1,\dots, m$}
\STATE $\bm{x}_{0,i}\sim p(\cdot \mid \bm{\theta})$
\STATE $\mathring{\mathcal{S}}_{i}\sim \textrm{U}(\bm{\mathcal{S}}_{o_{j}}) \textrm{ where } \bm{\mathcal{S}}_{o_{j}}=\{\mathcal{S}_{i}\subset \mathcal{S} : \abs{\mathcal{S}_{i}} = o_{j}\}$
\STATE $\tilde{\bm{x}}_{0,i} \sim \tilde{p}_{0}(\cdot \mid \bm{\mathring{x}}_{0,i}, \bm{\theta}) \textrm{ for } \bm{\mathring{x}}_{0,i} = (x_{0,i,j}: M_{j}(\mathring{\mathcal{S}}_{i})=1)' \textrm{ and NCS/FCS distr.\ } \tilde{p}_{0}(\cdot \mid \bm{\mathring{x}}_{0,i}, \bm{\theta})$
\STATE $\mathcal{D}_{\textrm{val}} \leftarrow \mathcal{D}_{\textrm{val}} \cup \{(\bm{\mathring{x}}_{0,i}',\tilde{\bm{x}}_{0,i}')'\}$
\ENDFOR
\STATE Return $\mathcal{D}_{\textrm{val}} = \{(\bm{\mathring{x}}_{0,i}',\tilde{\bm{x}}_{0,i}')'\}_{i \in [m]}$
\end{algorithmic}
\label{algo:linhartmethodsmall}
\end{algorithm}
\spacingset{2}

\paragraph*{Unconditional Validation Dataset (Parameter and Proportion U-Net)} We apply the process described in Section~\ref{sec:validation} and the previous paragraph but with varying parameters and observed proportions as shown in Algorithm~\ref{algo:linhartmethod}. We generate $m=4000$ NCS-approximated unconditional simulations using either varying range $\lambda \in \{1,2,3,4,5\}$, fixed smoothness $\nu = 1.5$, and fixed observed proportion $\rho=.05$ for the parameter U-Net, or varying observed proportion $\rho \in \{0.01,0.05,0.1,0.25,0.5\}$ with fixed range $\lambda = 3$ and smoothness $\nu = 1.5$ for the proportion U-Net.

\spacingset{1}
\begin{algorithm}[t!]
\caption{Generating NCS-approximated Unconditional Validation Dataset (Parameter and Proportion U-Nets)}
\begin{algorithmic}
\STATE Fix parameter $\bm{\theta}=(\lambda, \nu)'$ and observed proportion $\rho$ and set $\mathcal{D}_{\textrm{val}}=\emptyset$
\STATE Set the number of NCS-approximated unconditional replicates $m$.
\FOR{$i$ in $1,\dots, m$}
\STATE $\bm{x}_{0,i}\sim p(\cdot \mid \bm{\theta})$
\STATE $\bm{M}(\mathring{\mathcal{S}}_{i}) \sim \textrm{Bin}(n,\rho)$ with $n=32^{2}$
\STATE $\tilde{\bm{x}}_{0,i} \sim \tilde{p}_{0}(\cdot \mid \bm{\mathring{x}}_{0,i}, \bm{\theta}) \textrm{ for } \bm{\mathring{x}}_{0,i} = (x_{0i,j}: M_{j}(\mathring{\mathcal{S}}_{i})=1)'$ and NCS/LCS distr.\ $\tilde{p}_{0}(\cdot \mid \bm{\mathring{x}}_{0,i}, \bm{\theta})$
\STATE $\mathcal{D}_{\textrm{val}} \leftarrow \mathcal{D}_{\textrm{val}} \cup \{(\bm{\mathring{x}}_{0,i}',\tilde{\bm{x}}_{0,i}')'\}$
\ENDFOR
\STATE Return $\mathcal{D}_{\textrm{val}} = \{(\bm{\mathring{x}}_{0,i}',\tilde{\bm{x}}_{0,i}')'\}_{i \in [m]}$
\end{algorithmic}
\label{algo:linhartmethod}
\end{algorithm}
\spacingset{2}

\paragraph*{Multivariate Metrics}
The metrics in Section~\ref{supp:validationmetricsboth} such as the conditional univariate and bivariate densities, conditional mean fields, and conditional correlation metrics only compare low-dimensional properties of NCS to those of the true predictive distributions or standard approximations. Here, we discuss metrics that implicitly validate the high-dimensional patterns of NCS-approximated predictive distributions via NCS-approximated unconditional simulations. These metrics, referred to as multivariate metrics, are evaluated using the aforementioned unconditional validation datasets. For the Brown--Resnick case study, where these multivariate metrics are especially relevant, we considered the extremal correlation function and the spatial minimum, maximum, and absolute summation in any given full realization. Distributions of these quantities are shown in our results (e.g., Figure~\ref{fig:brextremalcoefficient}) for different parameters and missingness cases.

\subsubsection{Results}
\label{supp:brvalidation}
\subparagraph*{Visualizations (Small Conditioning Set)}

\begin{figure}[H]
\centering
\includegraphics[scale = .41]{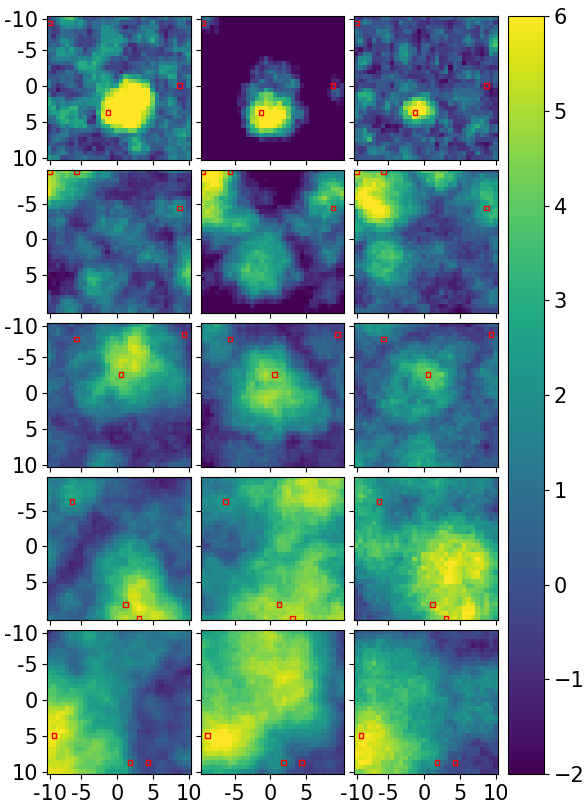}
\caption{Fully observed (left) fields and FCS (middle) and NCS (right) simulations of a Brown--Resnick Process with smoothness $\nu = 1.5$ and range $\ell = 1,2,3,4,5$ (top to bottom) on the Gumbel scale for three observed locations shown in red. Values outside $[-2,6]$ have been replaced with the closest value in this interval.}
\label{fig:brfcsvisualization}
\end{figure}

In Figure~\ref{fig:brfcsvisualization}, we show results for both FCS (middle) and NCS (right) simulations for the case when three locations are observed. Clearly, FCS suffers from negative bias for unobserved locations far away from the spatial domain center when the range parameter is small. This is due to the way MCMC is implemented for this approximation method. Due to this negative bias, the NCS simulations appear more representative overall, although NCS suffers from slightly increased graininess as the range parameter increases.

\begin{figure}[t!]
 \centering
    \includegraphics[scale = .077]{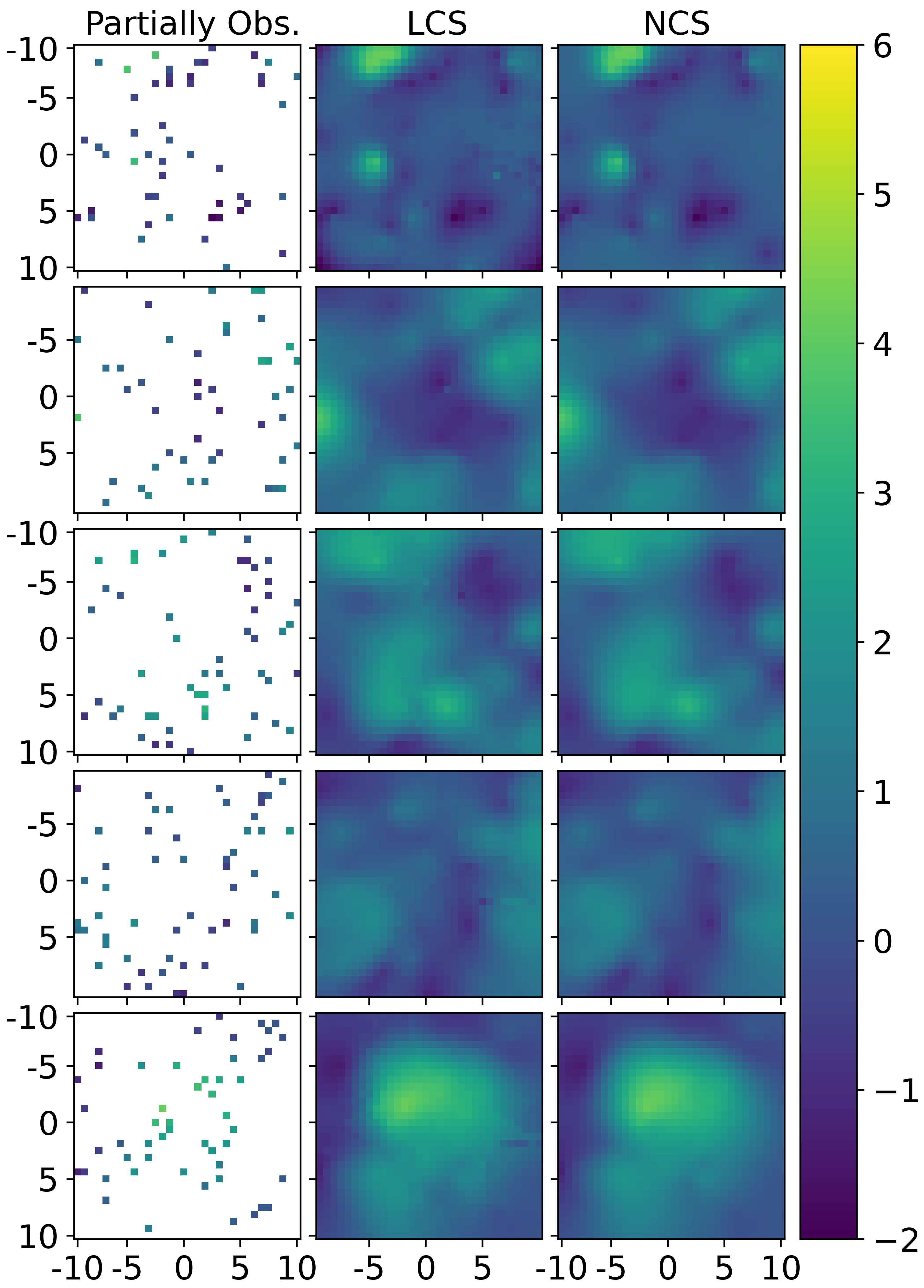}
    \includegraphics[scale = .077]{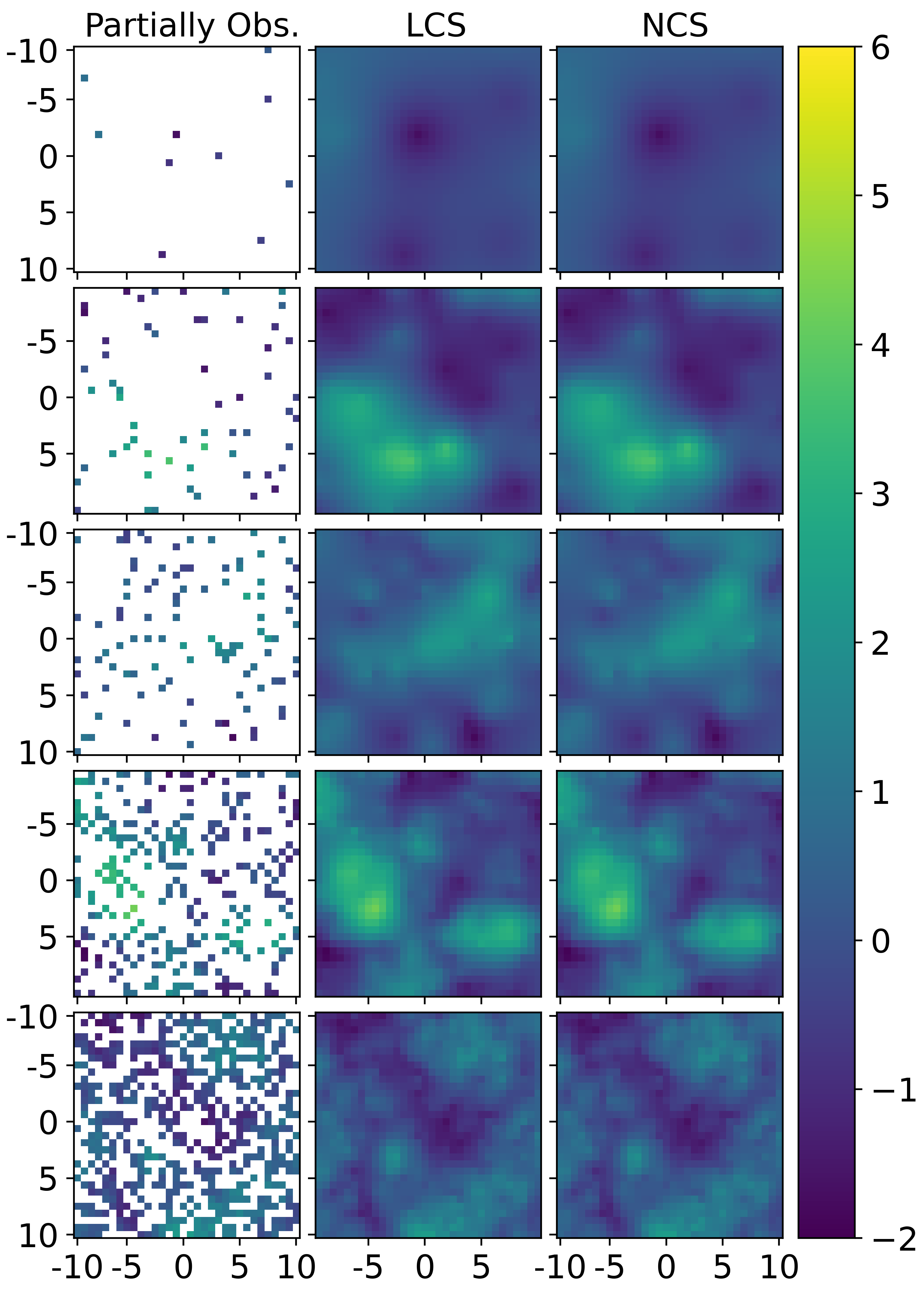}
    \caption{Partially observed spatial fields (left) and NCS (right) empirical conditional mean fields of a Brown--Resnick process with smoothness $\nu = 1.5$ and range $\lambda = 1,2,3,4,5$ (top to bottom) and observed proportions $\rho = .05$ (left panel) and $\lambda = 3$ and observed proportions $\rho = .01,.05,.1,.25,.5$ (top to bottom, right panel) on the Gumbel scale. Values outside $[-2,6]$ have been replaced with the closest value in this interval.}
\label{fig:brconditionalmean}
\end{figure}
\subparagraph*{Conditional Mean Fields (Small Conditioning Set)} In Figure~\ref{fig:brconditionalmean}, we show results for various spatial process parameter and proportion settings. We see no discernible differences between the univariate LCS and NCS empirical conditional mean fields.

\subsection{Data Application}
\label{supp:dasupp}

\subsubsection{Parameter Estimation using a Pairwise Likelihood}\label{sec:comp_lik}

We compute pairwise likelihood on a grid of range and smoothness parameters using all $31$ realizations at each location, resulting in a single gridded pairwise likelihood surface for each spatial location. We select the range and smoothness parameters which maximize the surface as the parameter estimate for each of the three locations. See Algorithm~\ref{algo:pwl} for details.

As mentioned in Section~\ref{sec:da}, the pairwise likelihood has a tuning parameter $\delta$ determining which pairs of locations are used for the bivariate likelihood contributions. Since the maximum distance between any two locations in the $2\degree \times 2 \degree$ window is approximately $2.8\degree$, we considered pairwise likelihood estimates for a range of cut-off distances, $\delta = 0.5\degree,1\degree,2\degree$. To choose $\delta$,  we conducted a simulation experiment where  we generate Brown--Resnick simulations with fixed parameter values and, for each value of $\delta$, compare the resulting parameter estimates to the true parameters via mean squared error. We then selected the cut-off distance that minimizes this metric. Based on this procedure, we choose $\delta = 0.5\degree$.

% \begin{figure}[t!]
% \centering
%     \includegraphics[scale = .3]{Figures/RS/results/RS_locations_0}
% \caption{Fully observed annual maxima residuals (Gumbel scale) for the three locations shown on the map (left subfigure) for 1985.}
% \label{fig:rswindow}
% \end{figure}

\spacingset{1}
\begin{algorithm}
\caption{Computing Pairwise Likelihood Estimates}
\begin{algorithmic}
\STATE Let $\Theta^{L}$ be a $40\times 40$ grid over the parameter space $ (0,2)\times (0,2)$ of the range and smoothness parameters of the Brown--Resnick model.
\STATE Set cut-off distance $\delta = 0.5\degree$.
\FOR{$j$ in $1:3$}
\STATE Let ${\bm z}_{j,t}$ be the annual maxima residuals in the $j$th $2\degree \times 2\degree$ window in year $t$.
\STATE Compute $\mathcal{L}_{\textrm{pwl}}(\boldsymbol{\theta} ; \{{\bm z}_{j,t}\}_{t\in[31]})$ for $\boldsymbol{\theta} \in \Theta^L$, with $\delta = 0.5\degree$.
\STATE Compute the estimate $\hat{\boldsymbol{\theta}}_{\textrm{pwl},j}=\argmax_{\boldsymbol{\theta} \in \Theta^{L}} \mathcal{L}_{\textrm{pwl}}(\boldsymbol{\theta} ; \{{\bm z}_{j,t}\}_{t\in [31]})$
\ENDFOR
\end{algorithmic}
\label{algo:pwl}
\end{algorithm}
\spacingset{2}

\subsubsection{NCS Simulation}\label{sec:data_NCS}
The key differences in NCS between the simulation studies in Sections~\ref{sec:casestudies},~\ref{supp:GPcasestudy}, and~\ref{supp:br}, and this data application are the incorporation of coastal boundaries, the smaller spatial domain and grid size ($2\degree \times 2\degree$ and $20\times 20$ versus $20\degree \times 20\degree$ and $32\times 32$), and amortization with respect to both range and smoothness as well as proportion of observed locations. 

\paragraph*{NCS with Coastal Boundaries}
During training and simulation, we combine two masks: (i) a simulation $\bm{M}_{1}\in \{0,1\}^{20\times 20}$ from a binomial distribution with $n=20^{2}$ locations and probability-of-presence parameter $p$,  and (ii) a simulation $M_{2} \in \{0,1\}^{20\times 20}$ of sea- and land-based locations in a $2\degree \times 2\degree$ spatial window. These masks are combined via a Hadamard product,
\begin{equation}
\bm{M} =\bm{M}_{1}\odot \bm{M}_{2}, \quad \bm{M}_{1}\sim \text{Bin}(n=400,p), \quad \bm{M}_{2} \sim \text{Unif}(\mathcal{M}_{\textrm{RS}}),
\label{eq:maskboundary}
\end{equation}
where $\mathcal{M}_{\textrm{RS}}$ is the set of masks representing sea- and land-based locations for each $2\degree \times 2\degree$ spatial window centered on the grid centroids of the $0.1\degree \times 0.1\degree$ gridding of the Red Sea.

\paragraph*{Parameter and Proportion Amortization}

As in Section~\ref{sec:brvalidation}, we amortize with respect to both the parameters and the missingness proportion. We therefore use three input channels---one for the spatial field and one for each of the two parameters. The channels representing each parameter also incorporate the mask indicating which locations are unobserved or land-based (i.e. the two input channels are $\theta_{1}\bm{M}$ and $\theta_{2}\bm{M}$), mirroring the process described in Section~\ref{sec:brvalidation}. As we are only generating local NCS simulations at three different locations in the Red Sea, we only need to train the U-Nets in a confined parameter space centered on the pairwise likelihood parameter estimates of those three regions. We hence train two U-Nets on the parameter spaces $\Theta_{1}=[0.5,1.0]\times [1.0,1.5]$ and $\Theta_{2}=[1.0,1.5]\times [1.0,1.5]$ and for the range of proportions $[.01,0.1]$. 

\paragraph*{Training Details}
We first simulated five proportions from $[0.01,0.1]$ and, for each of these, we simulated 100 parameters from the relevant subset of the parameter space $\Theta_1 \times \Theta_2$ via Latin Hypercube sampling. For each of those 100 parameters, we simulated 32 full spatial field realizations via the unconditional simulator. Finally, we simulated 200 masks for each full spatial field realization, where each mask is simulated according to \eqref{eq:maskboundary} and the corresponding missingness proportion. We refer to the simulated data as a ``data draw''---we used five data draws and trained the network using $10$ epochs per data draw with a mini-batch size of $512$.

\subsubsection{Results}

\begin{figure}[t!]
 \centering
  \begin{minipage}[c]{0.35\textwidth}
  \centering
    \includegraphics[width=\linewidth]{Figures/RS/results/locations_map}
  \end{minipage}
  \hfill
  \begin{minipage}[c]{0.6\textwidth}
  \centering
    \includegraphics[width=\linewidth]{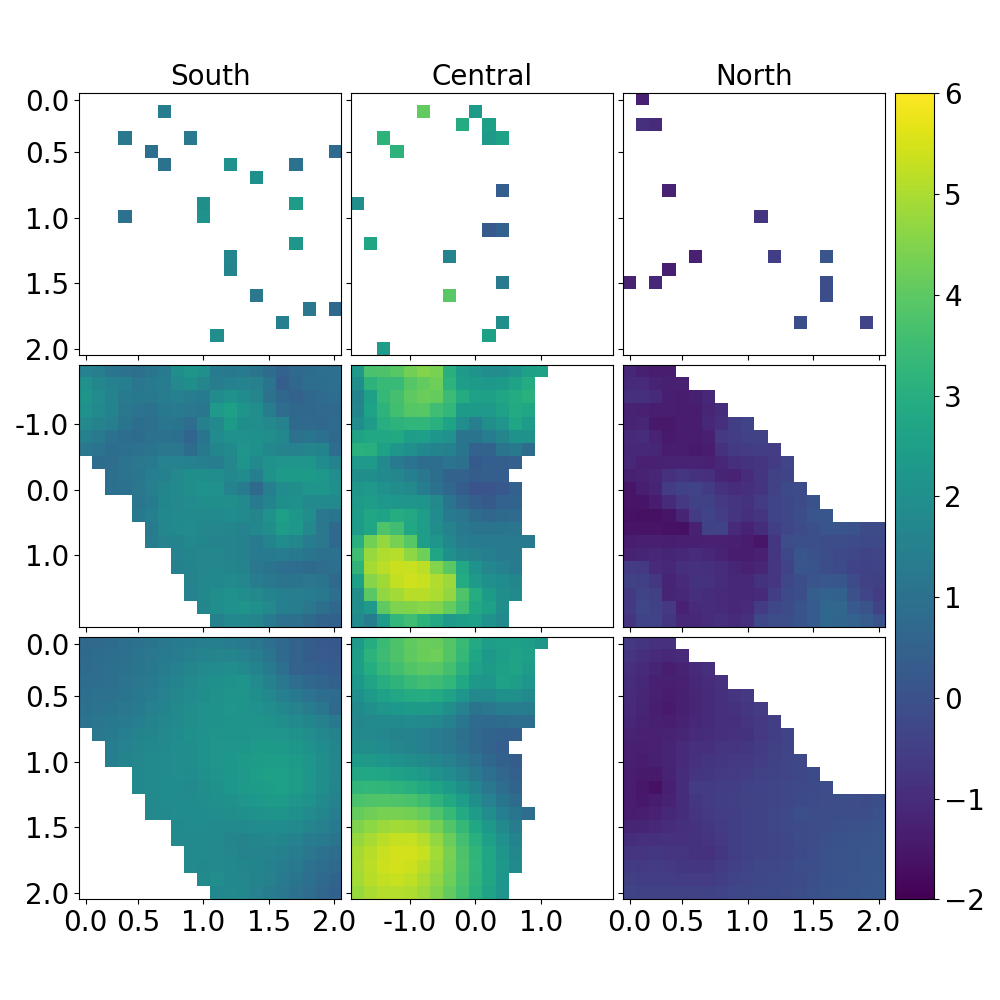}
  \end{minipage}
  \caption{Partially observed (right-top), fully observed (right-middle), and NCS conditional mean fields of (right-bottom) annual maxima residuals in the year 1985 on the Gumbel scale at three locations in the Red Sea (left panel) with pairwise estimates $\hat{\boldsymbol{\theta}}_{\textrm{pwl},j}=(0.8,1.4)$ (South), $\hat{\boldsymbol{\theta}}_{\textrm{pwl},j}=(0.45,1.55)$ (Central), and $\hat{\boldsymbol{\theta}}_{\textrm{pwl},j}=(0.55,1.25)$ (North). NCS is done using a Brown--Resnick process with parameter estimates $\hat{\bm{\theta}}$ estimated using a pairwise likelihood.}
  \label{fig:RSncsmean} 
\end{figure}

In Figure~\ref{fig:RSncsmean}, we show the empirical conditional mean fields (bottom row). First, the large-scale spatial patterns in the fully observed annual maxima residuals (middle row) are present in the NCS conditional mean fields. Secondly, the partially observed field is smoothly incorporated into the NCS conditional mean field. Thus, NCS can produce conditional mean fields, a standard spatial prediction tool, that are realistic.

\end{document}